\documentclass[acmsmall]{acmart}

\AtBeginDocument{%
  }

\setcopyright{cc}
\setcctype{by}
\acmDOI{10.1145/3798205}
\acmYear{2026}
\acmJournal{PACMPL}
\acmVolume{10}
\acmNumber{OOPSLA1}
\acmArticle{97}
\acmMonth{4}
\received{2025-09-22}
\received[accepted]{2026-02-17}

\usepackage{subcaption}
\usepackage{makecell}
\usepackage{enumitem}
\usepackage{multirow}
\usepackage{xcolor}
\usepackage{color}
\usepackage{colortbl}
\usepackage{cancel}
\usepackage{rotating}
\usepackage{pifont}
\usepackage{wrapfig}
\usepackage{floatflt}

\usepackage[most]{tcolorbox}
\usepackage{lipsum} 

\usepackage{listings}
\usepackage{algorithm}
\usepackage{algpseudocode}

\algdef{SE}[VARIABLES]{Variables}{EndVariables}
   {\algorithmicvariables}
   {\algorithmicend\ \algorithmicvariables}
\algnewcommand{\algorithmicvariables}{\textbf{global variables}}
\algnewcommand{\LineComment}[1]{\State \(\triangleright\) \textit{#1}}
\newcommand{\CommentIt}[1]{\Comment{\textit{#1}}}

\usepackage{tikz}
\newcommand*\circled[1]{\tikz[baseline=(char.base)]{
    \node[shape=circle,draw,inner sep=1pt, fill=black] (char) {\textcolor{white}{#1}};}}
\newcommand*\circledw[1]{\tikz[baseline=(char.base)]{
    \node[shape=circle,draw,inner sep=1pt, fill=white] (char) {{#1}};}}

\newcommand{\overallbudget}{120}

\newcommand{\yes}{\ding{55}}
\newcommand{\no}{\ding{51}}
\newcommand{\artifacturl}{\url{https://zenodo.org/records/18795126}}
\newcommand{\rangeline}{--}

\newcommand{\techname}{\textsc{PoCo}}

\newcommand{\ntarget}{eight}
\newcommand{\nseedsets}{seven}
\newcommand{\nbaseline}{seven}
\newcommand{\camphours}{24}
\newcommand{\nrepeat}{10}
\newcommand{\vga}{\hat{A}_{12}}
\newcommand{\aflppver}{version 4.10c}
\newcommand{\mg}{Magma}
\newcommand{\aflcmin}{\texttt{afl-cmin}}
\newcommand{\aflcc}{\texttt{afl-cc}}
\newcommand{\optimin}{\textsc{OptiMin}}
\newcommand{\cmin}{\textsc{Cmin}}

\newcommand{\techA}{\techname{}$_\text{A}$}
\newcommand{\cminpA}{\cminplus{}$_\text{A}$}
\newcommand{\tpstwoh}{$t_\mathcal{P}(|S|_\text{2h})$}
\newcommand{\techall}{\techname$_\text{A}$}

\newcommand{\rqtitle}[1]{RQ#1}
\newcommand{\rqline}[2]{\textbf{\rqtitle{#1}:} #2}
\newcommand{\rqbox}[1]{%
  \begin{tcolorbox}[
    colback=gray!10,
    colframe=gray!50,
    left=1mm,
    right=1mm,
    top=1mm,
    bottom=1mm,
    enhanced,
    sharp corners,
    boxrule=0pt,
    borderline west={2pt}{0pt}{gray!50},
    width=\textwidth
  ]
  #1
  \end{tcolorbox}
}

\newcommand{\figcap}[1]{Fig. #1}
\newcommand{\tabcap}[1]{Table #1}
\newcommand{\algcap}[1]{Algorithm #1}
\newcommand{\seccap}[1]{\S#1}

\newcommand{\equcap}[1]{Equation #1}
\newcommand{\linecap}[1]{line #1}
\newcommand{\linescap}[2]{lines #1\rangeline{}#2}

\newcommand{\lineno}[1]{(#1)}

\usepackage[normalem]{ulem}

\begin{document}

\title{Peeling Off the Cocoon: Unveiling Suppressed Golden Seeds for Mutational Greybox Fuzzing}

\author{Ruixiang Qian}
\email{qianrx@smail.nju.edu.cn}
\orcid{0009-0003-5040-3123}
\affiliation{%
  \institution{State Key Laboratory for Novel Software Technology}
  \city{Nanjing University}
  \country{China}
}

\author{Chunrong Fang}
\authornote{Chunrong Fang and Zhenyu Chen are the corresponding authors.}
\email{fangchunrong@nju.edu.cn}
\orcid{0000-0002-9930-7111}
\affiliation{%
  \institution{State Key Laboratory for Novel Software Technology}
  \city{Nanjing University}
  \country{China}
}

\author{Zengxu Chen}
\email{522025320022@smail.nju.edu.cn}
\orcid{0009-0003-7924-1295}
\affiliation{%
  \institution{State Key Laboratory for Novel Software Technology}
  \city{Nanjing University}
  \country{China}
}

\author{Youxin Fu}
\email{202005570318@smail.xtu.edu.cn}
\orcid{0009-0000-5048-7193}
\affiliation{%
  \institution{State Key Laboratory for Novel Software Technology}
  \city{Nanjing University}
  \country{China}
}

\author{Zhenyu Chen}
\email{zychen@nju.edu.cn}
\authornotemark[1]
\orcid{0000-0002-9592-7022}
\affiliation{%
  \institution{State Key Laboratory for Novel Software Technology}
  \city{Nanjing University}
  \country{China}
}

\begin{abstract}
Mutational greybox fuzzing (MGF) is a powerful software testing technique. 
Initial seeds are critical for MGF since they define the space of possible inputs and fundamentally shape the effectiveness of MGF.
Nevertheless, having more initial seeds is \textit{not} always better.
A bloated initial seed set can inhibit throughput, thereby degrading the effectiveness of MGF. 
To avoid bloating, modern fuzzing practices recommend performing \textit{seed selection} to maintain golden seeds (i.e., those identified as beneficial for MGF) while minimizing the size of the set. 
Typically, seed selection favors seeds that execute unique code regions and discards those that contribute stale coverage.
This coverage-based strategy is straightforward and useful, and is widely adopted by the fuzzing community.
However, coverage-based seed selection (CSS) is \textit{not} flawless and has a notable blind spot: it fails to identify golden seeds suppressed by unpassed coverage guards, even if these seeds contain valuable payload that can benefit MGF.
This blind spot prevents suppressed golden seeds from realizing their true values, which may ultimately degrade the effectiveness of downstream MGF. 

In this paper, we propose a novel technique named \techname{} to address the blind spot of traditional CSS. 
The basic idea behind \techname{} is to manifest the true strengths of the suppressed golden seeds by gradually disabling obstacle conditional guards.
To this end, we develop a lightweight program transformation to enable flexible disabling of guards and devise a novel guard hierarchy analysis to identify obstacle ones.
An iterative seed selection algorithm is constructed to stepwise select suppressed golden seeds.
We prototype \techname{} on top of the AFL++ utilities (\aflppver{}) and compare it with \nbaseline{} baselines, including two state-of-the-art tools \aflcmin{} and \optimin{}.
Compared with \aflcmin{}, \techname{} selects 3\rangeline{}40 additional seeds within a practical time budget of two hours.
To evaluate how effective the studied techniques are in seeding MGF, we further conduct extensive fuzzing (over 17,280 CPU hours) with \ntarget{} different targets from a mature benchmark named \mg{}, adopting the most representative fuzzer AFL++ for MGF.
The results show that the additional seeds selected by \techname{} yield modest improvements in both code coverage and bug discovery. 
Although our evaluation reveals some limitations of \techname{}, it also demonstrates the presence and value of suppressed golden seeds. 
Based on the evaluation results, we distill lessons and insights that may inspire the fuzzing community.

\end{abstract}




\maketitle

\section{Introduction}
\label{sec:intro}
Mutational greybox fuzzing (MGF) is one of the most popular software testing techniques \cite{marcel2025software}.
Using code coverage as feedback, MGF continuously runs the target program with test inputs generated by mutating seeds, driving executions towards deeper code regions, attempting to trigger latent vulnerabilities \cite{qian2022investigating, afl++paper}.  
Owing to its simple framework and strong cost-effectiveness, MGF has become a de facto standard for software security \cite{raj2024fuzz}, adapted to various types of software systems and responsible for a large number of bugs \cite{libfuzzer, pan2024edefuzz, afltrophy, nyx, nyx-net}.

Initial seeds are critical for MGF because they define the starting points for input generation \cite{klees2018evaluating}; they establish the effectiveness of MGF not only by providing knowledge about valid input structures but also by carrying data capable of driving MGF towards interesting states \cite{dipritosem, pham2020aflnet}. 
Before formally starting MGF, testers usually perform \textit{seed selection} to prepare premium initial seeds.
The general purpose of seed selection is to distill the given seed corpus as much as possible while retaining ``golden seeds'' of high quality \cite{shen2022drifuzz}.
As the most commonly used test adequacy indicator \cite{ammann2016introduction}, code coverage is naturally adopted by seed selection to identify golden seeds.
For instance, Rebert et. al. formulate the seed selection problem and conduct an extensive study on six selection strategies; they find that the \textsc{Unweighted Minset} strategy (corresponding to the standard coverage-based approach) performs best in terms of reduction ability and uncovers the second highest number of bugs \cite{rebert2014optimizing}.
Herrera et. al. systematically explore mainstream seed selection practices; they derive several insightful guidelines and propose a novel MaxSAT-based tool called \textsc{OptiMin}, which refines existing coverage-based approaches \cite{herrera2021seed}.
Another evergreen example is \aflcmin{}, the seed selection tool supplied by the famous AFL/AFL++ family \cite{afl,afl++};
\aflcmin{} performs coverage-based checks to discard redundant seeds that offer stale coverage, intending to create reduced seed sets while maintaining high effectiveness in bug discovery.

\begin{figure}
    \begin{minipage}[b]{.49\linewidth}
        \centering
        \includegraphics[width=\linewidth]{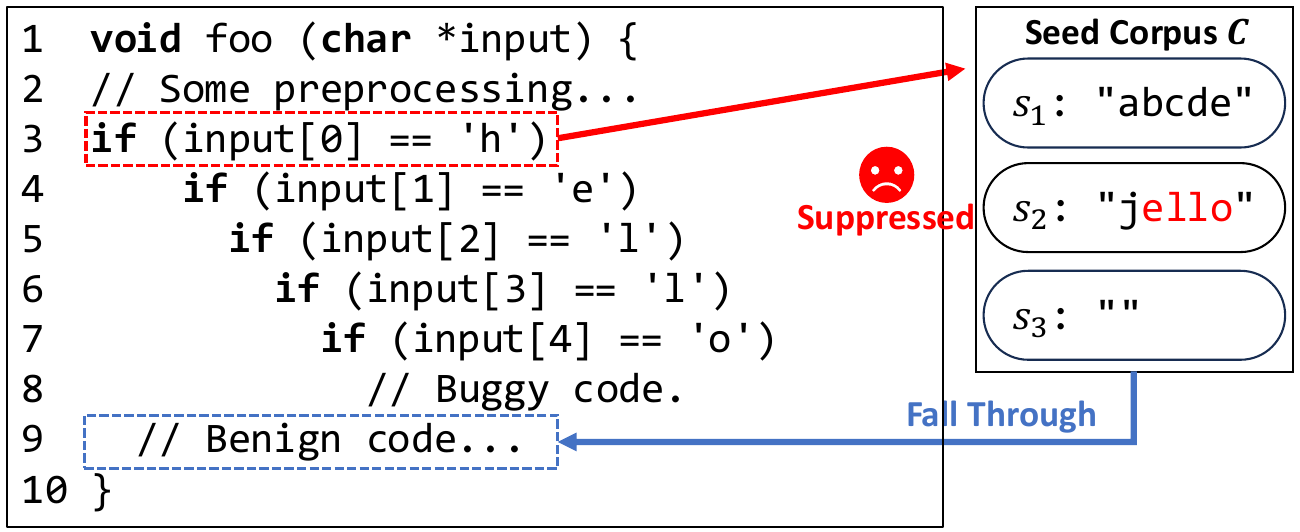}
        \subcaption{A golden seed $s_2$ suppressed by line 3.}
        \label{subfig:intro-mot-a}
    \end{minipage}
    \begin{minipage}[b]{.49\linewidth}
        \centering
        \includegraphics[width=\linewidth]{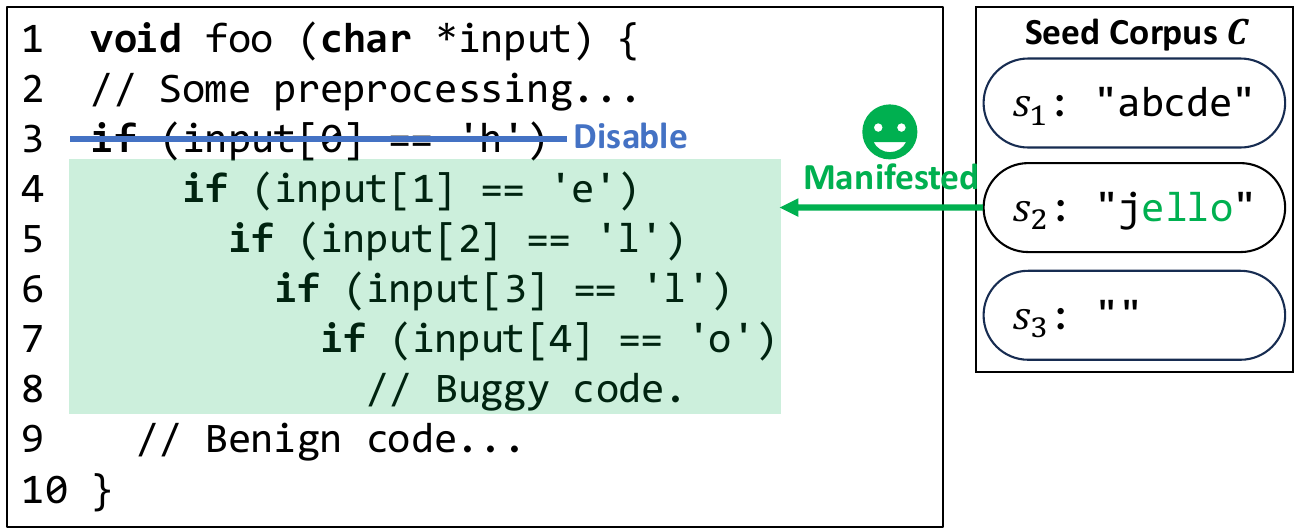}
        \subcaption{Unleashing $s_2$ by disabling line 3.}
        \label{subfig:intro-mot-b}
    \end{minipage}
    \caption{
        A motivating example that reveals how CSS might overlook suppressed golden seeds like $s_2$. 
    }
    \label{fig:motivation}
    \Description{Motivating example: start.}
\end{figure}

Coverage-based seed selection (CSS) favors seeds that reveal fresh coverage (e.g., unique paths) and will shed those that redundantly exercise previously explored code regions. 
This straightforward strategy is pretty useful and has contributed to the widespread adoption of CSS.
However, CSS is \textit{not} perfect and contains some drawbacks.
One particularly substantial issue is that CSS cannot identify golden seeds suppressed by dominant conditional guards, even though these seeds contain highly valuable data critical to the performance of downstream MGF. 
We use a simple example to illustrate this issue.
\figcap{\ref{subfig:intro-mot-a}} displays a fuzz target \texttt{foo()} requiring string inputs, where a crash-inducing bug lies in \linecap{8}.
To expose this bug, a fuzzer needs to generate string inputs that start with the specific string "\texttt{hello}" to pass successive checks guarding in \linescap{3}{7}.
Suppose that there is a seed corpus $C = \{s_1, s_2, s_3\}$, where $s_1$ is "\texttt{abcde}", $s_2$ is "\texttt{jello}", and $s_3$ is an empty string. 
Obviously, none of these seeds can directly reach the buggy location.
All three seeds will fail at the first check on \texttt{input[0]} and then fall through to the benign code.
From a coverage perspective, fuzzing with all three seeds in $C$ seems redundant because they exercise the same path.
To examine the coverage perspective in practice, we run \aflcmin{} to reduce $C$ against \texttt{foo()}, and it turns out that \aflcmin{} only preserves $s_3$ (the empty string).
The reasons why \aflcmin{} retains $s_3$ are twofold: 
\lineno{1} all three seeds achieve the same path, and \lineno{2} the size of $s_3$ is the smallest and AFL++ (the paired MGF of \aflcmin{}) opts for smaller seeds since they execute faster \cite{afl++}.
The result implies that \aflcmin{} considers $s_3$ to be \textit{equivalent} to $s_1$ and $s_2$, reflecting the general view of CSS.

\definecolor{brightgreen}{RGB}{0,176,80}
However, we argue that $s_1,s_2,s_3$ are in fact \textit{not equivalently} helpful to MGF.
Conceptually, generating a bug-exposing input from $s_2$ is much easier than $s_1$ or $s_3$, because $s_2$ contains the magic data "\textcolor{red}{\texttt{ello}}" and is much closer to the bug-triggering prefix "\texttt{hello}".
Suppose $\mathcal{F}$ is a greybox fuzzer that can generate test inputs by 
\lineno{1} flipping the characters of parent seeds in sequence or \lineno{2} synthesizing new characters from scratch.
Restrict the generated inputs to five characters from an electable set of \texttt{[a-z]}.
To generate "\texttt{hello}" from $s_1$ ("\texttt{abcde}"), $\mathcal{F}$ has to replace all the five characters, taking $25^5$ trials in the worst case.  
As for $s_3$ (the empty string), the worst-case number of trials is $26^5$, since $\mathcal{F}$ needs to synthesize "\texttt{hello}" from scratch.
In contrast, when seeding with $s_2$, $\mathcal{F}$ can crash \texttt{foo()} with at most 25 trials by flipping the first character "\texttt{j}" to "\texttt{h}", which is more efficient by orders of magnitude than $s_1$ and $s_3$.
Unfortunately, since CSS mainly focuses on coverage, it fails to detect the obscured magic data "\textcolor{red}{\texttt{ello}}", causing it to miss the golden seed $s_2$.

In this paper, we propose \techname{} to address the aforementioned blind spot of CSS. 
The basic idea behind \techname{} is very intuitive: disable the conditional guards that suppress the golden seeds, make the magic data of these seeds manifest their intrinsic values.
\figcap{\ref{subfig:intro-mot-b}} showcases the feasibility of this idea: by disabling the conditional guard in \linecap{3}, the effect of the magic suffix "\textcolor{brightgreen}{\texttt{ello}}" is manifested, pushing $s_2$ into the branches in \linescap{4}{8}, thus distinguishing $s_2$ from $s_1$ and $s_3$.
Similar ideas appeared in pioneering research.
For example, T-Fuzz dynamically eliminates hard-to-resolve checks during MGF to allow deeper explorations of the fuzz target \cite{peng2018t}.
MirageFuzz creates phantom versions of fuzz targets by dismantling the nested branches, exploiting the information from the phantom targets to guide MGF \cite{wu2023enhancing}.
In contrast to these works that retrofit fuzzers, \techname{} is dedicated to addressing the blind spot of modern CSS techniques, offering an orthogonal angle and enhancing MGF in an alternative way. 
Although the idea illustrated by \figcap{\ref{subfig:intro-mot-b}} is intuitive, employing it to enhance CSS is non-trivial and poses two challenges:

\begin{itemize}[leftmargin=*, topsep=3pt]
    \item \textbf{Challenge-1}:
    \textbf{How to disable conditional guards while keeping the target program valid?}
    Simply removing guards seems straightforward and feasible; however, such removal is inflexible and can probably invalidate the target program \cite{wu2023enhancing}. 
    More critically, as reported by Peng et al. \cite{peng2018t}, simply removing the guards can induce false positives, thus compromising the effectiveness of downstream MGF.

    \item \textbf{Challenge-2}:
    \textbf{How to identify obstacle guards precisely and efficiently?}
    Obstacle guards refer to conditional statements that suppress the golden seeds by preventing their magic data from manifesting.
    Recognizing obstacle guards requires not only understanding the control flows, but also grasping the semantics of the target program, which turns out to be tricky.
\end{itemize}

We propose two key techniques to overcome the two challenges.
Specifically, to cope with \textbf{Challenge-1}, we devise a lightweight program transformation that inserts toggles for conditional guards. 
By switching the toggles, we can effectively disable guards through short-circuiting, which preserves the core functionalities of the fuzz targets while providing significant flexibility.
To handle \textbf{Challenge-2}, we develop a guard hierarchy analysis to reveal the coverage boundary of seeds and reflect potential obstacle guards (\seccap{\ref{subsec:GHA}}).
We also devise heuristics to identify and recover reckless guards---the guards that can undermine the performance of \techname{} by forcing fuzz targets to behave unexpectedly (\seccap{\ref{subsubsec:reckless-guard}}).
Based on the two techniques, we formulate an iterative seed selection to progressively \textbf{\underline{p}}eels \textbf{\underline{o}}ff the \textbf{\underline{co}}coon-like obstacle guards and incrementally incorporate the unleashed golden seeds until reaching the fixed point (\seccap{\ref{subsubsec:iss}}).  

We prototype \techname{} atop AFL++ utilities (i.e., \aflcmin{} and \aflcc{}), and evaluate it against
\nbaseline{} different baselines (including state-of-the-art tools \aflcmin{} and \optimin{}) on \ntarget{} targets chosen from a widely used benchmark named \mg{} \cite{hazimeh2020magma}.
We prepare high-quality seeds from a well-examined dataset \cite{herrera2021seed} and open-source projects \cite{luaprojectlist}, and employ AFL++ to perform MGF.
The experimental results show that, with a practical time budget of two hours, \techname{} can select 3\rangeline{}40 additional seeds compared with the base CSS tool \aflcmin{}.
The seed sets produced by \techname{} can not only enable MGF to achieve 0.2 to 51.4 more edges (averaged over \nrepeat{} repetitions) than \aflcmin{} does, but also find the second most bugs among all the studied techniques.
\techname{} (under two configurations) also helps the downstream MGF expose two unique bugs that are missed by the baselines.
Our evaluation demonstrates the presence of suppressed golden seeds and highlights their potential in improving MGF.
However, although \techname{} yields moderate improvements by including additional seeds, its practical applicability is limited due to the extra time overhead and its inability to continuously identify golden seeds over long runs.
Based on these results, we highlight both the dark and bright sides and distill lessons to inform the fuzzing community (\seccap{\ref{subsec:lessons}}).
The contributions of this paper are summarized as follows:

\begin{itemize}[leftmargin=*, topsep=3pt]

    \item \textbf{Innovative technique.}
    We propose \techname{}, a novel CSS enhancement technique that offers a new perspective on seed selection, focusing on strategic inclusion rather than mere reduction.
 
    \item \textbf{Extensive evaluation.}
    We experiment \techname{} against \nbaseline{} baselines and \ntarget{} fuzz targets. 
    The results demonstrate the effectiveness of \techname{} and support our insights into CSS’s blind spot.

    \item \textbf{Practical tool.}
    We prototype \techname{} and release all code to support open-source science.
    Our artifacts can be found at: \artifacturl{}.
    
\end{itemize}


\section{Background and Motivation}
In this section, we introduce the general workflow of MGF (\seccap{\ref{subsec:gf}}) and further discuss the blind spot of CSS with empirical data (\seccap{\ref{subsec:example-empirical}}).
We also present an example to show how \techname{} works (\seccap{\ref{subsec:example}}). 

\subsection{Mutational Greybox Fuzzing and Seed Selection}
\label{subsec:gf}

MGF is a vital branch of modern fuzzing techniques.
Fed with a fuzz target (i.e., the program under test) and a seed corpus, MGF performs \lineno{1} lightweight instrumentation to get ready for runtime coverage tracking and \lineno{2} seed selection to prepare promising initial seeds.
After preprocessing, MGF enters the main fuzzing loop:
First, it conducts seed scheduling to prioritize a parent seed, and mutates the selected parent to spawn offspring inputs.
Second, it executes every offspring input against the fuzz target and collects the execution results, such as code coverage and crashes. 
Third, based on the execution results, MGF can \lineno{1} save the offspring input as a potential vulnerability proof-of-concept if it crashes the target, or \lineno{2} adds it to the queue if it is considered interesting (e.g, attaining fresh coverage), or \lineno{3} discards it if it neither crashes nor finds fresh coverage.

\textbf{Seed selection for MGF.}
Seed selection, an important preprocessing for MGF, is highly recommended and widely practiced by academia and industry \cite{afl++,serebryany2017oss}. 
In a broad sense, seed selection comprises two steps \cite{rebert2014optimizing}:
First, collecting seeds of certain types (e.g., PNG and XML) from sources such as the Internet to construct seed corpora.
Second, distilling the seed corpus obeying specific quality indicators (e.g., code coverage) to obtain a refined seed set as initial seeds.
The second step, also known as \textit{corpus minimization}, is often synonymous with seed selection in recent research \cite{herrera2021seed}.
Therefore, in this paper, we follow this convention and use these two terms interchangeably.
Generally, seed selection aims to minimize the size of the seed set while ensuring quality, where the most commonly used quality indicator is code coverage.
Although practical, a coverage-based seed selection tool (e.g., \aflcmin{}) is generally blind to golden seeds suppressed by unpassed conditional guards, limiting its effectiveness.
In the following subsections, we first discuss the blind spot of CSS in-depth with empirical data and then illustrate how \techname{} resolves this blind spot with an example.

\begin{table}
\small
  \centering
  \caption{
  Results of using seeds $s_1$\rangeline{}$s_7$ to crash the example shown in \figcap{\ref{fig:motivation}}.
"\boldmath{}\textbf{$R_\mathit{crash}$}\unboldmath{}" is the crash ratio across 30 repetitions.
"\boldmath{}\textbf{$t_\mathit{min}$}\unboldmath{}", "\boldmath{}\textbf{$t_\mathit{max}$}\unboldmath{}", and "\boldmath{}\textbf{$t_\mathit{ave}$}\unboldmath{}" are minimum, maximum, and average time-to-crash (in seconds).
  }
  \label{tab:bug-expose-example}%
  
\begin{tabular}{cl|rrrr}
\hline
\textbf{SID} & \textbf{Seed} & \boldmath{}\textbf{$R_\mathit{crash}$}\unboldmath{} & \boldmath{}\textbf{$t_\mathit{min}$}\unboldmath{} & \boldmath{}\textbf{$t_\mathit{max}$}\unboldmath{} & \boldmath{}\textbf{$t_\mathit{ave}$}\unboldmath{} \\
\hline
$s_1$ & "\texttt{abcde}" & 10/30 & 3455 & 73953 & 31805 \\
\hline
$s_2$ & "\texttt{jello}" & 30/30 & 3     & 616   & 135 \\
\hline
$s_3$ & "\texttt{}" & 9/30  & 807   & 79726 & 40605 \\
\hline
$s_4$ & "\texttt{hbcde}" & 30/30 & 53    & 12247 & 3953 \\
\hline
$s_5$ & "\texttt{hecde}" & 30/30 & 31    & 607   & 178 \\
\hline
$s_6$ & "\texttt{helde}" & 30/30 & 3     & 545   & 128 \\
\hline
$s_7$ & "\texttt{helle}" & 30/30 & 3     & 300   & 126 \\
\hline
\end{tabular}%
\end{table}%

\subsection{Blind Spot in Coverage-based Seed Selection: Empirical Evidence}
\label{subsec:example-empirical}

To support our conceptual inference presented in \seccap{\ref{sec:intro}}, we further experiment with \texttt{foo()} and $s_1$\rangeline{}$s_3$ shown in \figcap{\ref{fig:motivation}}. 
Specifically, we leverage AFL++ \cite{afl++paper} (\aflppver{}) to perform MGF, fuzzing \texttt{foo()} with each seed from $C$.
To compare how fast $s_1$\rangeline{}$s_3$ can help AFL++ reach the buggy location, we set AFL++ to terminate when it \lineno{1} triggers a crash or \lineno{2} exhausts the given time budget, which is set to 24 hours.
Each fuzz campaign is repeated 30 times to build statistical significance.
\tabcap{\ref{tab:bug-expose-example}} exhibits the experimental results. 
As expected, $s_2$ performs the best by significantly exceeding $s_1$ and $s_3$.
Specifically, $s_2$ consistently triggers the crash by reaching the buggy location in all 30 runs, while $s_1$ and $s_2$ succeed only nine and ten times.
Moreover, by achieving an average crashing time (i.e., $t_\mathit{ave}$) of only 1/200th of $s_1$ and $s_3$, $s_2$ is observed faster in driving AFL++ to a crashing state, revealing its superiority in facilitating MGF.  

\textbf{The unrecognized value of~$s_2$.}
We further evaluate the value of $s_2$ by extensively experimenting with seeds $s_4$\rangeline{}$s_7$.
Compared with $s_1$\rangeline{}$s_3$, seeds $s_4$\rangeline{}$ s_7$ cover deeper codes of \texttt{foo()} by including prefixes closer to "\texttt{hello}", which are namely "\texttt{h}", "\texttt{he}", "\texttt{hel}", and "\texttt{hell}".
From a coverage perspective, $s_4$\rangeline{}$s_7$ are naturally preferred to $s_2$ because they unveil more codes.
However, our experiment demonstrates that $s_2$ is comparable to $s_4$\rangeline{}$s_7$ in helping the fuzzer crash \texttt{foo()}.
As illustrated in \tabcap{\ref{tab:bug-expose-example}}, the $R_\mathit{crash}$ of $s_2$ is equivalent to those of $s_4$\rangeline{}$s_7$, suggesting a consistent ability to expose crashes.
For $t_\mathit{ave}$, $s_2$ performs comparably to $s_6$ and $s_7$ (which are the seeds with the highest coverage), demonstrating that $s_2$ remains competitive even when evaluated with high-coverage seeds.
The results show that the value of $s_2$ has been suppressed, causing it to be underestimated by modern CSS.

\begin{figure}
    \centering
    \includegraphics[width=.95\linewidth]{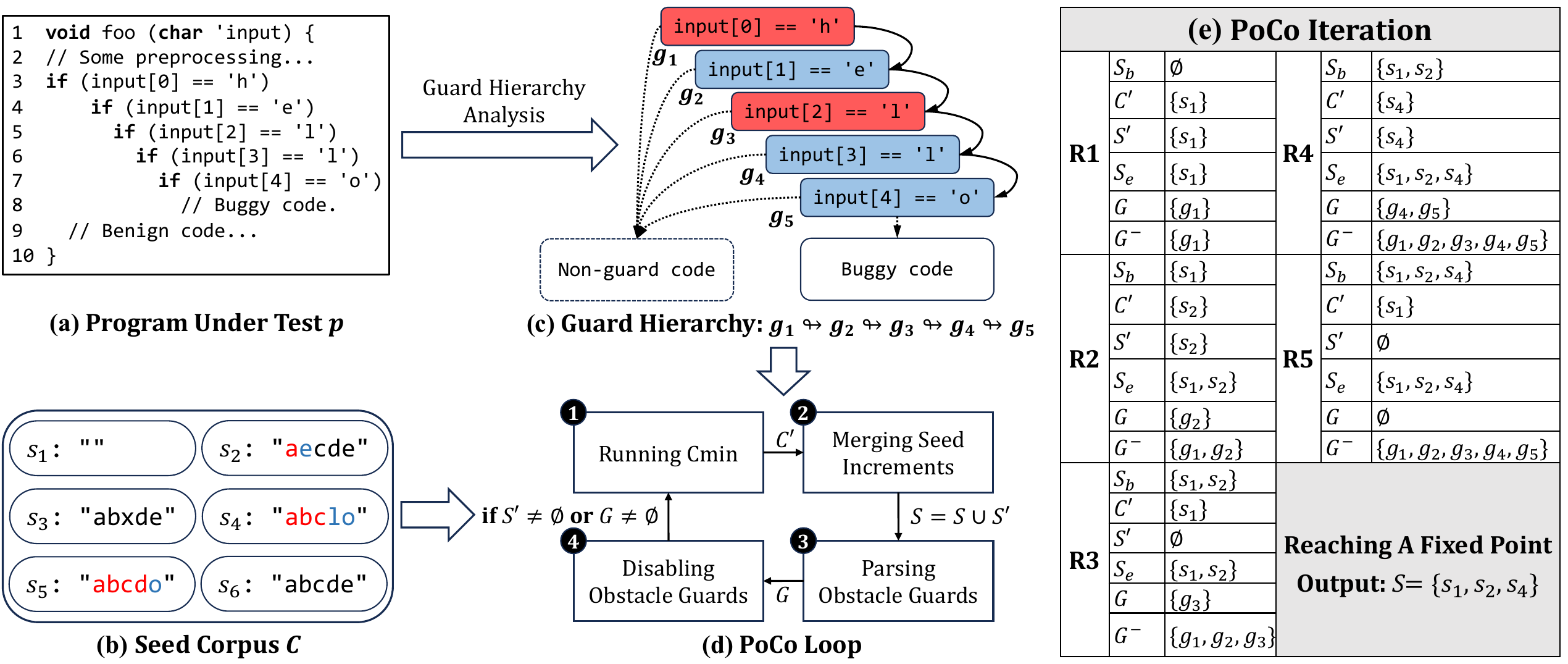}
    \caption{
    An illustration about how \techname{} works.
    \figcap{\ref{fig:example-process}}(a) and (b) display the inputs of \techname{} (i.e., the fuzz target $p$ and the seed corpus $C$). 
    \figcap{\ref{fig:example-process}}(c) displays and the guard hierarchies extracted from $p$, of which the guards in \linescap{3}{7} are marked as $g_1$\rangeline{}$g_5$; the codes that are not guarded by any conditional statements are unified into a ``\texttt{Non-guard code}'' node for clarity.
    \figcap{\ref{fig:example-process}}(d) displays the simplified workflow of \techname{}.
    \figcap{\ref{fig:example-process}}(e) displays a tabular overview of the manipulations in each round, where $S_b$ and $S_e$ are seed sets at the beginning and end of each round, $S'$ denotes the incremental seeds (i.e., $C'$$\setminus$$S_b$), and $G^-$ denotes the conditional guards disabled.}
    \label{fig:example-process}
    \Description{An illustration of \techname{}'s working process.}

\end{figure}

\subsection{The \techname{} Solution}
\label{subsec:example}
\newcommand{\guarddom}{\looparrowright}
\definecolor{darkblue}{RGB}{46, 117, 182}
\definecolor{brightblue}{RGB}{66, 115, 177}
We propose \techname{} to address the blind spot of modern CSS tools.
As shown in \figcap{\ref{fig:motivation}}, the primary culprit suppressing $s_2$ is the conditional statement guarding in \linecap{3} of \texttt{foo()}.
Hence, to resolve the blind spot, a basic idea is to disable the obstacle guard (e.g, \linecap{3} in \figcap{\ref{fig:motivation}}), liberate the suppressed seeds (e.g., $s_2$), and manifest the due effects of the valuable data (e.g., "\texttt{ello}" in $s_2$).

Given a fuzz target $p$, a seed corpus $C$, and a base CSS tool \cmin{}, \techname{} iteratively performs the following four steps to construct a seed set $S$ until reaching the fixed point (discussed in \seccap{\ref{subsubsec:iss}}):
\circled{1} it runs \cmin{} against $p$ and $C$ to get a set $C'$ minimized by coverage.
\circled{2} it analyzes the difference between $S$ and $C'$, merging incremental seeds $S'$ into $S$. 
\circled{3} it parses the obstacle conditional guards $G$ regarding $p$ and $C$.
\circled{4} it disables the guards within $G$, pushing seed selection forward.
\figcap{\ref{fig:example-process}} illustrates how \techname{} copes with the scenario shown in \figcap{\ref{fig:motivation}}; it first extracts the guard hierarchy from $p$ and then operates as follows:
\begin{itemize}[leftmargin=*, topsep=3pt]
    \item \textbf{R1}:
    No guards are disabled, and no seeds in $C$ can pass $g_1$.
    The result of \cmin{} is $C'=\{s_1\}$ because $s_1$ is the smallest in size, leading to $S_e=\{s_1\}$.
    \techname{} then checks the guard hierarchy (i.e., $g_1\guarddom g_2\guarddom g_3\guarddom g_4\guarddom g_5$) and appends the obstacle guard $g_1$ to $G^-$, so that it can be disabled in the next round.
    Note that for better visualization, the guards (e.g., $g_1$) that are disabled because no seeds can pass through are painted red in \figcap{\ref{fig:example-process}}; the seed data that are blocked by guards like $g_1$ are also colored red to indicate their inner associations (e.g., the first "\textcolor{red}{\texttt{a}}" of $s_2$). 

    \item \textbf{R2}:
    By disabling $g_1$, all seeds in $C$ can now get into the first true branch, but only $s_2$ can pass through $g_2$ and enter the second true branch, resulting in $C'=\{s_2\}$.
    To ensure passing through $g_2$ in subsequent rounds, we treat $g_2$ as an obstacle guard and merge it into $G^-$.
    To distinguish from $g_1$, we mark the obstacle guards (e.g., $g_2$) that can be triggered by the data of newly unleashed seeds (e.g., $s_2$) as blue in \figcap{\ref{fig:example-process}}; again, the seed data responsible for passing through guards like $g_2$ are also highlighted in blue to indicate the associations (e.g., the "\textcolor{brightblue}{\texttt{e}}" of $s_2$).
    
    \item \textbf{R3}:
    By disabling $G^-=\{g_1, g_2\}$, seeds of $C$ can all enter the branch of $g_2$, but no seed can pass through $g_3$, resulting in $C'=\{s_1\}$ again.
    Although the incremental seed set is $S=\emptyset$ in this round, the obstacle guards $G$ are \textit{not} empty yet.
    Since the guard hierarchy is not exhausted, we mark the outermost guard $g_3$ as the next obstacle to disable, analogous to how $g_1$ is handled.
    As such, the fixed point (i.e., $S' = \emptyset$ and $G = \emptyset$) is not reached, and the iteration will continue to move.
    Note that the fixed point here is a simplified version; the complete condition is discussed in \seccap{\ref{subsubsec:iss}}.
    
    \item \textbf{R4}:
    By disabling $G^-=\{g_1, g_2, g_3\}$, seeds of $C$ can now enter the branch guarded by $g_3$.
    Thanks to the suffix "\textcolor{brightblue}{\texttt{lo}}", $s_4$ now covers the most codes, thus being selected by \cmin{}.
    Since $g_4$ and $g_5$ have been satisfied, they are marked as obstacles and will be disabled in the next round. 
    
    \item \textbf{R5}:
    At this round, guards $g_1$\rangeline{}$g_5$ are all disabled, and every seed of $C$ can reach the true branch of $g_5$ through the same path.
    The resultant $C'$ is again $\{s_1\}$ thus $S'=\emptyset$.
    Moreover, since no guards are satisfied by suppressed seeds and no candidate obstacles are left in the hierarchy, the resultant $G$ is $\emptyset$.
    So far, \techname{} reaches a fixed point and will terminate by outputting $S=\{s_1, s_2, s_4\}$.
\end{itemize}

Compared with the base \cmin{}, \techname{} additionally picks $s_2$ and $s_4$, which contain payloads (i.e., the "\textcolor{brightblue}{\texttt{e}}" of $s_2$ and the "\textcolor{brightblue}{\texttt{lo}}" of $s_4$) that can pass through $g_2$, $g_4$, and $g_5$.
Just like the seed "\texttt{jello}" discussed in \seccap{\ref{subsec:example-empirical}}, $s_2$ and $s_4$ can also provide downstream MGF with valuable payload, which can potentially enable fuzzers to explore deeper codes and detect bugs. 
The whole iteration involves a stepwise disabling of obstacle guards (detailed in \seccap{\ref{subsubsec:iss}}), which is somewhat similar to the process of \textbf{\underline{P}eeling \underline{\textbf{o}}ff a \underline{\textbf{Co}}coon}. 
This similarity inspires us to name the proposed technique \techname{}.

\section{Methodology}
\label{sec:method}
In this section, we elaborate on the methodology behind \techname{}.
We start by defining the problem scope that \techname{} focuses on (\seccap{\ref{subsec:problem-def}}) and then delve into the technical details, namely toggle insertion (\seccap{\ref{subsec:toggle-insertion}}), guard hierarchy analysis (\seccap{\ref{subsec:GHA}}), and iterative seed selection (\seccap{\ref{subsec:iterative-seed-selection}}).  

\subsection{Problem Definition}
\label{subsec:problem-def}
\newcommand{\minfunc}{\phi}
\newcommand{\improve}{\mathrel{\triangleright}}

Traditional CSS hypothesizes that minimizing seeds by code coverage can speed up the downstream MGF while preserving the effectiveness of $C$ to the greatest extent.
Suppose $C$ is a seed corpus containing seeds of a specific file format (e.g., XML, PNG, and ELF) and $\mathbb{C}~(C\in\mathbb{C})$ is the universe of such corpora. 
Let $P$ denote the domain of fuzz targets $p$ that require input files of the same format as $C$.
The purpose of CSS is to develop a function $\minfunc: \mathbb{C}\times P \to \mathbb{C}$ that can produce a minimized seed set $S_\minfunc$ from $C$ ($S_\minfunc\subseteq C$) without losing initial coverage.

Based on the above, we devise \techname{} as a CSS enhancement technique that improves $\minfunc$ to $\minfunc': C\times P \to C$.
The main objective of \techname{} is to reveal the golden seeds suppressed by obstacle guards.
Let $\Delta = \minfunc'(p, C)\setminus\minfunc(p, C)$ denote the set of seeds additionally selected by $\minfunc'$. 
To achieve the goal of \techname{}, the following condition should hold:
\begin{equation}
\label{eq:problem}
    \Delta \neq \emptyset \implies \exists s\in\Delta, s\improve\mathcal{F}
\end{equation}
where $s$ denotes an individual seed and $\mathcal{F}$ denotes a mutational greybox fuzzer.
The notation $\improve$ stands for the \textit{Seed-Improvement} relation, defined below:

\begin{definition}
\label{def:seed-improve}
    \textit{Seed-Improvement} ($\improve$). 
    Suppose $S_\minfunc$ is a minimized seed set produced by a CSS map $\minfunc$. 
    Let $\mathcal{F}(S, p, t)$ denote the findings (e.g., code regions and crashes) of a fuzz campaign, where $\mathcal{F}$ is the fuzzer, $S$ is the initial seed set, $p$ is the fuzz target, and $t$ is the duration.
    If $\mathcal{F}(S_\minfunc\cup\{s\}, p, t)\setminus\mathcal{F}(S_\minfunc, p, t) \neq \emptyset$, then \textit{Seed-Improvement}($s,\mathcal{F}$) holds, i.e., $s\improve\mathcal{F}$.
\end{definition}

\newcommand{\examplehead}[1]{\textit{\underline{Example}.}}
\newcommand{\exampleparam}[1]{\textit{\underline{Example #1}}}
\newcounter{mycounter} 
\newcommand{\autonum}{%
  \stepcounter{mycounter}
  \arabic{mycounter}
}
\newcommand{\examplecounter}{\exampleparam{\autonum}.}

\examplecounter{}
The seed $s_2$ in \figcap{\ref{fig:motivation}} is an instance of the \textit{Seed-Improvement} relation, where the fuzz target is \texttt{foo()} and the minimized seed set is $S_\minfunc=\{s_3\}$.
Suppose $\mathcal{F}(S_\minfunc,p,t)$ and $\mathcal{F}(S_\minfunc\cup\{s_2\},p,t)$ are two fuzz campaigns whose findings are $\{\mathit{line}_{3-5}\}$ and $\{\mathit{line}_{3-8},\mathit{crash}_{8}\}$, where $\mathit{line}_{i-j}$ informally represents the code coverage from line $i$ to line $j$ and $\mathit{crash}_{i}$ represents the crash on line $i$.  
Since $\mathcal{F}(S_\minfunc\cup\{s_2\},p,t) \setminus \mathcal{F}(S_\minfunc,p,t) = \{\mathit{line}_{6-8},\mathit{crash}_{8}\}\neq\emptyset$ holds, the relation $s_2\improve\mathcal{F}$ establishes.

\subsection{Toggle Insertion}
\label{subsec:toggle-insertion}

This step acts as a preprocessing of \techname{} and is designed to handle the \textbf{Challenge-1}.
Given a fuzz target $p$, toggle insertion operates to \lineno{1} identify conditional guards within $p$ and \lineno{2} add a toggle condition to each of the guards so that they can be flexibly switched on and off. 
Let $\mathit{cond}$ denote a conditional guard.
The transformation $\tau$ behind guard toggle insertion can be formalized as:
\begin{equation}
    \tau: \mathit{cond} \mapsto \mathit{tog}\lor\mathit{cond}  
\end{equation}
where $\mathit{tog}$ is the inserted toggle condition.
Essentially, we take advantage of the short-circuit feature of conditional statements to realize flexible control of guards.
The toggles are set to off (i.e., false) by default so that the semantics of $p$ are preserved.
Once a $\mathit{cond}$ is identified as an obstacle, the corresponding $\mathit{tog}$ condition will be set to true, allowing \techname{} to dive deeper into the code.

\begin{figure}
    \centering
    \begin{minipage}[b]{.329\linewidth}
        \centering
        \includegraphics[width=\linewidth]{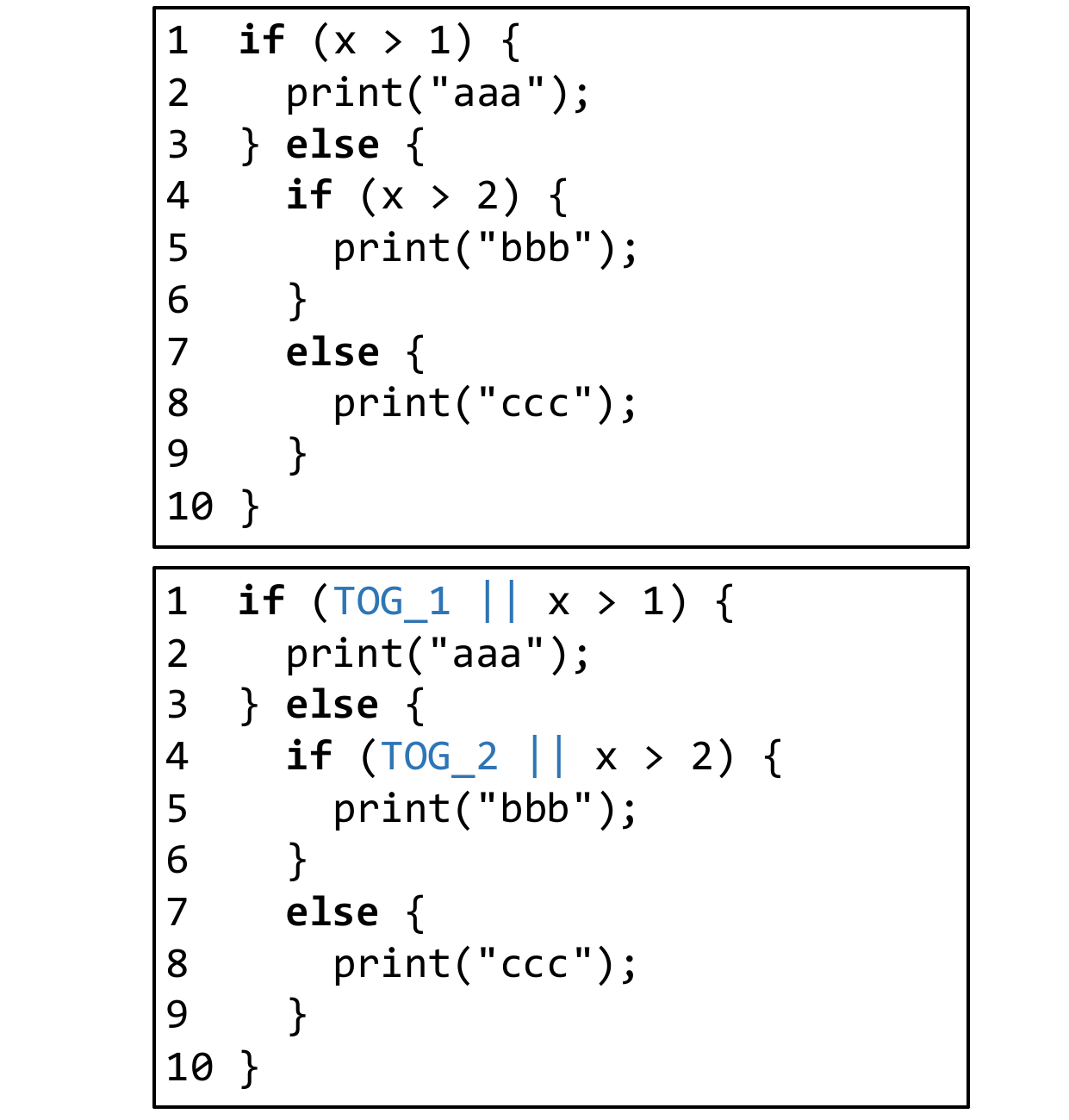}
        \subcaption{A code snippet with two guards.}
        \label{subfig:example-tog-insert-code}
    \end{minipage}
    \begin{minipage}[b]{.329\linewidth}
        \centering
        \includegraphics[width=\linewidth]{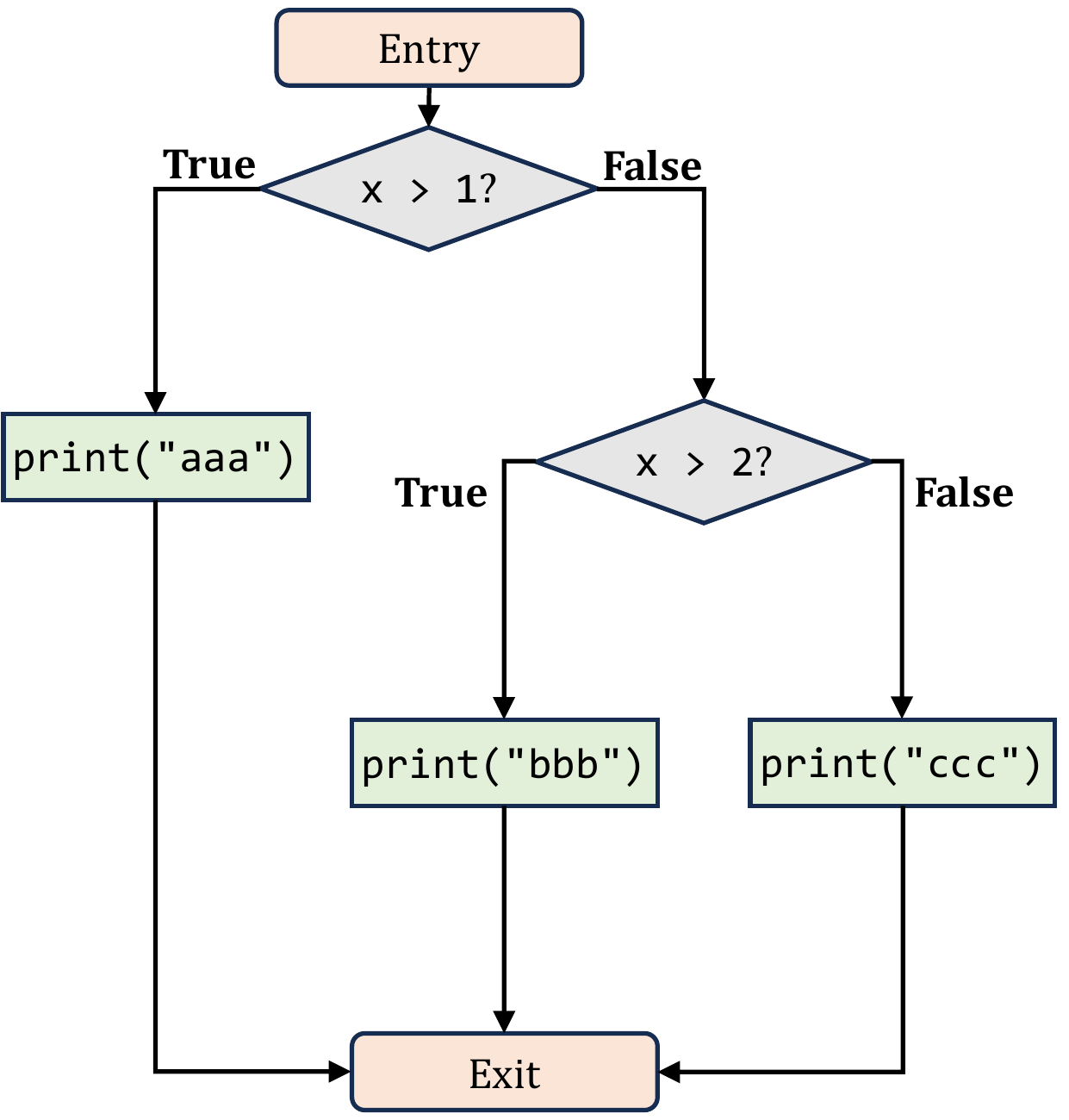}
        \subcaption{CFG before inserting toggles.}
        \label{subfig:example-tog-insert-orig}
    \end{minipage}
    \begin{minipage}[b]{.329\linewidth}
        \centering
        \includegraphics[width=\linewidth]{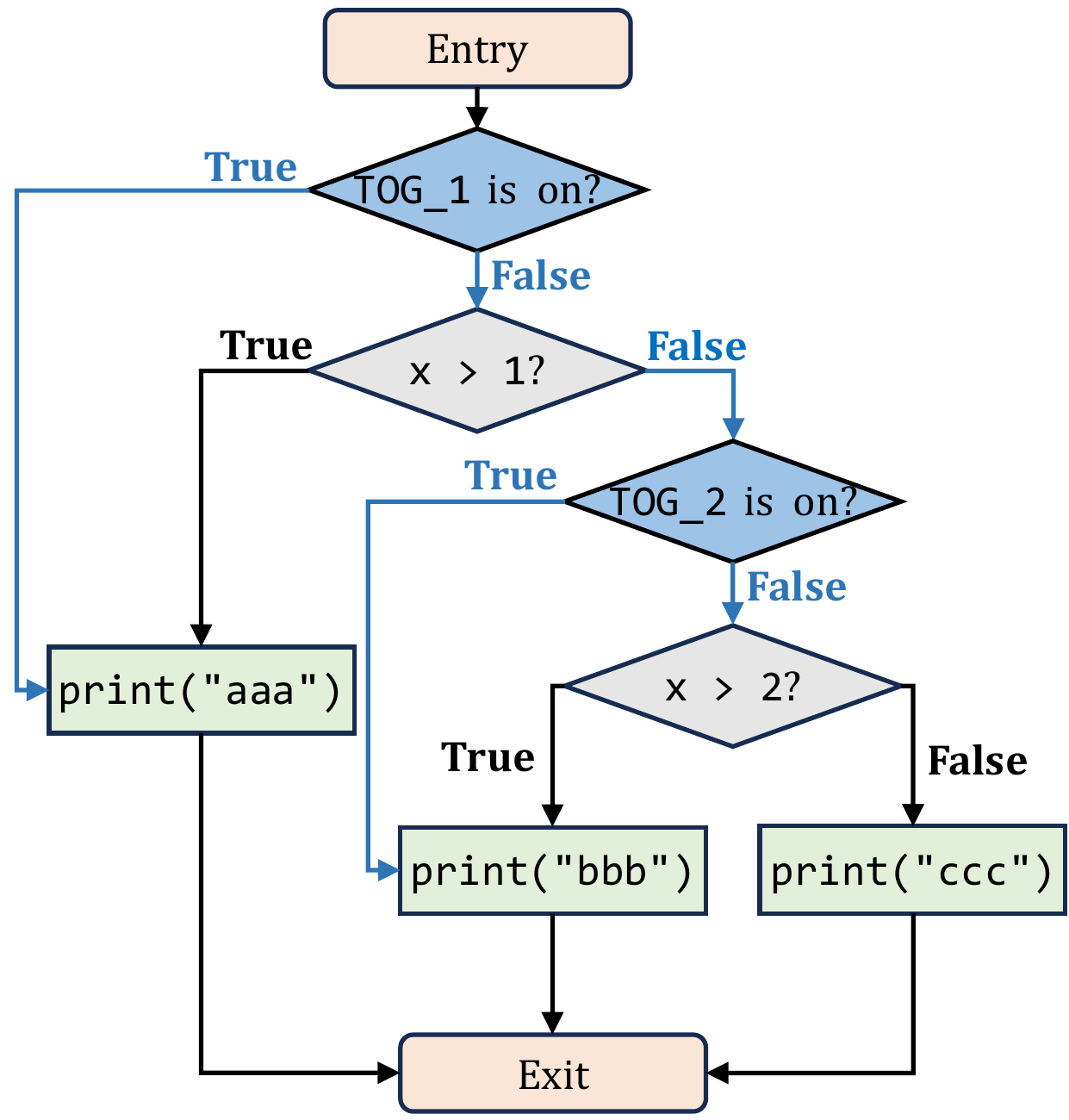}
        \subcaption{CFG after inserting toggles.}
        \label{subfig:example-tog-insert-trans}
    \end{minipage}
    \caption{An illustration of the flexible, semantic-preserving program transformation for toggle insertion.}
    \label{fig:toggle-insertion}
    \Description{The illustration of the semantic-preserving program transformation for toggle insertion.}
\end{figure}

\examplecounter{}
The upper half of \figcap{\ref{subfig:example-tog-insert-code}} shows a simple code snippet containing two conditional guards $\mathit{cond}_1:\texttt{x > 1}$ and $\mathit{cond}_2:\texttt{x > 2}$.
Let the illustrated code snippet be the fuzz target $p$, \techname{} insert toggles for $p$ by transforming $\mathit{cond}_1$ and $\mathit{cond}_2$ into $(\mathit{tog}_1\vee\mathit{cond}_1)$ and $(\mathit{tog}_2\vee\mathit{cond}_2)$, resulting in the transformed $p'$ shown in the lower half of \figcap{\ref{subfig:example-tog-insert-code}}.
The control flow graphs (CFGs) of $p$ and $p'$ are shown in \figcap{\ref{subfig:example-tog-insert-orig}} and \figcap{\ref{subfig:example-tog-insert-trans}}, where the nodes and edges that are added or modified are colored blue.  
At first, conditions $\mathit{tog}_1$ and $\mathit{tog}_2$ are set to the off status by default; the executions of $p'$ follow the false branches of the toggle conditions \texttt{TOG\_1} and  \texttt{TOG\_2}, driving them to the original conditional nodes (painted gray in \figcap{\ref{fig:toggle-insertion}}) and making $p'$ behave the same as $p$.
If a guard, say $\mathit{cond}_2$, needs to be disabled, then the corresponding toggle $\mathit{tog}_2$ is switched on, gravitating executions towards the true branch of \texttt{TOG\_2} and let them reach the guarded statement \texttt{print("aaa")}.

\subsection{Guard Hierarchy Analysis}
\label{subsec:GHA}
\newcommand{\HAshort}{GHA}
\newcommand{\tuple}[1]{\langle#1\rangle}
This step aims to clarify the relationships among guards and their corresponding toggles; it is a fundamental mechanism for handling \textbf{Challenge-2} and also a critical module of iterative seed selection (elaborated in \seccap{\ref{subsec:iterative-seed-selection}}).
In the following, we adopt guards as the primitive elements of the descriptions. 
Suppose $G$ is the set of guards that are identified through toggle insertion (\seccap{\ref{subsec:toggle-insertion}}), the task of guard hierarchy analysis is to \lineno{1} organize the hierarchical relationship among guards in $G$ and \lineno{2} identify the \textit{outermost guards} that are likely to be obstacles to subsequent seed selection.

\subsubsection{Organizing guard hierarchy}
\label{subsubsec:organize-guard-hierarchy}
\newcommand{\guardop}[1]{\xrightarrow{#1}}
\newcommand{\action}{\alpha}
\newcommand{\actop}{a}
\newcommand{\actenb}{\actop^+}
\newcommand{\actdis}{\actop^-}
\newcommand{\actnon}{\varepsilon}
\newcommand{\guardopdef}{\xrightarrow{\mathit{\action}}}
\newcommand{\enabled}{s^+}
\newcommand{\disabled}{s^-}

A guard hierarchy is a triple $H=\tuple{G, E, \Sigma}$, where $E$ is the set of edges among the guards and $\Sigma$ is the set of guard statuses.
An edge between two guards $g_1$ and $g_2$ implies that the status of the predecessor $g_1$ determines the next-round operation on the successor $g_2$, i.e., $g_1$ is dominant over $g_2$.
The status of a guard $g$ is a Boolean value $\sigma\in\{\enabled,\disabled\}$, where $\enabled$ denotes the enabled status and $\disabled$ denotes disabled.
By default, any $g$ is enabled and has $\sigma=\enabled$; once the corresponding toggle is activated, the status of $g$ will thus change to $\disabled$.
For simplicity, we use $g.\sigma$ to denote the status of $g$, and use $g^+$ or $g^-$ to denote when $g$ is enabled or disabled.
We refer to the manipulation that alters the status of a guard as \textit{Guard-Operation}, defined below.

\begin{definition}
    \textit{Guard-Operation} ($\guardopdef$).
    Let $g$ be a guard. 
    A guard operation is a transition $\guardopdef$ that changes $g.\sigma$ to $g.\sigma'$ using an action $\action$, where $\sigma,\sigma'\in\Sigma$ and $\action\in A.$
    The domain of actions is a finite set $A=\{\actenb, \actdis, \actnon\}$, where $\actenb$ represents enabling a guard $g$ (i.e., toggling off) and $\actdis$ represents disabling (i.e., toggling on); $\actnon$ is a special action representing no transition.
\end{definition}

We define the \textit{Guard-Domination}$(g, g')$ relation (denoted as $g\guarddom g'$) based on control flows, where $g$ is the predecessor and $g'$ is the successor.
The rationale behind this definition is twofold:
\lineno{1} 
\textit{Guard-Domination} relations can generally be derived from control flows, as a guard is only manipulated when it is identified as \textit{outermost}---that is, when all of its control-flow predecessors are disabled; we formally define \textit{Outermost-Guard} later in \seccap{\ref{subsubsec:outermost-guard}}.
\lineno{2}
Identifying \textit{Guard-Domination} relations based on the successors' next-round operations is inapplicable, since these operations have not yet been applied when the guard hierarchy is being constructed.

\begin{figure}
    \centering
    \begin{minipage}[b]{.49\linewidth}
        \centering
        \includegraphics[width=.7\linewidth]{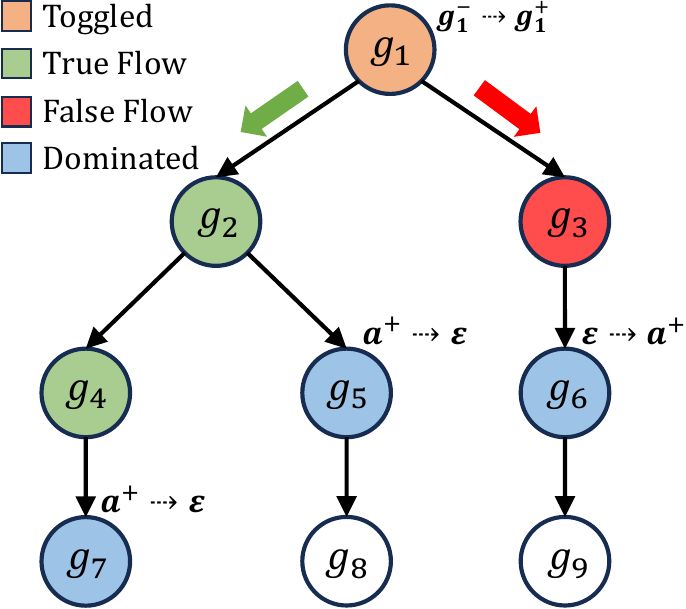}
        \subcaption{Indirect \textit{Guard-Domination} controlled by $g_1$.}
        \label{subfig:guard-dom-enb}
    \end{minipage}
    \begin{minipage}[b]{.49\linewidth}
        \centering
    \includegraphics[width=.7\linewidth]{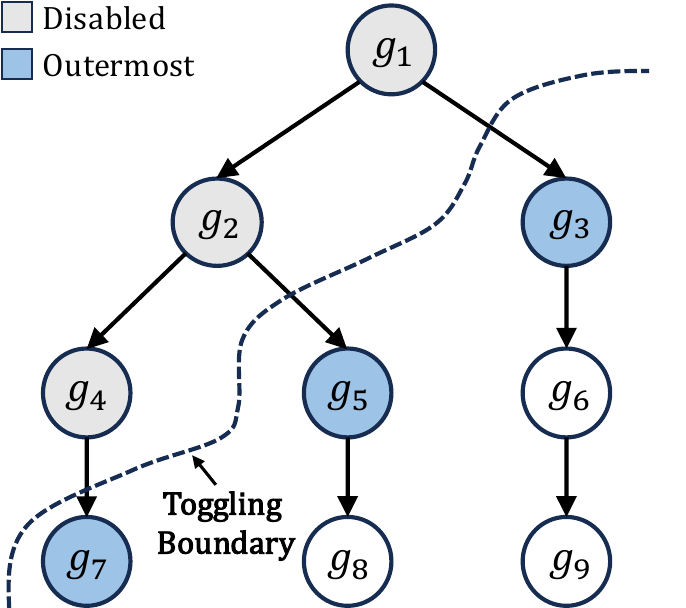}
        \subcaption{Outermost guards on the toggling boundary.}
        \label{subfig:outermost-by-cov}
    \end{minipage}
    \caption{Illustrations of guard hierarchy analysis.}
    \label{fig:GHA-examples}
    \Description{How we analyze the hierarchies of conditional guards.}
\end{figure}

\examplecounter{}
The motivating example shown in \figcap{\ref{fig:example-process}} exhibits several instances of the \textit{Guard-Domination} relation, such as $g_1\guarddom{}g_2$ and $g_3\guarddom{}g_4$.
We illustrate \textit{Guard-Domination} with $g_1\guarddom{}g_2$, where $g_2$ is control-dependent on $g_1$.
As described in our motivating example, in the first round of \techname{}, $g_1$ is identified as an obstacle guard to remove while $g_2$ is not (i.e., no operation); $g_1$ is in the enabled status at this moment, and the action applied to $g_2$ is $\action(g_2,g_1^+)=\actnon$.
After entering the second round, $g_1$ is disabled (i.e., $g_1^+\guardop{\actdis}g_1^-$) and the control-flow path to $g_2$ is unlocked; $g_2$ is now identified as an obstacle guard so that the next action applied to $g_2$ is $\action(g_2,g_1^-)=\actdis$.
The action applied to $g_2$ depends on its control-flow dominator $g_1$, which is consistent with our control-flow-based definition of \textit{Guard-Domination}; hence, $g_1\guarddom{}g_2$ holds.

\examplecounter{}
\figcap{\ref{subfig:guard-dom-enb}} exemplifies a case of the indirect \textit{Guard-Domination} relation, where the control-flow dominator $g_1$ determines the next-round operations of other guards, such as $g_5$, $g_6$, and $g_7$.
The flow directions and the guard nodes covered by the true and false flows are colored green and red, respectively; the transitions of the guard statuses or actions are informally denoted as ``$\dashrightarrow$''. 
Before toggling, $g_1$ is disabled and $g_2$ and $g_4$ are covered; $g_5$ and $g_7$ are identified as outmost and will be disabled next, i.e., $\action(g_5)=\action(g_7)=\actdis$. 
The false branch is not covered and $g_6$ will be unchanged in the next round, i.e., $\action(g_6)=\actnon$.
After toggling $g_1^-\dashrightarrow g_1^+$, the execution flow switches to the false branch; $g_3$ is now covered by the execution, and the next-round actions on $g_5$ ($\actdis\dashrightarrow \actnon$), $g_7$ ($\actdis\dashrightarrow \actnon$) and $g_6$ ($\actnon\dashrightarrow \actdis$) are changed.
These action changes establish indirect \textit{Guard-Domination} relations $g_1\guarddom g_5$, $g_1\guarddom g_6$, and $g_1\guarddom g_7$, consistent with the control-flow-based definition.

\subsubsection{Recognizing outermost guards}
\label{subsubsec:outermost-guard}
Outermost guards are crucial for the identification of obstacle guards.
They establish the boundary of the deepest coverage that the given seed corpus $C$ can reach, thus deserving to be considered as obstacles when CSS encounters a bottleneck. 
For clarity, we refer to the boundary established by the outermost guards as the toggling boundary. 
We define the \textit{Outermost-Guard} relation below.

\newcommand{\outermost}{\top}
\begin{definition}
    \textit{Outermost-Guard} ($\outermost$).
    Given a guard hierarchy $H=\tuple{G, E, \Sigma}$ and a guard $g\in H.G$.
    Let $D=\{g_d~|~\forall g_d\in H.G \land g_d\guarddom g\}$ be the set of guards that dominate $g$.
    If $\forall g_d \in D \land g_d.\sigma =\disabled$, then $g$ is outermost in $H$ and \textit{Outermost-Guard}($g,H$) holds, i.e., $g~\outermost~H$.
\end{definition}

\examplecounter{}
\figcap{\ref{subfig:outermost-by-cov}} illustrates how the outermost guards (marked blue) are identified based on the disabled guards that dominate them (marked gray).
Suppose $H$ is the guard hierarchy.
Let $D_3=\{g_1\},D_5=\{g_1,g_2\},D_7=\{g_1,g_2,g_4\}$ denote the dominant sets of $g_3,g_5$ and $g_7$, respectively.
Since the dominant guards $g_1,g_2,g_4$ have been disabled in previous rounds of seed selection, for $i\in\{3,5,7\}$, we have $g_d\in D_i \land g_d.\sigma=s^-$ satisfied.
The \textit{Outermost-Guard} relation holds for each of $g_3$, $g_5$, and $g_7$; the toggling boundary is thus established, and we visualize it with a dotted line.

\begin{algorithm}[t]
	\caption{Outermost Guard Collection.} 
	\label{alg:collect-outermost-guards} 
	\begin{algorithmic}[1] 
		\Require {Fuzz Target $p$, New Obstacle Guards $O_\text{new}$.}
		\Ensure {Outermost Guards $\Omega$.}
            \State $dQ\gets$ \textsc{AsDeque}($O_\text{new}$) \label{line:outermost-init-deque}
            \State $\mathit{checked}\gets\emptyset$ \label{line:outermost-init-checked}
            \While{$dQ$ is not empty} \label{line:outermost-loop-start}
                \State $g \gets $$dQ$.\textsc{PopHead}()
                \If{$g$ not in $\mathit{checked}$} \label{line:outermost-verify-checked}
                    \State $G_\succ \gets$ 
 \textsc{GetSuccessorGuards}($p,g$) \label{line:outermost-get-successor}
                    \For{$g_\succ$ in $G_\succ$}
                        \If{$g_\succ$ not in $O_\text{new}$ \textbf{and} $g_\succ.\sigma=\enabled$ } \label{line:outermost-decide}
                            \State $\Omega \gets \Omega~\cup~\{g_\succ\}$ \label{line:outermost-decide-start}
                        \Else
                            \State $dQ$.\textsc{AppendRear}($g_\succ$) \label{line:outermost-broaden}
                        \EndIf \label{line:outermost-decide-end}
                    \EndFor
                    \State $\mathit{checked} \gets \mathit{checked}~\cup~\{g\}$ \label{line:outermost-mark-checked}
                \EndIf
            \EndWhile \label{line:outermost-loop-end}
            \State \Return{$\Omega$}
	\end{algorithmic} 
\end{algorithm}

\algcap{\ref{alg:collect-outermost-guards}} displays how \techname{} collects the outermost guards. 
Given the fuzz target $p$ and the obstacle guards $O_\text{new}$ identified since the last outermost guard recognition, \techname{} first creates a deque $dQ$ by $O_\text{new}$ to control the collection loop (\linecap{\ref{line:outermost-init-deque}}) and prepares a set $\mathit{checked}$ to track guards that have been processed (\linecap{\ref{line:outermost-init-checked}}).
We use $O_\text{new}$ as the starting point for collecting outermost guards because the guards in it naturally reside on the previous toggling boundary, which can make the identification process more efficient.
As long as $dQ$ is not empty, \techname{} keeps looking for outermost guards in the successors of the head element $g$ of $dQ$ (\linescap{\ref{line:outermost-loop-start}}{\ref{line:outermost-loop-end}}).
Specifically, for every $g$, \techname{} first verifies whether it has been checked (\linecap{\ref{line:outermost-verify-checked}}); if not, \techname{} proceeds to extract the set of successors $G_\succ$ of $g$ from the program $p$ (\linecap{\ref{line:outermost-get-successor}}).
For each successor $g_\succ$ in $G_\succ$, \techname{} treats it as outermost if \lineno{1} $g_\succ$ is not in $O_\text{new}$ and \lineno{2} $g_\succ$ is in the enabled status (\linecap{\ref{line:outermost-decide}}).
If $g_\succ$ is not considered outermost, \techname{} then moves the toggling boundary by including $g_\succ$ in $dQ$ (\linecap{\ref{line:outermost-broaden}}).
As the final step, \techname{} marks the head element $g$ as \textit{checked} after all manipulations are finished (\linecap{\ref{line:outermost-mark-checked}}).

\subsection{Iterative Seed Selection}
\label{subsec:iterative-seed-selection}

In this subsection, we first describe the types of obstacle guards in detail, and then present their definitions.
After that, we discuss the recognition of \textit{reckless guards}---the guards that can prevent a fuzz target from unexpected behaviors and thus should \textit{not} be handled too \textit{recklessly}.
We finally define the fixed point of \techname{} and formalize the process of iterative seed selection.

\subsubsection{Defining obstacle guards}
\label{subsubsec:obstacle-guard}
\newcommand{\passed}{\bot}
\newcommand{\unpassed}{\bcancel{\bot}}
\newcommand{\obstacle}{\blacktriangle}

The essence of identifying obstacle guards is to find the guards that probably suppress golden seeds.
As illustrated in \figcap{\ref{fig:example-process}}, two types of guards are considered obstacles: 
\lineno{1} The guards newly \textit{passed} by the seeds picked in the last round of seed selection; we mark and disable these guards to prevent them from suppressing unselected seeds in subsequent seed selection.
\lineno{2} The outermost guards that no seeds in the given corpus can pass; we mark and disable these guards because they may hinder the ``magic data'' of seeds from manifesting true effects. 
To summarize above, we define the \textit{Passed-Guard} and \textit{Obstacle-Guard} below.

\begin{definition}
    \textit{Passed-Guard} ($\passed$). 
    Given a seed corpus $C$ and a fuzz target $p$ whose guard hierarchy is $H=\tuple{G,E,\Sigma}$.
    Let $g\in H.G$ be a guard within the hierarchy.
    Let $p(s,g) \in\{\mathit{BB}^t,\mathit{BB}^f\}$ be the basic blocks (BBs) that can be reached from $g$ by executing $p$ with a seed $s$, where $\mathit{BB}^t$ denotes the true branch BB and $\mathit{BB}^t$ denotes the false branch BB.
    If $\exists s\in C$ s.t. $p(s,g)=\mathit{BB}^t$, then we say that $g$ is already passed in $H$ and \textit{Passed-Guard}($g,H$) holds, i.e., $g~\passed~H$.  
\end{definition}

\examplecounter{}
Consider the code snippet in the upper of \figcap{\ref{subfig:example-tog-insert-code}} to be the fuzz target $p$, and its CFG is shown in \figcap{\ref{subfig:example-tog-insert-orig}}.
There are two guards $g_1$: \texttt{x > 1} and $g_2$: \texttt{x > 2} included in the guard hierarchy $H$ of $p$, (i.e., $g_1,g_2\in H.G$).
Let $C=\{s_1, s_2\}$ be the seed corpus prepared for fuzzing $p$, where $s_1$ is \texttt{x = 2} and $s_2$ is \texttt{x = -2}.
There exists $s_1 \in C$ that satisfies the condition of $g_1$ and makes $p(s_1,g)=\mathit{BB}_1^t$, where $\mathit{BB}_1^t$ is the true branch BB of $g_1$ (i.e., \texttt{printf("aaa")}).
Therefore, $g_1$ is passed in $H$, thus $g_1~\passed~H$ holds.
In contrast, since no $s \in C=\{s_1,s_2\}$ can satisfy the condition \texttt{x > 2}, no seeds in $C$ can reach the true branch BB of $g_2$; therefore, $g_2$ is unpassed, and $g_2~\passed~H$ fails.

\begin{definition}
    \textit{Obstacle-Guard} ($\obstacle$). 
    Given a fuzz target $p$ whose guard hierarchy is $H=\tuple{G,E,\Sigma}$.  
    A guard $g\in H.G$ is considered an obstacle if both conditions hold: \lineno{1} it is enabled, i.e., $g.\sigma=\enabled$; and \lineno{2} it has been recognized as \textit{passed} or \textit{outermost} in $H$, i.e., ($g~\passed~H~\lor~g~\outermost~H$). 
    Let $O~(O\subset H.G)$ denote a subset of guards.
    We say $O$ is the set of obstacle guards that satisfies \textit{Obstacle-Guard}($O,H$) (denoted as $O~\obstacle~H$) if the following condition holds: 
    \begin{equation}
        O~\obstacle~H\stackrel{\mathrm{def}}{\Longleftrightarrow} \forall g\in O,  g.\sigma = \enabled \land( g~\passed~H~\lor~g~\outermost~H)
    \end{equation}
\end{definition}

Note that when performing iterative seed selection, \techname{} adopts an aggressive strategy that indiscriminately removes all guards identified as obstacles to ensure efficiency.
However, since some guards are responsible for behaviors such as error handling and out-of-bounds checking, they should not be disabled ``recklessly'', or unwanted behaviors can occur and disrupt \techname{}.
Next, we describe reckless guards in detail, along with the lazy strategies developed to manage them.  

\subsubsection{Recognizing reckless guards.}
\label{subsubsec:reckless-guard}
\newcommand{\reckless}{\triangledown}
Reckless guards refer to those who can protect a fuzz target from performing unexpected behaviors.
Properly managing reckless guards is vital for the handling of \textbf{Challenge-2} because it can avoid undesired consequences and ensure precision in recognizing obstacle guards.
Specifically, disabling reckless guards can lead to two undesired consequences:
\begin{itemize}[leftmargin=*, topsep=3pt]
    \item \textbf{Conseq-1 (Crashing-Reckless)}:
    Removing reckless guards can unexpectedly disable certain sanity checks, allowing invalid inputs to reach unintended program regions and cause crashes.

    \item \textbf{Conseq-2 (Converging-Reckless)}:
    Removing reckless guards may omit too many control flows or steer execution into unusual branches, forcing the iterative seed selection to converge prematurely, as the test coverage of most seeds becomes identical after disabling the guards.
\end{itemize}

\begin{figure*}
 \centering
    \begin{minipage}[b]{.49\linewidth}
        \centering
        \includegraphics[width=.9\linewidth]{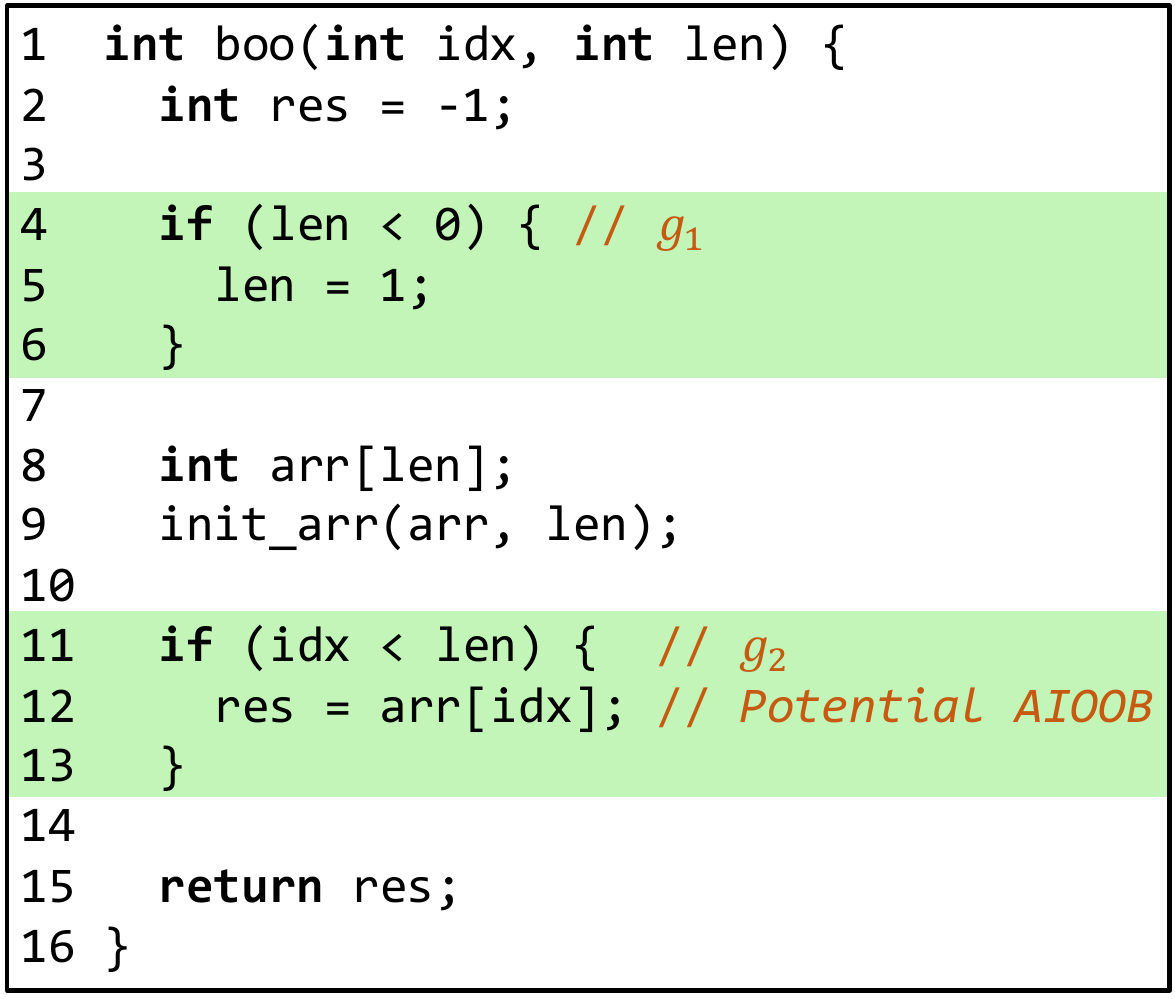}
        \subcaption{A code snippet exemplifying \textbf{Cond-1}.}
        \label{subfig:example-reckless-crash}
    \end{minipage}
    \begin{minipage}[b]{.49\linewidth}
        \centering
        \includegraphics[width=.9\linewidth]{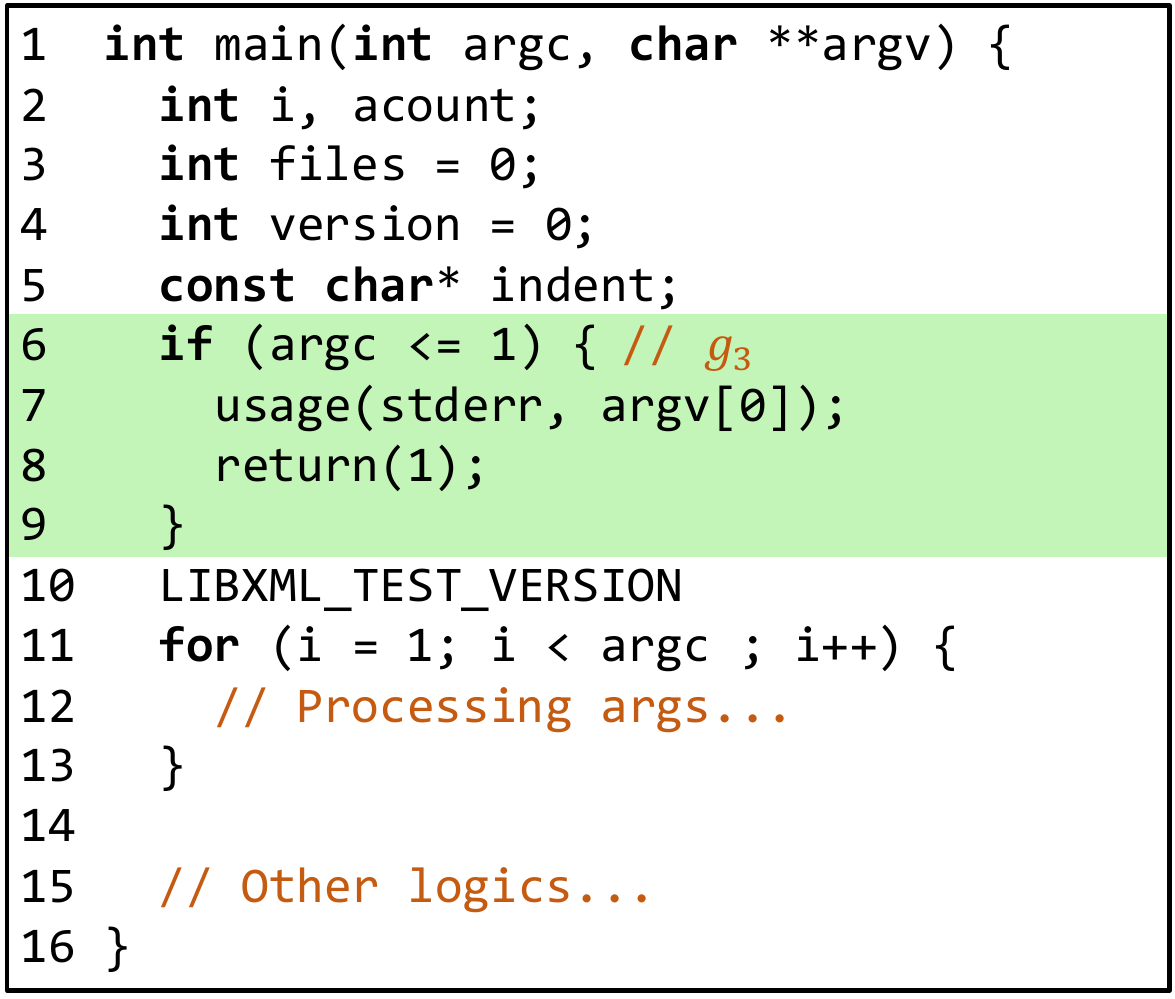}
        \subcaption{An excerpt of \texttt{xmllint.c} exemplifying \textbf{Cond-2}.}
        \label{subfig:example-reckless-xmllint}
    \end{minipage}
    \caption{Two examples of reckless toggles. The code lines relevant to the reckless guards are colored in green.}
    \label{fig:reckless-toggls-examples}
    \Description{How we analyze the hierarchies of conditional guards.}   
\end{figure*}

\examplecounter{}
\figcap{\ref{subfig:example-reckless-crash}} shows a code snippet \texttt{boo()} that contains Crashing-Reckless, where a potential array index out-of-bounds (AIOOB) crash lies on \linecap{12} and $g_2$ is the guard responsible for preventing the crash.
Let \texttt{boo()} be the fuzz target $p$ and denote its guard hierarchy as $H$.
Suppose $g_2$ is now disabled in the guard hierarchy and $G^-=\{g_2\}$. 
Given a seed corpus $C=\{s_1\}$ where $s_1$ is (\texttt{idx=4},\texttt{len=2}). 
An execution of $s_1$ via $p(C,H,G^-)$ will erroneously flow to \linecap{12} and cause an AIOOB crash, guiding us to identify $g_2$ as a reckless guard.

\examplecounter{}
\figcap{\ref{subfig:example-reckless-xmllint}} displays an example of Converging-Reckless, where the code fragment is excerpted from \texttt{xmllint.c}, the main source file of a well-known fuzz target \texttt{xmllint} from the open source project LibXML2 \cite{libxml2}. 
The if statement in \linecap{7} is a reckless guard located (denoted as $g_3$) at the very beginning of the \texttt{main()}, responsible for checking whether the number of arguments is insufficient.
Let $C$ be the seed corpus for fuzzing \texttt{xmllint}.
Normally, most seeds in $C$ have sufficient arguments; they will be blocked by $g_3$, skip the error handling codes in \linescap{8}{9}, and then flow to execute other functional code.
Once $g_3$ is disabled, the situation changes.
Because $g_3$ is at the very beginning of \texttt{main()}, seeds in $C$ will now involuntarily enter the code block guarded by $g_3$ and reach \texttt{return(1)}, resulting in similar coverage.
Since \lineno{1} the results of seed selection become identical after $g_3$ is disabled, which implies that a plausible fixed point is reached, and \lineno{2} re-enabling $g_3$ can break that convergence, we thus identify $g_3$ as a reckless guard.

\begin{algorithm}
	\caption{Crashing-Reckless Collection.} 
	\label{alg:collect-crashing-reckless} 
	\begin{algorithmic}[1] 
		\Require {Fuzz Target $p$, Seed Corpus $C$, Disabled Guards $G^-$.}
		\Ensure {Reckless Guards $R$.}
            \State $R\gets\emptyset, G^-_\text{tmp}\gets \emptyset$, $\mathit{res}\gets\mathit{none}, \mathit{pos}\gets-1, \mathit{len}\gets1$ \label{line:reck-init-start}
            \State $L\gets$ \textsc{AsList}($G^-$) \label{line:reck-init-end}
            \While{$\mathit{pos} + 1 < |G^-|$}
                \State $\tilde{R}\gets$ \textsc{AsSet}($L[\mathit{pos}+1 \ldots \mathit{pos}+1+\mathit{len}]$) \label{line:crash-candi-start}
                \State $G^-_\text{tmp}\gets G^-_\text{tmp}~\cup~\tilde{R}$ \label{line:crash-candi-end}
                \State $\mathit{res} \gets p(C, G^-_\text{tmp})$
                \If{$res$ is $\mathit{crash}$} \label{line:run-crash-candi}
                    \State $G^-_\text{tmp}\gets G^-_\text{tmp}\setminus \tilde{R}$ 
                    \If{$\mathit{len} = 1$}
                        \State $R\gets R~\cup~\tilde{R}$ , $\mathit{pos} \gets \mathit{pos} + 1$ \label{line:determine-crash}
                    \Else
                        \State $\mathit{len} \gets \mathit{len} / 2$ \label{line:half-window}
                    \EndIf
                \Else   
                    \State $\mathit{pos} \gets \mathit{pos} + \mathit{len}$, $\mathit{len} \gets \mathit{len} \times 2$ \label{line:enlarge}
                \EndIf
            \EndWhile
            \State \Return{$R$}
	\end{algorithmic} 
\end{algorithm}

We observe that recognizing reckless guards is challenging and often requires handling with some delay. 
The reasons can be concluded as twofold: 
\lineno{1} The judgment of recklessness is intrinsically delayed. 
The existence of reckless guards can only be realized once the fuzz target is crashed (\textbf{Conseq-1}) or a plausible fixed point is reached (\textbf{Conseq-2}).   
\lineno{2} The synergistic effect among reckless guards.
Some guards may not appear reckless in isolation, but they can become reckless under specific combinations. 
Considering fuzzing the \texttt{boo()} in \figcap{\ref{subfig:example-reckless-crash}} with a seed corpus $C=\{s_1\}$, where $s_1$ is (\texttt{idx=4},\texttt{len=8}).
Disabling either $g_1$ or $g_2$ will not induce the crash, so neither $g_1$ nor $g_2$ will be recognized as reckless alone.
However, when $g_1$ and $g_2$ are disabled together, the value of \texttt{len} will be modified to \texttt{1} in \linecap{5}; at the same time, since the comparison of \texttt{idx} and \texttt{len} in \linecap{11} is disabled, the execution of $s_1$ is now allowed to reach \linecap{12} and trigger the potential crash.
In this example, it may seem feasible to recognize such recklessness using a method like enumerating all possible combinations of $g_1$ and $g_2$.
However, such a brute-force strategy is generally inapplicable in practice, since real-world targets often contain thousands of guards. 

We develop two strategies to tackle the issues mentioned above:
\lineno{1} To address the issue induced by the delay, we adopt a lazy strategy to troubleshoot when signals of recklessness (i.e., crashes or plausible fixed points) occur.
\lineno{2} To address the issue induced by the synergistic effect, we develop two heuristics to recognize the two types of reckless guards accurately and efficiently.

\algcap{\ref{alg:collect-crashing-reckless}} presents the heuristic for collecting Crashing-Reckless, where a binary search-based strategy is adopted.
First, \techname{} prepares for the binary search (\linescap{\ref{line:reck-init-start}}{\ref{line:reck-init-end}}) by \lineno{1} initializing search control variables, including the temporary set of disabled guards $G_\text{tmp}^-$, the result of execution $\mathit{res}$, the starting position of the search $\mathit{pos}$, and the size of the search window $\mathit{len}$; \lineno{2} casting the set of the disabled guards into a variable $L$ of list type to allow for a subscript-based guard location.
After preparation, \techname{} detects reckless guards by repeatedly \lineno{1} adjusting the search window, where $\tilde{R}$ denotes the candidates of Crashing-Reckless (\linescap{\ref{line:crash-candi-start}}{\ref{line:crash-candi-end}}) and \lineno{2} checking the result $\mathit{res}$ of executing $p$ with guards in $G_\text{tmp}^-$ temporarily disabled (\linecap{\ref{line:run-crash-candi}}). 
If the execution is crashed ($\mathit{res}=\mathit{crash}$), the candidates $\tilde{R}$ are then decided reckless depending on whether the window size $\mathit{len}$ is 1:
if so, $\tilde{R}$ is merged into $R$ and $\mathit{pos}$ is moved forward by 1 (\linecap{\ref{line:determine-crash}}); if not, the window is halved to pave the way for a more precise search of Crashing-Reckless candidates (\linecap{\ref{line:half-window}}).
If the execution is not crashed, the guards in the current window are \textit{not} considered reckless; the search is next restarted from the last end position and the window is enlarged twice (\linecap{\ref{line:enlarge}}).

\begin{algorithm}
	\caption{Converging-Reckless Collection.} 
	\label{alg:collect-converging-reckless} 
	\begin{algorithmic}[1] 
		\Require {Fuzz Target $p$, Newly Selected Seeds $S_\text{new}$, Disabled Guards $G^-$.}
		\Ensure {Reckless Guards $R$.}
            \State $R \gets \emptyset$, $M_\pi\gets p(S_\text{new},G^-)$ \label{line:passed-run}
                \For{$g$ in $G^-$} \label{line:conv-travers-start}
                    \If{$M_\pi[g]=1$} \label{line:conv-cond}
                        \State $R \gets R~\cup~\{g\}$
                    \EndIf
                \EndFor \label{line:conv-travers-end}
            \State \Return{$R$}
	\end{algorithmic} 
\end{algorithm}

\algcap{\ref{alg:collect-converging-reckless}} presents the heuristic for collecting Converging-Reckless.
As exemplified in \figcap{\ref{subfig:example-reckless-xmllint}}, converging program behaviors are usually protected by certain guards.
Reaching such behaviors requires first passing the corresponding guards, which establishes a necessary condition (denoted as $\beta$) of being Converging-Reckless. 
Compared with seeking crashing candidates from all disabled guards (which usually requires executing the fuzz target with many different guard combinations), adopting the necessary condition $\beta$ to recognize reckless guards is much more efficient, inspiring us to develop a heuristic based on it.
In \algcap{\ref{alg:collect-converging-reckless}},
\techname{} first executes the fuzz target $p$ with the last selected seeds $S'$, tracking the passing statues of guards as a bitmap $M_\pi$ (\linecap{\ref{line:passed-run}}) and deciding reckless guards by traversing every disabled guard in the set $G^-$ (\linescap{\ref{line:conv-travers-start}}{\ref{line:conv-travers-end}}), where the bits of guards are set to 1 for passed and 0 for unpassed. 

\subsubsection{Formalizing iterative Seed Selection.}
\label{subsubsec:iss}
For iterative seed selection, the basic condition to reach the fixed point is that the given CSS map $\minfunc$ can no longer select any new seed incrementally.
Let us denote the condition as $\rho:S'=S_\text{new}$, where $S'$ is the seeds selected in the last round, and $S_\text{new}$ is the current round.
As discussed in \seccap{\ref{subsubsec:reckless-guard}}, the condition $\rho$ is insufficient; it misses three implicit conditions required by the real fixed point of \techname{}: 
\lineno{1} no reckless guards, since they are false positives and can disrupt the decision of the fixed point; \lineno{2} no newly passed guards and \lineno{3} no outermost guards, because the latter two types of guards are obstacle candidates that imply possible further advancement in iterative seed selection.
To address this, we define \textit{\techname{}-Fixed-Point} below.

\begin{definition}
    \textit{\techname{}-Fixed-Point}. 
    Let $S'$ and $S_\text{new}$ be the set of seeds selected in the last and current rounds.
    Let $R$, $\Pi$, and $\Omega$ denote the sets of reckless guards, passed guards, and outermost guards recognized in the current round.
    \techname{} reaches the fixed point in iterative seed selection if the following condition $\rho'$ is satisfied:
    \begin{equation}
    \label{eq:fixed-point}
        \rho': (S'=S_\text{new}) \land (R=\emptyset) \land (\Pi=\emptyset) \land (\Omega=\emptyset)
    \end{equation}
\end{definition}

\begin{algorithm}
	\caption{Iterative Seed Selection} 
	\label{alg:iterative-seed-selection} 
	\begin{algorithmic}[1] 
		\Require {Seed Corpus $C$, Fuzz Target $p$, CSS map $\minfunc$}
		\Ensure {Selected Seeds $S$}
        \State $S \gets \emptyset, S' \gets \emptyset$, $R \gets \emptyset, G^- \gets \emptyset, O_\text{new} \gets \emptyset$ \label{line:init-seeds-guards}
        \Repeat
            \State $R' \gets \emptyset, \Omega \gets \emptyset$
            \State $S_\text{new}, \Pi, \mathit{res}\gets \minfunc(p(G^-),C)$ \CommentIt{Incremental seed selection.} \label{line:inc-CSS}
            \If{$\mathit{res}$ is $\mathit{crash}$} \CommentIt{Recognizing reckless guards} \label{line:iss-reck-start}
                \State $R' \gets $ \textsc{CollectCrashingReckless}($p,C,G^-$)
            \ElsIf{$S'=S_\text{new}$} 
                \State $R' \gets $ \textsc{CollectConvergingReckless}($p,S_\text{new},G^-$)
            \EndIf \label{line:iss-reck-end}
            \If{$R' \ne \emptyset$}
                \State $R\gets R~\cup~R'$, $G^-\gets G^-\setminus R$ \label{line:iss-update-R}
            \Else
                
                \State $S \gets S~\cup~S_\text{new}$, $S' \gets S_\text{new}$ \CommentIt{Adding newly selected seeds.} \label{line:iss-update-S}
                \State $\Pi \gets\Pi~\setminus(R~\cup~G^-)$ \CommentIt{Recognizing newly passed guards.} \label{line:iss-recognize-new-passed}
                \If{$\Pi \ne \emptyset$} \label{line:iss-handle-passed-start}
                    
                    \State $G^- \gets G^-~\cup~\Pi$, $O_\text{new} \gets O_\text{new}~\cup~\Pi$ \label{line:iss-handle-passed-end}
                \Else   \CommentIt{Recognizing outermost guards}
                    \State $\Omega \gets $ \textsc{CollectOutermost}($p,O_\text{new}$)$~\setminus~R$ \label{line:iss-outermost-start}
                    \If{$\Omega \ne \emptyset$}
                        \State $G^-\gets~G^-\cup~\Omega$, $O_\text{new} \gets \Omega$ \label{line:iss-outermost-refresh}
                    \EndIf \label{line:iss-outermost-end}
                \EndIf \label{line:iss-obstacle-end}
            \EndIf            
        \Until{\textit{timeout} \textbf{or} $\rho'$ is satisfied}
        \State \Return {$S$} 
	\end{algorithmic} 
\end{algorithm}

\algcap{\ref{alg:iterative-seed-selection}} shows the process of iterative seed selection, where \techname{} takes the seed corpus $C$, the fuzz target $p$, and the basic CSS map $\minfunc$ as inputs and outputs a set $S$ of selected seeds once \lineno{1} the time budget is exhausted or \lineno{2} the fixed point $\rho'$ is reached.
To begin with, \techname{} initializes seed sets $S$ and $S'$ as well as guard sets $R,G^-$ and $O_\text{new}$ (\linecap{\ref{line:init-seeds-guards}}).
Specifically, $S$ contains seeds selected through all iterations, $S'$ contains seeds selected in the last round, $R$ contains reckless guards seen so far, $G^-$ contains disabled seeds, and $O_\text{new}$ contains obstacle guards recognized since the last outermost guard recognition.
After that, \techname{} incrementally selects seeds by invoking the given CSS $\minfunc$ (\linecap{\ref{line:inc-CSS}}).
Regarding the results of $\minfunc$, \techname{} checks whether any new reckless guards $R'$ are detected (\linescap{\ref{line:iss-reck-start}}{\ref{line:iss-reck-end}}). 
If so, \techname{} merges $R'$ into $R$, removes the updated $R$ from the obstacle candidates, and advances to the next round (\linecap{\ref{line:iss-update-R}}).  
If not, \techname{} expands $S$ with $S_\text{new}$ (\linecap{\ref{line:iss-update-S}}) and turns to recognize new obstacle candidates.
\techname{} first identifies fresh, non-reckless passed guards by excluding those in $R$ or $G^-$ from $\Pi$ (\linecap{\ref{line:iss-recognize-new-passed}}); if $\Pi\ne\emptyset$, \techname{} recognizes the remains in $\Pi$ as obstacles and then moves to the next round (\linescap{\ref{line:iss-handle-passed-start}}{\ref{line:iss-handle-passed-end}}).
Otherwise, if no passed guards remain, \techname{} turns to handle outermost guards (\linescap{\ref{line:iss-outermost-start}}{\ref{line:iss-outermost-end}}), where the guards in $R$ are excluded from $\Omega$ to prevent recklessness (\linecap{\ref{line:iss-outermost-start}}) and $O_\text{new}$ is refreshed (\linecap{\ref{line:iss-outermost-refresh}}).

\section{Implementation}
\label{sec:impl}
\newcommand{\toolccloc}{1,550}
\newcommand{\toolpycloc}{1,700}
We prototype \techname{} on top of AFL++ (\aflppver{}) and implement it as five components, namely a toggle instrumentor, a metadata extractor, a guard hierarchy analyzer, a reckless guard recognizer, and a seed selection booster.
The toggle instrumentor and guard hierarchy analyzer are implemented as two separate LLVM passes \cite{llvmpass} of around \toolccloc{} lines of C/C++ code, and the other three components are implemented with about \toolpycloc{} lines of Python code \cite{pocoartifact}.

\section{Evaluation}
Our evaluation aims to answer the following research questions (RQs):
\begin{itemize}[leftmargin=*, topsep=3pt]
    \item \rqline{1 (Additional Seeds selected by \techname{})}{Can \techname{} select additional seeds compared with the base CSS tool? Can \techname{} reach the fixed point within the given time budget?}
    \item \rqline{2 (Practical Effectiveness of \techname{})}{With practical time budgets, can \techname{} seeds improve the effectiveness of downstream MGF in terms of code coverage and bug discovery?}
    \item \rqline{3 (Effectiveness with Longer \techname{})}{Can \techname{} produce better seed sets for greybox fuzzing under longer time budgets compared with shorter but practical ones?}
    \item \rqline{4 (Adopting \techname{} vs. Fuzzing Longer)}{Can shorter fuzz campaigns using \techname{} seeds achieve comparable or better results than longer campaigns using seeds selected by basic \cmin{}?}
    \item \rqline{5 (Time Cost of \techname{})}{What are the compositions of the time used by \techname{} in iterative seed selection? How much overhead is caused by each composition?}
\end{itemize}

\subsection{Experimental Setup}
\newcommand{\nmgtargets}{21} 
\newcommand{\ncsstarget}{16} 
\newcommand{\ncovbadtarget}{eight}
\newcommand{\showmap}{\texttt{afl-showmap}}
\subsubsection{Fuzzers \& fuzz targets}
We evaluate \techname{} on \mg{}, a bug-based benchmark that integrates mainstream fuzzers (e.g., AFL++) and various real-world programs \cite{hazimeh2020magma}.
\mg{} is widely adopted in fuzzing research \cite{dipritosem, li2023accelerating, kokkonis2025rosa}; it provides bug-finding ground truths by injecting targets with real vulnerabilities \cite{magmabugs}.
We run experiments on \ntarget{} targets from \mg{} and conduct MGF using AFL++, a state-of-the-art and representative greybox fuzzer.  
The details of targets are shown in \tabcap{\ref{tab:targets}}, where the first column is the IDs of the evaluated targets and the second to fourth columns are project names, versions, and target names obtained from pioneer research \cite{hazimeh2020magma, li2023accelerating}.

\begin{table}
    \small
  \centering
  \caption{Information of the \mg{} fuzz targets and seeds used in the experiments.}
  \label{tab:targets}%
\begin{tabular}{c|l|l|l|r|l}
\hline
\textbf{TID} & \textbf{Project} & \textbf{Version} & \textbf{Fuzz Target} & \boldmath{}\textbf{$|C|$}\unboldmath{} & \textbf{File Type} \\
\hline
T1   & LibSndfile & 1.0.31 & \texttt{sndfile\_fuzzer} & 1223  & Audio \\
\hline
T2   & \multirow{1}[4]{*}{LibXML2} & \multirow{1}[4]{*}{2.9.12} & \texttt{xmllint} & 1268  & \multirow{1}[4]{*}{XML} \\
\cline{1-1}\cline{4-5}T3   &       &       & \texttt{libxml2\_xml\_read\_memory\_fuzzer} & 1268  &  \\
\hline
T4   & SQLite3 & 3.37.0 & \texttt{sqlite3\_fuzz} & 1068  & SQL \\
\hline
T5   & Lua   & 5.4.3 & \texttt{lua} & 1305  & LUA \\
\hline
T6   & Libpng & 1.6.38 & \texttt{libpng\_read\_fuzzer} & 1004  & PNG \\
\hline
T7   & \multirow{1}[4]{*}{LibTIFF} & \multirow{1}[4]{*}{4.3.0} & \texttt{tiffcp} & 1021  & \multirow{1}[4]{*}{TIFF} \\
\cline{1-1}\cline{4-5}T8   &       &       & \texttt{tiff\_read\_rgba\_fuzzer} & 1021  &  \\
\hline
\end{tabular}%

\end{table}%

\subsubsection{Seed corpora}
\label{subsubsec:seed-corpora}
We collect files from two resources to build high-quality seed corpora: 
\lineno{1} \textbf{\mg{}}. We include the seeds that \mg{} integrate for the experimental targets.
\lineno{2} \textbf{External}. For targets with less than 1,000 seeds in \mg{}, we expand their corpora by randomly selecting 1,000 seeds from the dataset published by \citet{herrera2021seed}.
Since this dataset does not include LUA seeds, we prepare a seed corpus for \texttt{lua} by collecting files from five popular LUA projects \cite{luaprojectlist}.
The corpus sizes (i.e., the number of seeds included) are shown in the \textbf{$|C|$} column of \tabcap{\ref{tab:targets}}.

\newcommand{\ntech}{five}
\newcommand{\allss}{\textsc{All}}
\newcommand{\cminplus}{\cmin{}+}
\subsubsection{Seed selection techniques}
Our evaluation involves \ntech{} seed selection techniques, namely \allss{}, \cmin{}, \cminplus{}, \optimin{}, and \techname{}. 
Specifically, \allss{} represents a non-selection strategy that retains all seeds in a given corpus; we include \allss{} to see whether seed selection always brings benefits.
\cmin{} stands for \aflcmin{}, a well-known and state-of-the-art CSS tool integrated in AFL++; we choose \cmin{} as a baseline because it is the base CSS tool of our \techname{} implementation.
\cminplus{} is a variant of \cmin{}, supplemented with $|\Delta|$ randomly selected seeds from the initial corpora, where $|\Delta|$ denotes the number of seeds that are additionally selected by \techname{}; we devise \cminplus{} to examine whether \techname{} actually reveals golden seeds.
\optimin{} is another state-of-the-art CSS tool that solves the seed selection problem using an EvalMaxSAT solver \cite{issta21seed}; we incorporate \optimin{} to explore whether \techname{} is competitive with other state-of-the-art CSS techniques.

\subsubsection{Effectiveness Metrics}
\label{subsubsec:metric}
The effectiveness of a seed selection technique is largely reflected in the performance of the MGF it supports \cite{issta21seed}.
Since existing research commonly evaluates MGF by code coverage and bug discovery \cite{zheng2025mendelfuzz, li2023accelerating, funfuzztosem}, we also adopt them as indicators of effectiveness.
The details are as follows:

\textbf{Code coverage metrics}.
We adopt edge coverage, the built-in coverage measurement of AFL++ \cite{aflreadme}, for coverage analysis.
Specifically, we extract the final edges and average them by \nrepeat{} repetitions, and calculate the Vargha-Delaney $\hat{A}_{12}$ and $p$-values to assess the significance of coverage differences between techniques.
Vargha-Delaney $\hat{A}_{12}$ is a nonparametric statistic that shows the probability of superiority between two independent groups, which is highly recommended and widely adopted in fuzzing \cite{bohme2022benchmarking, metzman2021fuzzbench,schloegel2024sok} and other fields of software engineering \cite{arcuri2014hitchhiker, vargha2000critique}.

\textbf{Bug discovery metrics}.
We evaluate the bug-finding capability of each technique from three perspectives: the number of unique bugs found, the time required to find bugs, and the success rate of bug discovery.
To this end, we first extract raw bug-finding data (e.g., bug IDs and time-to-bugs in every repetition) using a benchmarking tool provided by \mg{} \cite{magmatools}, and then perform further analysis. 
We also analyze the overlaps of the unique bugs triggered by different seed sets. 
Due to the page limit, we only report the bug overlap results in the main text.
Complete bug-finding results can be found in Appendix \cite{pocoappendix}.

\subsubsection{Environments}
We conduct experiments on a workstation with 32 cores (13th Gen Intel(R) Core(TM) i9-13900K) and 32 GB of memory, running Ubuntu 22.04.
Each fuzz campaign is repeated \nrepeat{} times to build statistical significance.
Specific campaign setups are described in \seccap{\ref{subsec:rq2}}\rangeline{}\seccap{\ref{subsec:rq4}}.

\subsection{\rqline{1}{Additional Seeds Selected by \techname{}}}
\label{subsubsec:rq1}

We set the time budget of \techname{} to \overallbudget{} hours and particularly study the number of seeds selected at \lineno{1} two hours, which is a more reasonable budget for practical use, and \lineno{2} end of the \overallbudget{}-hour run.
The seed selection results of \techname{}, \cmin{}, and \optimin{} are displayed in \tabcap{\ref{tab:stats-seed-selection}}.
\techname{} achieves the fixed point $\rho'$ (defined in \equcap{\ref{eq:fixed-point}}) on five of the \ntarget{} targets.
It selects 2\rangeline{}40 more seeds after running two hours on the \ntarget{} targets, and 2\rangeline{}78 more seeds after exhausting all \overallbudget{} hours.
These results show that \techname{} can find additional seeds compared with the base CSS tool \cmin{}. 

According to $\%|S|_\text{2h}$, \techname{} selects 59.6\%\rangeline{}100.0\% of all those it selects in two hours.
Among all \ntarget{} targets, half of the $|S|_\text{2h}$ are achieved  in less than 20 minutes (i.e., $t_\mathcal{P}(|S|_\text{2h}) < 20$ min).
These results suggest that \techname{} can be stopped early in practice to balance efficiency and effectiveness.

\rqbox{
\textbf{Answers to \rqtitle{1}:} 
Given a time budget of \overallbudget{} hours, \techname{} can select 2\rangeline{}78 additional seeds on the \ntarget{} studied targets compared with \cmin{}; it reaches the fixed point on five targets and can achieve 59.6\%\rangeline{}100.0\% of its final selection in less than two hours, implying practical usability. }

\begin{table}
\small
  \centering
  \caption{
  Statistics on the seed sets constructed by different seed selection techniques. 
  \yes{} and \no{} show whether \techname{} reaches the fixed point $\rho'$ within the given time budget.
  $|S|$ shows the number of seeds.
  $|\Delta|$ displays the number of additional seeds compared with \cmin{}.
  The subscript ``A'' denotes all the seeds obtained by \techname{} within the full time budget, while ``2h'' denotes \techname{} seeds at two hours.
  $t_\mathcal{P}(S)$ represents the time \techname{} cost to achieve the seed set $S$.
  $\%|S|_\text{2h}$ is the ratio of seeds obtained in two hours, calculated as $\frac{|S|_\text{2h}}{|S|_\text{A}}\times 100\%$.}
\begin{tabular}{c|c|r|r|r|r|r|r|r|r|r}
\hline
\multirow{1}[4]{*}{\textbf{TID}} & \multicolumn{8}{c|}{\textbf{\techname{}}}                     & \textbf{\cmin{}} & \textbf{\optimin{}} \\
\cline{2-11}      & \boldmath{}\textbf{$\rho'$}\unboldmath{} & \boldmath{}\textbf{$t_\mathcal{P} (|S|_\text{A}$)}\unboldmath{} & \boldmath{}\textbf{$|S|_\text{A}$}\unboldmath{} & \boldmath{}\textbf{$|\Delta|_\text{A}$}\unboldmath{} & \boldmath{}\textbf{$t_\mathcal{P} (|S|_\text{2h}$)}\unboldmath{} & \boldmath{}\textbf{$|S|_\text{2h}$}\unboldmath{} & \boldmath{}\textbf{$|\Delta|_\text{2h}$}\unboldmath{} & \boldmath{}\textbf{$\% |S|_\text{2h}$}\unboldmath{} & \boldmath{}\textbf{$|S|$}\unboldmath{} & \boldmath{}\textbf{$|S|$}\unboldmath{} \\
\hline
T1   & \ding{55} & 1:16:58:03 & 172   & 49    & 00:00:54 & 126   & 3     & 73.3\% & 123   & 84 \\
\hline
T2   & \ding{55} & 4:05:01:56 & 378   & 74    & 00:14:57 & 311   & 7     & 82.3\% & 304   & 170 \\
\hline
T3   & \ding{51} & 4:09:03:09 & 424   & 78    & 01:49:10 & 374   & 28    & 88.2\% & 346   & 206 \\
\hline
T4   & \ding{55} & 0:00:01:51 & 84    & 2     & 00:01:51 & 84    & 2     & 100.0\% & 82    & 14 \\
\hline
T5   & \ding{51} & 0:08:12:01 & 311   & 10    & 00:48:22 & 305   & 4     & 98.1\% & 301   & 109 \\
\hline
T6   & \ding{51} & 0:09:57:31 & 133   & 49    & 00:17:47 & 124   & 40    & 93.2\% & 84    & 26 \\
\hline
T7   & \ding{51} & 0:04:37:30 & 46    & 16    & 00:36:12 & 35    & 5     & 76.1\% & 30    & 8 \\
\hline
T8   & \ding{51} & 0:20:55:07 & 47    & 22    & 01:48:21 & 28    & 3     & 59.6\% & 25    & 12 \\
\hline
\end{tabular}%
  \label{tab:stats-seed-selection}%
\end{table}%

\subsection{\rqline{2}{Practical Effectiveness of \techname{}}}
\label{subsec:rq2}

As implied in \seccap{\ref{subsubsec:rq1}}, \techname{} selects most seeds in two hours, which can be considered a reasonable time budget in practice.
Therefore, to answer \rqtitle{2}, we fuzz each target with the seed sets that \techname{} produces in two hours; the duration of fuzzing is set to 24 hours following pioneer works \cite{bohme2022benchmarking, klees2018evaluating}. 
To make a fair comparison, we prepare \cminplus{} seeds by randomly adding $|\Delta|_\text{2h}$ (displayed in \tabcap{\ref{tab:stats-seed-selection}}) seeds to \cmin{}; the random seeds are from the seed corpus of each target (\seccap{\ref{subsubsec:seed-corpora}}).
Fuzz campaigns with \allss{}, \optimin{}, \cmin{}, and \cminplus{} are also set to 24 hours.

\tabcap{\ref{tab:cov-final}} shows the final edge coverage averaged across \nrepeat{} repetitions.
The results of \cmin{} are placed to the leftmost of the table, serving as a primary baseline for comparisons.
The column \techname{} lists the results of fuzzing using the seed sets created by \techname{} in two hours. 
Cells with results better than \cmin{} are colored gray.
As shown in \tabcap{\ref{tab:cov-final}}, \techname{} outperforms \cmin{} by covering 0.2 to 51.4 more edges (i.e., 0.002\%\rangeline{}0.716\% increase) on seven of the \ntarget{} targets; it also ranks highest among \allss{}, \optimin{}, \cmin{}, and \cminplus{} on five of the targets.

\tabcap{\ref{tab:a12-stats}} shows the Vargha-Delaney $\vga$ and $p$-values obtained under the hypothesis that \allss{}, \optimin{}, \cminplus{}, and \techname{} outperform \cmin{}.
Results indicating small or greater effects are highlighted in gray.
According to \tabcap{\ref{tab:a12-stats}}, \techname{} presents small to medium effects ($0.71 > \vga > 0.56$) compared with \cmin{} on four targets; it only underperforms \cmin{} with a small effect on T6 and performs similarly to \cmin{} on the remaining three targets.  
In contrast, \cminplus{} shows negligible or even negative effects on seven of the \ntarget{} targets. 
These observations reflect that, compared with the additional seeds randomly selected by \cminplus{}, the additional seeds selected by \techname{} are more valuable for improving the performance of MGF, implying the moderate effectiveness of \techname{}.

\begin{figure}
    \centering
    \begin{minipage}[b]{.329\linewidth}
        \centering
        \includegraphics[width=\linewidth]{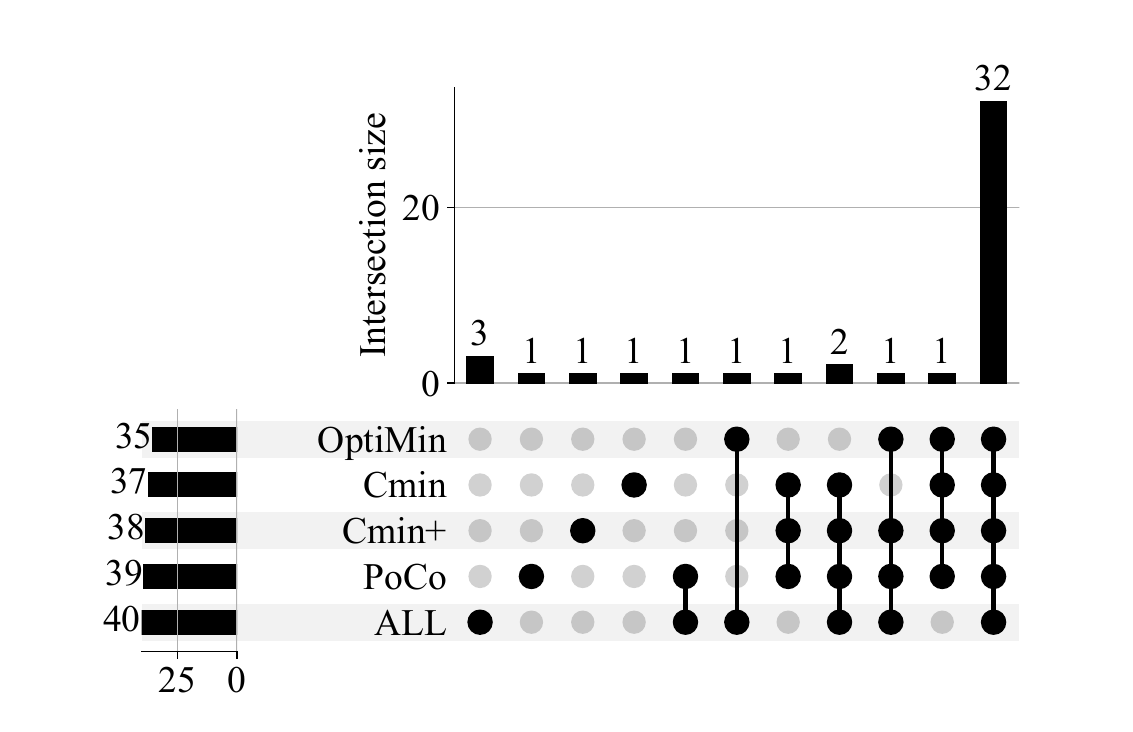}
        \subcaption{
        $t_\mathcal{P} = \text{2h}, t_\mathcal{F} = \text{24h}$
        }
        \label{subfig:bug-rq2}
    \end{minipage}
    \begin{minipage}[b]{.329\linewidth}
        \centering
        \includegraphics[width=\linewidth]{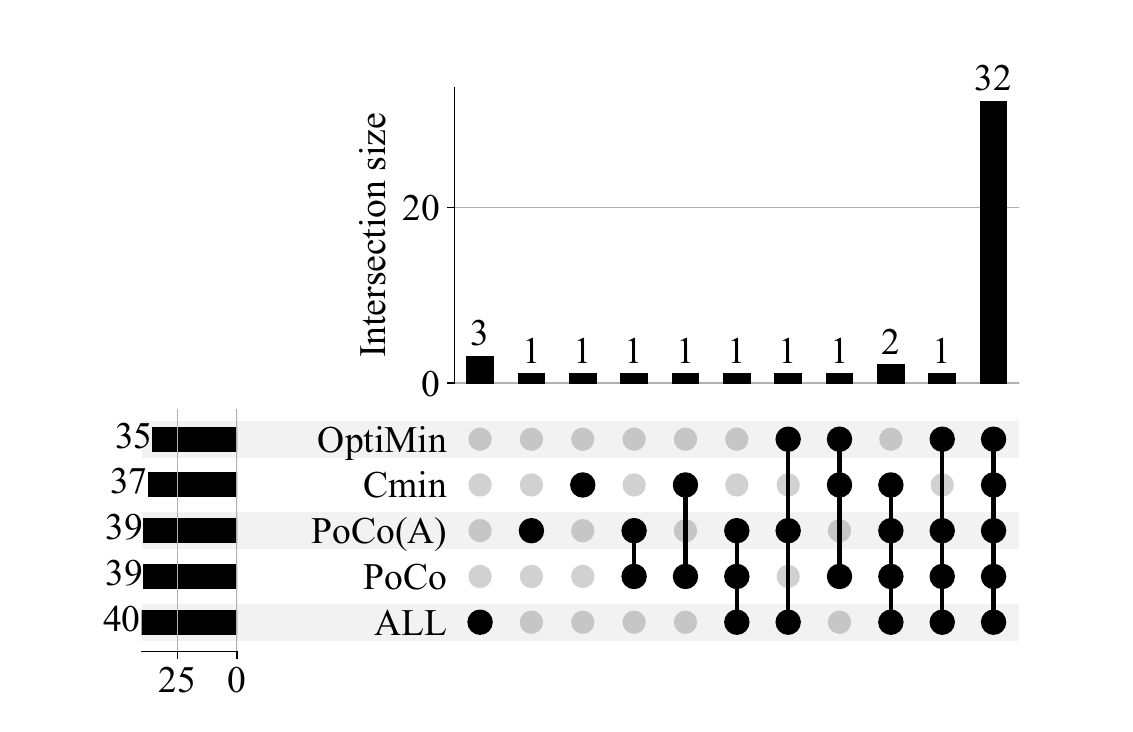}
        \subcaption{
        $t_\mathcal{P} = \text{\overallbudget{}h}, t_\mathcal{F} = \text{24h}$
        }
        \label{subfig:bug-rq3}
    \end{minipage}
    \begin{minipage}[b]{.329\linewidth}
        \centering
        \includegraphics[width=\linewidth]{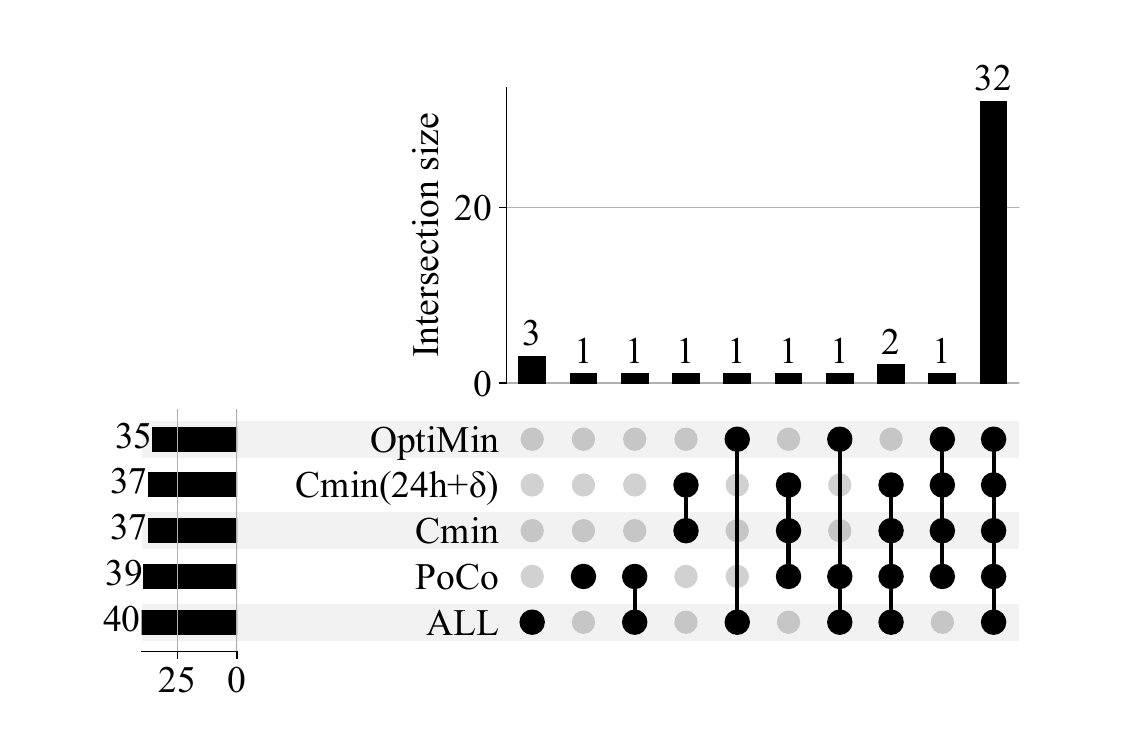}
        \subcaption{
        $t_\mathcal{P} = \text{2h}, \delta=t_\mathcal{F}(|S|_\text{2h})$
        }
        \label{subfig:bug-rq4}
    \end{minipage}
    \caption{UpSet plots illustrating the triggered bugs.
    Results of fuzzing \cmin{} for ``24h$+\delta$'' are denoted as Cmin($\delta$).}
    \label{fig:bugs-upset}
    \Description{Bug overlaps.}
\end{figure}

\figcap{\ref{subfig:bug-rq2}} visualizes the overlaps of the bugs found by \allss{}, \optimin{}, \cmin{}, \cminplus{}, and \techname{} in \nrepeat{} repetitions.
It shows that \techname{} (39 bugs), \allss{} (40 bugs), and \cminplus{} (38 bugs) reveal more bugs than \cmin{} does (37 bugs), including unique bugs that \cmin{} fails to uncover.
These results suggest that the inclusion of additional seeds can benefit MGF in bug discovery. 
In contrast, although \optimin{} minimizes the seed corpora the most (see the $|S|$ column in \tabcap{\ref{tab:stats-seed-selection}}), it ranks the lowest among the studied techniques by only finding 35 bugs. This result implies that \techname{} may make an ``over-minimization'' that impairs bug-finding capabilities of the downstream MGF.

\begin{table}
  \small
  \centering
  \caption{Final edge coverage achieved on the \ntarget{} targets, averaged across \nrepeat{} repetitions.
  $t_\mathcal{F}$ denotes the fuzzing duration.
  Columns 2\rangeline{}7 are results of 24-hour fuzzing, where the results outperforming \cmin{} are colored gray.
  Columns 8 shows results of fuzzing \cmin{} for a longer duration $24\text{h}+\delta$, where the extra duration $\delta = t_\mathcal{P}(|S|_\text{2h})$.
  The results in column 9 that underperform \techname{} are highlighted red.}
\begin{tabular}{c|r|r|r|r|r|r|r}
\hline
\multirow{1}[4]{*}{\textbf{TID}} & \multicolumn{6}{c|}{\boldmath{}\textbf{$t_\mathcal{F} = \text{24h}$}\unboldmath{}} & \multicolumn{1}{c}{\boldmath{}\textbf{$t_\mathcal{F} = \text{24h} + \delta$}\unboldmath{}} \\
\cline{2-8}      & \textbf{\cmin{}} & \textbf{\allss{}} & \textbf{\optimin{}} & \textbf{\cminplus{}} & \textbf{\techname{}} & \boldmath{}\textbf{\techname{}$_\text{A}$}\unboldmath{} & \textbf{\cmin{}} \\
\hline
T01   & 2972.7 & 2953.4 & \cellcolor[rgb]{0.906, 0.902, 0.902} 2976.9 & 2967.3 & \cellcolor[rgb]{0.906, 0.902, 0.902} 2989.1 & 2939.9 & \cellcolor[rgb]{0.961, 0.776, 0.773} 2967.2 \\
\hline
T02   & 8017.8 & 7946.4 & 7891.4 & 7998.0 & \cellcolor[rgb]{0.906, 0.902, 0.902} 8018.0 & \cellcolor[rgb]{0.906, 0.902, 0.902} 8019.1 & \cellcolor[rgb]{0.961, 0.776, 0.773} 8005.6 \\
\hline
T03   & 8738.1 & 8650.9 & 8620.9 & 8736.0 & \cellcolor[rgb]{0.906, 0.902, 0.902} 8755.6 & \cellcolor[rgb]{0.906, 0.902, 0.902} 8744.0 & \cellcolor[rgb]{0.961, 0.776, 0.773} 8749.3 \\
\hline
T04   & 12614.1 & \cellcolor[rgb]{0.906, 0.902, 0.902} 14896.2 & 11161.1 & 12538.1 & \cellcolor[rgb]{0.906, 0.902, 0.902} 12665.5 & \cellcolor[rgb]{0.906, 0.902, 0.902} 12665.5 & 12828.1 \\
\hline
T05   & 4908.9 & 4857.0 & 4778.3 & 4885.0 & \cellcolor[rgb]{0.906, 0.902, 0.902} 4930.9 & \cellcolor[rgb]{0.906, 0.902, 0.902} 4936.6 & \cellcolor[rgb]{0.961, 0.776, 0.773} 4916.6 \\
\hline
T06   & 1462.2 & 1441.9 & 1450.4 & 1459.9 & 1460.6 & \cellcolor[rgb]{0.906, 0.902, 0.902} 1464.9 & 1463.6 \\
\hline
T07   & 4246.5 & 4209.1 & 4174.7 & \cellcolor[rgb]{0.906, 0.902, 0.902} 4264.0 & \cellcolor[rgb]{0.906, 0.902, 0.902} 4276.9 & 4210.5 & \cellcolor[rgb]{0.961, 0.776, 0.773} 4237.7 \\
\hline
T08   & 3850.2 & \cellcolor[rgb]{0.906, 0.902, 0.902} 3894.2 & \cellcolor[rgb]{0.906, 0.902, 0.902} 3854.0 & 3832.3 & \cellcolor[rgb]{0.906, 0.902, 0.902} 3871.5 & \cellcolor[rgb]{0.906, 0.902, 0.902} 3862.4 & \cellcolor[rgb]{0.961, 0.776, 0.773} 3862.5 \\
\hline
\end{tabular}%
  \label{tab:cov-final}%
\end{table}%

\begin{table}
\small 
  \centering
  \caption{
  Vargha–Delaney $A$ measures and one-sided $p$-values (proposed > baseline).
  Each cell is a pair of ``$\hat{A}_{12}$ \tiny{($p$-val)}\small''.
  Columns 2\rangeline{}6 are results of 24-hour fuzzing, where \cmin{} is the baseline method, and others are the proposed methods.
  Columns 7 shows the $\vga$ results of fuzzing \cmin{} for 24h+$\delta$ ($\delta=t_\mathcal{P}(|S|_\text{2h})$), where \cmin{} results are the baselines and \techname{} is the proposed method.
  Results showing small or greater effects (i.e., $\hat{A}_{12}> 0.56$ \cite{arcuri2014hitchhiker}) are shaded gray.}
\begin{tabular}{c|r|r|r|r|r|r}
\hline
\multirow{1}[4]{*}{\textbf{TID}} & \multicolumn{5}{c|}{\boldmath{}\textbf{$t_\mathcal{F}(\cmin{}) = \text{24h}$}\unboldmath{}} & \boldmath{}\textbf{$t_\mathcal{F} = \text{24h} + \delta$}\unboldmath{} \\
\cline{2-7}      & \textbf{\allss{}} & \textbf{\optimin{}} & \textbf{\cminplus{}} & \textbf{\techname{}} & \boldmath{}\textbf{\techname{}$_\text{A}$}\unboldmath{} & \textbf{\cmin{}} \\
\hline
T01   & 0.46 \tiny{(0.63)} & \cellcolor[rgb]{0.906, 0.902, 0.902} 0.57 \tiny{(0.31)} & 0.49 \tiny{(0.53)} & \cellcolor[rgb]{0.906, 0.902, 0.902} 0.68 \tiny{(0.10)} & 0.29 \tiny{(0.94)} & \cellcolor[rgb]{0.906, 0.902, 0.902} 0.74 \tiny{(0.03)} \\
\hline
T02   & 0.21 \tiny{(0.99)} & 0.02 \tiny{(1.00)} & 0.35 \tiny{(0.88)} & 0.53 \tiny{(0.43)} & 0.54 \tiny{(0.40)} & \cellcolor[rgb]{0.906, 0.902, 0.902} 0.68 \tiny{(0.10)} \\
\hline
T03   & 0.03 \tiny{(1.00)} & 0.00 \tiny{(1.00)} & 0.42 \tiny{(0.74)} & \cellcolor[rgb]{0.906, 0.902, 0.902} 0.61 \tiny{(0.21)} & 0.54 \tiny{(0.41)} & 0.54 \tiny{(0.40)} \\
\hline
T04   & \cellcolor[rgb]{0.906, 0.902, 0.902} 1.00 \tiny{(0.00)} & 0.00 \tiny{(1.00)} & 0.42 \tiny{(0.74)} & 0.52 \tiny{(0.45)} & 0.52 \tiny{(0.45)} & 0.35 \tiny{(0.88)} \\
\hline
T05   & 0.40 \tiny{(0.79)} & 0.23 \tiny{(0.98)} & 0.49 \tiny{(0.55)} & 0.47 \tiny{(0.59)} & \cellcolor[rgb]{0.906, 0.902, 0.902} 0.58 \tiny{(0.29)} & 0.48 \tiny{(0.56)} \\
\hline
T06   & 0.07 \tiny{(1.00)} & 0.15 \tiny{(1.00)} & 0.45 \tiny{(0.68)} & 0.43 \tiny{(0.70)} & 0.54 \tiny{(0.41)} & 0.38 \tiny{(0.84)} \\
\hline
T07   & 0.41 \tiny{(0.76)} & 0.29 \tiny{(0.95)} & \cellcolor[rgb]{0.906, 0.902, 0.902} 0.56 \tiny{(0.34)} & \cellcolor[rgb]{0.906, 0.902, 0.902} 0.64 \tiny{(0.16)} & 0.40 \tiny{(0.79)} & \cellcolor[rgb]{0.906, 0.902, 0.902} 0.66 \tiny{(0.12)} \\
\hline
T08   & \cellcolor[rgb]{0.906, 0.902, 0.902} 0.66 \tiny{(0.13)} & 0.53 \tiny{(0.43)} & 0.44 \tiny{(0.69)} & \cellcolor[rgb]{0.906, 0.902, 0.902} 0.57 \tiny{(0.31)} & 0.55 \tiny{(0.37)} & 0.54 \tiny{(0.41)} \\
\hline
\end{tabular}%
  \label{tab:a12-stats}%
\end{table}%

\rqbox{
\textbf{Answers to \rqtitle{2}:} 
With a practical time budget of two hours, \techname{} produces seed sets that can help downstream MGF \lineno{1} achieve slightly higher coverage than the base CSS tool \cmin{}, with $\vga$ effect sizes ranging from small to medium (0.56\rangeline{}0.71); and \lineno{2} find the second most bugs (behind \allss{}) among the five techniques, including one bug missed by other seed sets.
Overall, \techname{} exhibits moderate but limited effectiveness under a practical time budget of two hours.}

\subsection{\rqline{3}{Effectiveness with Longer \techname{}}}
\label{subsec:rq3}

To answer \rqtitle{3}, we adopt all the seeds selected by \techname{} after running \overallbudget{} hours, and name the seed set as \techall{} for distinction.
Note that we also experiment with \cminpA{}, a variant of \cminplus{} that supplements \cmin{} sets with $|\Delta|_\text{A}$ (displayed in \tabcap{\ref{tab:stats-seed-selection}}) random seeds.
The results of \cminpA{} are reported in Appendix \cite{pocoappendix}.

As shown in \tabcap{\ref{tab:cov-final}}, \techA{} covers slightly more edges than \cmin{}, performing similarly to \techname{} in terms of the final coverage.
However, according to the $\vga$ results displayed in \tabcap{\ref{tab:a12-stats}}, \techA{} outperforms \cmin{} on one target, underperforms on two targets, and achieves comparable results on the remaining five; overall, it performs worse than \techname{} in terms of edge coverage.

\figcap{\ref{subfig:bug-rq3}} illustrates the intersections of the bugs found by \allss{}, \optimin{}, \cmin{}, \techname{}, and \techA{} in \nrepeat{} repetitions.
As shown, \techA{} performs as well as \techname{} in the total number of bugs found, with minor differences in the specific bugs identified.
Although \techA{} identifies one bug less than \allss{} in total, it exposes two bugs that \allss{} fails to discover, including one bug that \techname{} misses; these results demonstrate the value of the extra seeds that \techA{} selects beyond \techname{}.
However, considering that \techA{} consumes 60$\times$ the time budget of \techname{} but finds only one additional unique bug, it seems neither worthwhile nor practical. 

\rqbox{
\textbf{Answers to \rqtitle{3}:} 
A longer time budget (e.g., 120 hours) helps \techname{} include additional seeds that sometimes enable MGF to find new bugs.
However, it does not yield seed sets that significantly improve coverage or bug counts over the two-hour time budget.
These results indicate that very long runs provide only limited marginal benefits for \techname{}, making the process not cost-effective.
}


\subsection{\rqline{4}{Adopting \techname{} vs. Fuzzing Longer}}
\label{subsec:rq4}

Given the extra time overhead brought by \techname{}, a natural question arises: Will it be better to use this time budget for longer fuzzing rather than running \techname{}? 
To answer this question, we design ``24h+$\delta$'' campaign setups for \cmin{}, where the extra duration $\delta$ is set to \tpstwoh{}, i.e., the time \techname{} takes to produce the seed sets at the time of two hours (see \tabcap{\ref{tab:stats-seed-selection}}).
In our full evaluation, we also experiment with $\delta$ set to two hours; this part of the results can be found in Appendix \cite{pocoappendix}.
The setups for \allss{}, \optimin{}, \cminplus{}, and \techname{} are consistent with \rqtitle{2} (\seccap{\ref{subsec:rq2}}).  

Column 8 of \tabcap{\ref{tab:cov-final}} reports the final coverage achieved under the ``24h+$\delta$'' campaign setup using \cmin{} corpus, where the results lagging behind \techname{} are marked red.
As shown, \techname{} outperforms \cmin{}(24h+$\delta$) on five targets in terms of the final coverage. 
\tabcap{\ref{tab:a12-stats}} extends the analysis with $\vga$, which shows that \techname{} outperforms \cmin{}(24h+$\delta$) on three targets, performs similarly to it on three other targets, and underperforms it on the remaining two targets.
These results suggest that, under a practical two-hour budget, \techname{} can produce seed sets that enable MGF to achieve final coverage comparable to longer fuzzing campaigns.

\figcap{\ref{subfig:bug-rq4}} displays the bug overlaps achieved by \allss{}, \optimin{}, \techname{}, \cmin{}, and \cmin{}(24h+$\delta$).
The bug-finding results of \cmin{}(24h+$\delta$) are identical to those obtained by fuzzing \techname{} for 24 hours, indicating that the limited additional fuzzing time does not further contribute to bug discovery.
The results also reveal that the seed sets produced by \techname{} enable 24-hour MGF to outperform a longer MGF seeded by \cmin{} by finding two additional bugs, reflecting \techname{}'s effectiveness.

\rqbox{
\textbf{Answers to \rqtitle{4}:} 
compared with the fuzz campaign \cmin{}(24h+$\delta$), \techname{} presents \lineno{1} comparable performance on code coverage and \lineno{2} better performance on bug discovery by fuzzing for only 24 hours. 
These results suggest that allocating a practical time budget to \techname{} can lead to additional bug discoveries, despite no significant improvement in coverage.
}

\subsection{\rqline{5}{Time Cost of \techname{}}}
\label{subsec:rq5}

The extra time overhead is the primary performance bottleneck and also the key optimization opportunity of \techname{}.
To better understand how the time budget is consumed by \techname{}, we conduct additional 24-hour runs of \techname{} and record the time consumed at different stages, including \circledw{1} running the base CSS tool \cmin{}, \circledw{2} handling the Crashing-Reckless, \circledw{3} handling the Converging-Reckless, \circledw{4} parsing the guard hierarchy, and \circledw{5} operating obstacle guards.

\figcap{\ref{fig:timecopo}} visualizes the compositions of \techname{} time.
The main occupancies are \circledw{1} and \circledw{2} (colored blue and orange), which together account for 96.5\%\rangeline{}99.9\% of the total time.
In contrast, the other three time compositions \circledw{3}\rangeline{}\circledw{5} (colored gray) are minor, which only occupy 0.1\%\rangeline{}3.5\% of the total.
\circledw{2} can be further divided into two parts: \lineno{1} the time spent operating the Crashing-Reckless, and \lineno{2} the time spent executing the targets and locating the Crashing-Reckless; the latter dominates, accounting for more than 95.7\% of \circledw{2} and 22.8\%\rangeline{}95.2\% of the total.
compared with the time consumed by running \cmin{} (4.6\%\rangeline{}76.8\%), the time spent executing the targets to handle Crashing-Reckless is even more on six targets.  
This is because when encountering a Crashing-Reckless, the fuzz target may have crashed or fallen into an infinite loop due to disabling guards.  
In contrast to \cmin{}, which terminates execution once \lineno{1} the target crashes or \lineno{2} a short timeout is reached, we assign a significantly longer timeout (i.e., 20 seconds) when handling the Crashing-Reckless.  
As a result, executions for handling Crashing-Reckless can take considerably more time.
The above analysis reveals that optimizing the handling of the Crashing-Reckless (e.g., statically analyzing the Crashing-Reckless beforehand to avoid part of the executions) could significantly improve \techname{}.

\rqbox{\textbf{Answers to \rqtitle{5}:} 
The time overhead of \techname{} comprises five parts, namely running \cmin{}, handling the Crashing-Reckless, handling the Converging-Reckless, parsing the guard hierarchy, and operating obstacle guards.
Most of the time is consumed by the first two parts, which occupy 4.6\%\rangeline{}76.8\% and 23.0\%\rangeline{}95.3\% of the total time, respectively.}

\definecolor{timeblue}{HTML}{1f77b4}
\definecolor{timeorange}{HTML}{ff7f0e}
\definecolor{timegray}{HTML}{7f7f7f}
\newcommand{\colorsquare}[1]{\textcolor{#1}{\rule{1.25ex}{1.25ex}}}
\begin{figure*}
 \centering
    \begin{minipage}[b]{.12\linewidth}
        \centering
        \includegraphics[width=\linewidth]{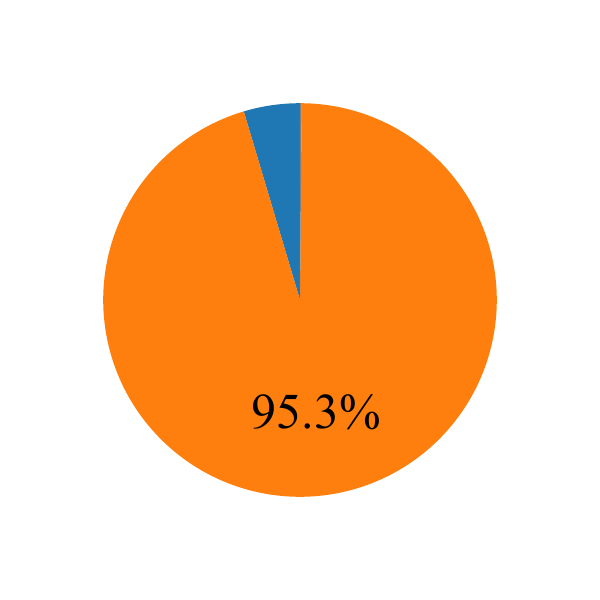}
        \subcaption{T1}
    \end{minipage}
    \begin{minipage}[b]{.12\linewidth}
        \centering
        \includegraphics[width=\linewidth]{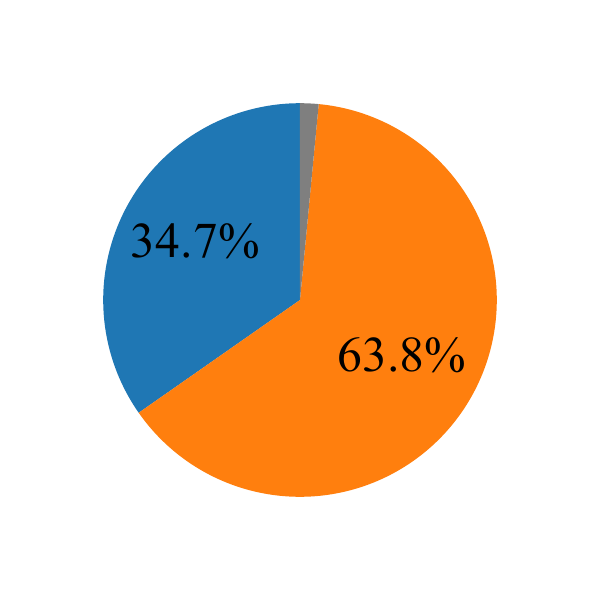}
        \subcaption{T2}
    \end{minipage}
    \begin{minipage}[b]{.12\linewidth}
        \centering
        \includegraphics[width=\linewidth]{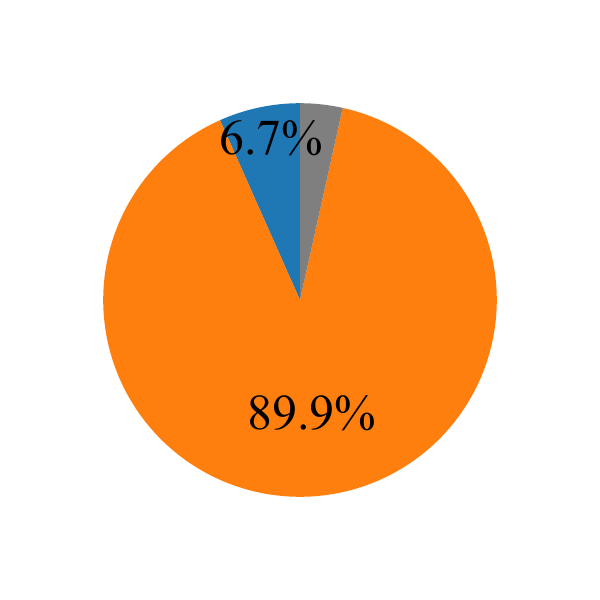}
        \subcaption{T3}
    \end{minipage}
    \begin{minipage}[b]{.12\linewidth}
        \centering
        \includegraphics[width=\linewidth]{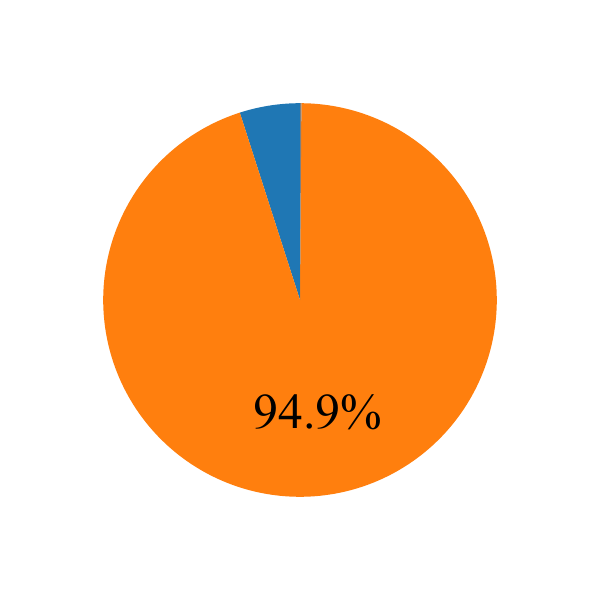}
        \subcaption{T4}
    \end{minipage}
    \begin{minipage}[b]{.12\linewidth}
        \centering
        \includegraphics[width=\linewidth]{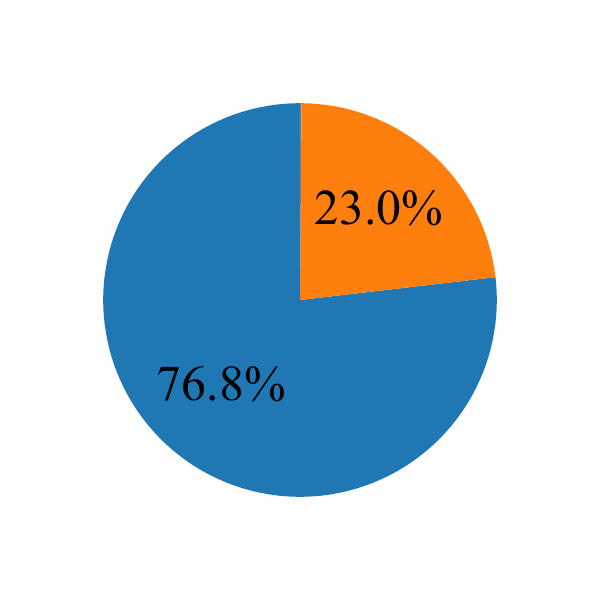}
        \subcaption{T5}
    \end{minipage}
    \begin{minipage}[b]{.12\linewidth}
        \centering
        \includegraphics[width=\linewidth]{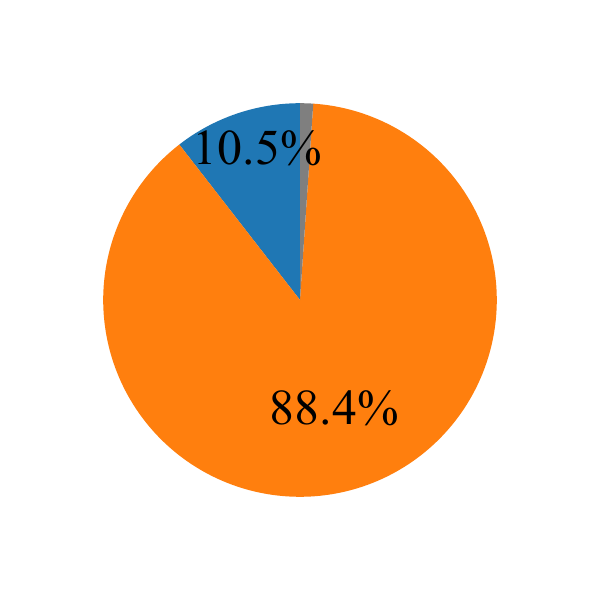}
        \subcaption{T6}
    \end{minipage}
    \begin{minipage}[b]{.12\linewidth}
        \centering
        \includegraphics[width=\linewidth]{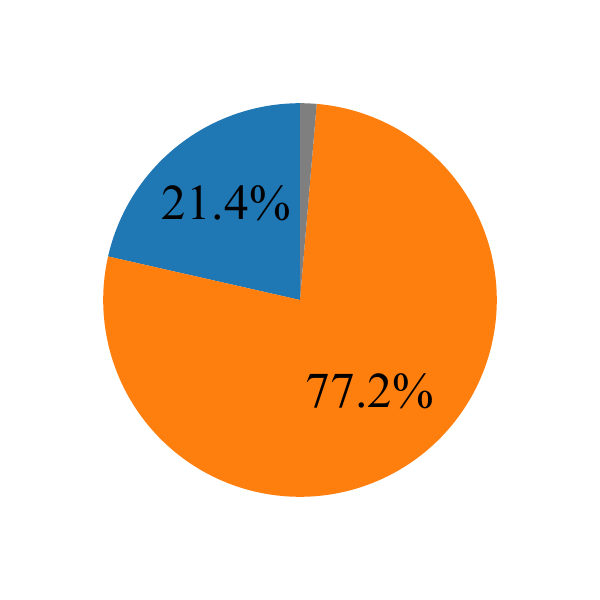}
        \subcaption{T7}
    \end{minipage}
    \begin{minipage}[b]{.12\linewidth}
        \centering
        \includegraphics[width=\linewidth]{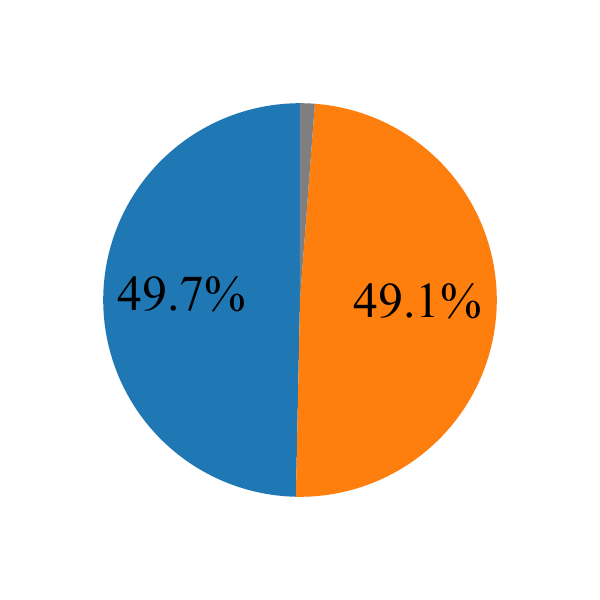}
        \subcaption{T8}
    \end{minipage}
    \caption{
    Time compositions of running \techname{} for 24 hours.
    Blue segments (\colorsquare{timeblue}) denote the time spent running \cmin{}, the base CSS tool of our \techname{} prototype.
    Orange segments (\colorsquare{timeorange}) denote the time used for handling the Crashing-Reckless.
    Gray segments (\colorsquare{timegray}) denote other time usage, including handling the Converging-Reckless, parsing the guard hierarchy, and operating obstacle guards. 
    Labels for segments < 5\% are omitted.}   
    \label{fig:timecopo}
    \Description{\techname{} time compositions.}
\end{figure*}

\section{Discussion}
In this section, we first analyze the early plateaus that are exhibited by \techname{} on some of the targets, followed by a discussion of the lessons learned.
We finally discuss threats to validity.

\subsection{Early Plateaus in \techname{} Seed Selection}
\begin{figure*}
 \centering
    \begin{minipage}[b]{\linewidth}
        \centering        \includegraphics[width=.6\linewidth]{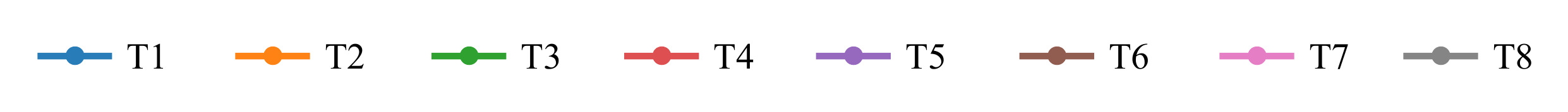}
    \end{minipage}
    \begin{minipage}[b]{.49\linewidth}
        \centering
        \includegraphics[width=.87\linewidth]{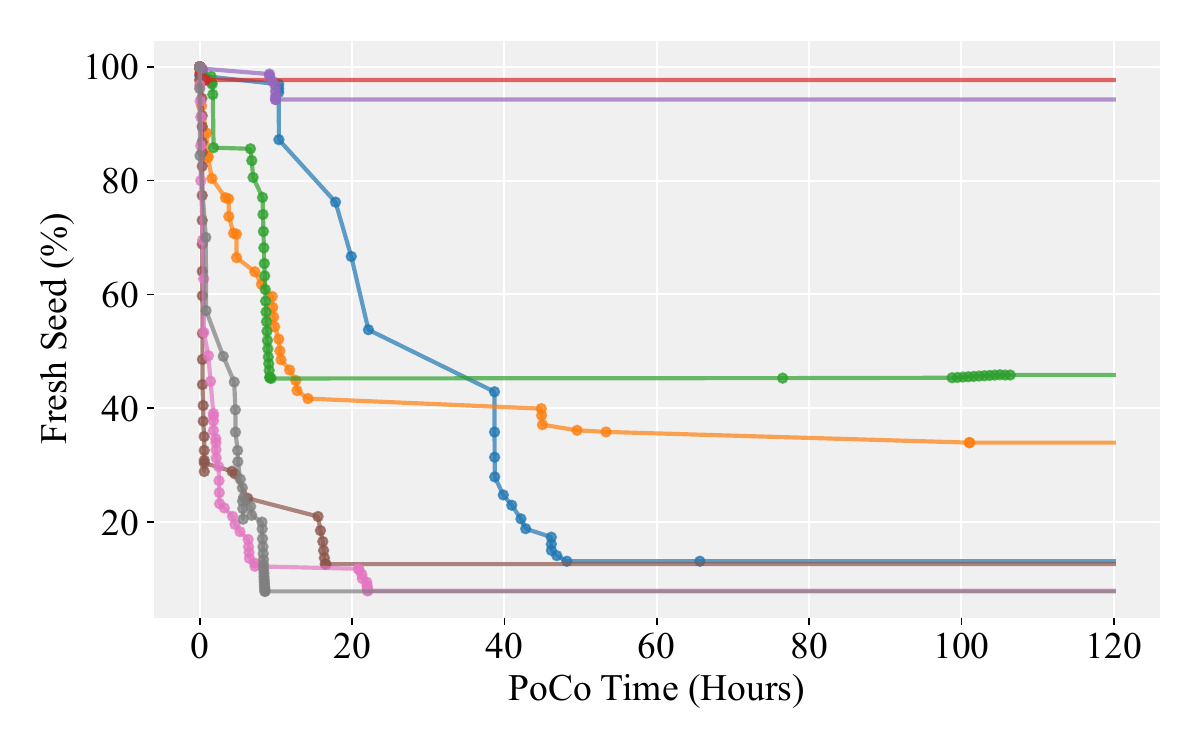}
        \subcaption{
        Ratio of selected fresh seed over time.}
        \label{subfig:fresh-seeds}
    \end{minipage}
    \begin{minipage}[b]{.49\linewidth}
        \centering
        \includegraphics[width=.87\linewidth]{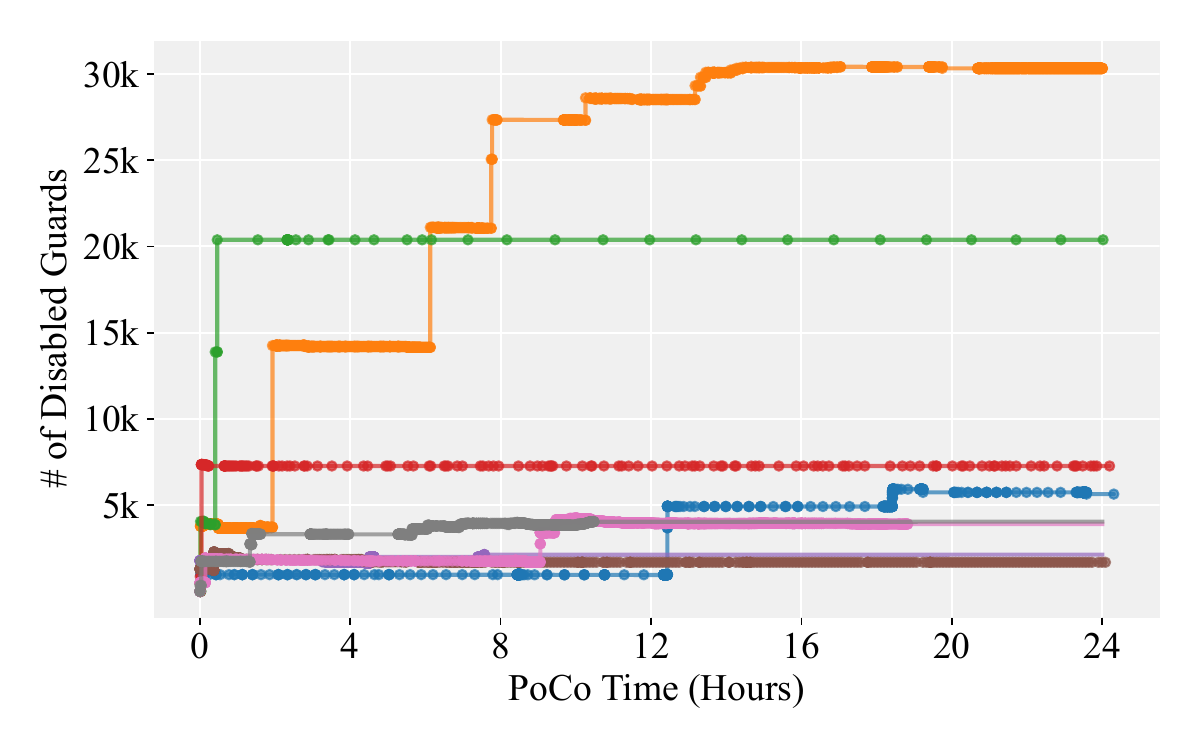}
        \subcaption{
        Number of guards checked over time.}
        \label{subfig:disable-guards}
    \end{minipage}
    \caption{Visualizations and analyses of the plateau phenomenon in \techname{} seed selection.}
    \Description{How we analyze the hierarchies of conditional guards.}   
\end{figure*}

We observe that \techname{} can suffer from slow selection increments or even early plateaus in some targets through our evaluation.
By “early plateaus,” we refer to the phenomenon in which \techname{} stops making progress for several hours after reaching a peak point, such as $t_\mathcal{P}(|S|_\text{2h})$.
The most representative case is T4 (\texttt{sqlite3\_fuzz}), where \techname{} selects all additional seeds within the first two minutes and then remains stagnant for the following hours without reaching the fixed point.
To investigate the early plateaus, we further analyze the seed selection process of \techname{} and examine the characteristics of the studied targets.
We identify two main reasons:

\textbf{Repeated selection of already-included seeds.}
Let $S_1,...,S_n$ be the sets of seeds \techname{} selects at round 1 to $n$.
The fresh seed ratio at round $n$ is calculated as $\frac{|\bigcup_{i=1}^nS_i|}{\Sigma_{i=1}^n|S_i|}$.
\figcap{\ref{subfig:fresh-seeds}} shows the trends of fresh seed ratios, which generally decline over time.
This suggests that when \techname{} disables certain obstacle guards and performs a new round of seed selection, it can select seeds that overlap with those from previous rounds.
One key reason is that some seeds remain dominant even after a new group of obstacle guards is disabled, causing them to be repeatedly selected by the base CSS tool \aflcmin{}.
As a result, \techname{} may repeatedly wander around already-included seeds, leading to slow progress in seed selection or even forming an early plateau.

\textbf{Overhead growth due to increasing disabled guards.}
\figcap{\ref{subfig:disable-guards}} illustrates the number of disabled guards over time, which is generally positively correlated.
By expanding the search space for reckless guards, an increase in disabled guards may also reduce \techname{}'s efficiency in identifying them --- especially the Crashing-Reckless, whose detection is inherently more complex and sensitive to such changes (see \algcap{\ref{alg:collect-crashing-reckless}}).
In addition, the impacts of the increase in disabled guards can vary depending on the characteristics of the targets.
For example, compared with T4, T2 (\texttt{xmllint}) has a larger number of disabled guards and a more significant growth over time.
However, as shown in \figcap{\ref{fig:timecopo}}, \techname{} spends much less time handling Crashing-Reckless on T2 than on T4 (63.8\% vs. 94.9\%).
One explanation for this discrepancy is that the backend libraries of T2 and T4 (i.e, LibXML2 \cite{libxml2} and SQLite3 \cite{sqlite3}) differ significantly in structural complexity.
Specifically, SQLite3 builds upon deeply nested loops (e.g., JOINs) and index scans that may execute hundreds of thousands to millions of times per query\footnote{\url{https://sqlite.org/optoverview.html}}. 
Disabling a guard inside such loops can significantly widen the dynamic execution paths, inflating runtime overhead and thus extending the identification of the Crashing-Reckless. 
In contrast, LibXML2’s loops, mainly for tree traversal and string handling, lack such depth and frequency. 
Therefore, although more guards are disabled in T2 than in T4, the impact shown on \techname{} time is significantly lower for T2.

\subsection{Lessons Learned}
\label{subsec:lessons}
\noindent
Through our extensive evaluation of \techname{}, we have identified both clear limitations and promising aspects. 
We summarize these findings as lessons learned, organized into dark and bright sides.

\textbf{Dark side: \techname{}'s practical use remains limited.}
Our evaluation identifies several limitations of \techname{}.
First, \techname{} does not always select golden seeds.
As shown in RQ3 (\seccap{\ref{subsec:rq3}}), \techname{} does not produce better seed sets (i.e., \techA{}) with a substantially longer time budget (i.e., 120 hours); in fact, the \techA{} seed sets sometimes result in slightly lower code coverage.
This suggests that the usefulness of suppressed golden seeds is inconsistent and requires more fine-grained identification, which is beyond the capability of \techname{}.
Second, \techname{} incurs additional time overhead to select golden seeds. 
As shown in RQ4 (\seccap{\ref{subsec:rq4}}), the benefits of these additional seeds are somewhat limited and could also be achieved by allocating more fuzzing time in some cases.
Despite these limitations, \techname{} still has its place, particularly given that its additional seeds can aid in bug discovery.
Since fuzzing is often deployed as a long-term service (e.g., OSS-Fuzz \cite{serebryany2017oss}) that lasts days to months, practitioners may be willing to allocate two hours to \techname{} to evolve initial seeds.

\textbf{Bright side: Suppressed golden seeds can improve MGF.}
Our evaluation also reveals positive findings.
As shown in RQ2 (\seccap{\ref{subsec:rq2}}) and RQ4 (\seccap{\ref{subsec:rq4}}), the additional seeds selected by \techname{} enable MGF to discover unique bugs that \cmin{} fails to uncover; this effect can even occur with additional seeds that are randomly selected (i.e., \cminplus{}).
This observation confirms that some golden seeds are indeed suppressed because their strengths are not captured by the coverage metric, which aligns with our hypothesis.
Based on these observations, an open question is how to identify and incorporate seeds that appear to be coverage-redundant but are in fact beneficial for fuzzing (see \seccap{\ref{subsec:problem-def}}).
As a complement to existing efforts in constructing coverage-based minimal seed sets, this could represent a potentially interesting direction for the fuzzing community to explore.

\newcommand{\superemph}[1]{\textbf{\textit{#1}}.}
\subsection{Threats to Validity}
\subsubsection*{Internal validity}
The internal validity correlates with the completeness and fairness of the experimental setups \cite{ampatzoglou2019identifying, funfuzztosem}. 
In our evaluation, we set the duration of fuzz campaigns to at least \camphours{} hours to obtain adequate results \cite{bohme2022benchmarking, klees2018evaluating}, repeat each fuzz campaign for \nrepeat{} times to build statistical significance, and prepare high-quality seed corpora by collecting a large number of seeds from well-curated datasets and popular open-source repositories.
All of these efforts greatly alleviate threats to internal validity.

\subsubsection*{External validity}
The external validity is about the generality of the study results \cite{ampatzoglou2019identifying, funfuzztosem}.
In this paper, we \lineno{1} experiment with the representative fuzzer AFL++, \lineno{2} compare \techname{} with seven different baselines, including state-of-the-art techniques like \aflcmin{} and \optimin{}, and \lineno{3} choose \ntarget{} real-world targets of different file formats. 
All of these efforts have consolidated the generality of this evaluation to a large extent.
A more extensive evaluation with distinct-source seeds, more targets, longer fuzzing, and heterogeneous fuzzers (e.g., T-Fuzz \cite{peng2018t}) can further alleviate external threats; we leave these for future work.

\section{Related Work}

\subsubsection*{Fuzzing and its optimization}
Fuzzing is a prevailing technique for software quality assurance \cite{marcel2025software}; it has been widely employed to test various software systems, such as compilers \cite{li2024ubfuzz}, standard libraries \cite{cheng2025rug}, mobile applications \cite{su2021fully}, web browsers \cite{zhou2023towards}, and network protocols \cite{wu2024logos}.  
In addition to applying fuzzing to specific domains, researchers also endeavor to optimize fuzzing by improving its theory and components, such as search strategy \cite{zheng2023fishfuzz}, seed prioritization \cite{dipritosem}, seed mutation \cite{wu2025tumbling}, power scheduling \cite{bohme2020boosting}, and fuzzing throughput \cite{li2023accelerating}. 
Some optimized fuzzers share similar ideas with \techname{}.
For example, T-Fuzz removes hard-to-satisfy checks from binaries to explore deeper program paths \cite{peng2018t};
RedQueen also recognizes the value of input data beyond mere coverage and leverages input-to-state correspondence to guide fuzzing \cite{aschermann2019redqueen};
Laf-Intel performs a program transformation on the fuzz target by splitting multi-byte comparisons into byte-wise checks to increase fuzzer effectiveness \cite{lafintel}.
In contrast to the works above, \techname{} focuses on seed selection---a preprocessing step typically carried out before fuzzing, which represents an orthogonal direction.

\subsubsection*{Preparing seeds for MGF}
Striking a balance between efficiency and effectiveness, mutational greybox fuzzing (MGF) has become a vital branch of the fuzzing family \cite{zhu2022fuzzing}.
The initial seeds are critical for MGF since they act as starting points for MGF, drawing much attention from both academia and industry \cite{meng2024large, liyanage2024extrapolating, qian2023dipri, aflsmart, serebryany2017oss}.
Generally, there are two mainstream methods for preparing initial seeds for MGF.
One is seed generation, where researchers endeavor to obtain input format and create structure-valid seeds using techniques such as grammar learning \cite{wang2017skyfire}, machine learning \cite{lyu2018smartseed}, and large language model \cite{shi2024harnessing}. 
Another more common one is seed selection, where researchers tend to first collect various files from open source data \cite{rebert2014optimizing} and then use tools (which usually are coverage-based) to distill and adopt high-quality subsets as initial seeds \cite{herrera2021seed, afl-cmin}. 
Unlike these works, we investigate the blind spot of coverage-based seed selection and develop an enhancement technique named \techname{} to tackle the blind spot; our work is supplementary to the mentioned works in this respect. 

\section{Conclusion}
In this paper, we present \techname{}, an enhancement technique designed for coverage-based seed selection (CSS). 
We analyze the blind spot of modern CSS both qualitatively and empirically and develop \techname{} to cope with it. 
We also conduct extensive experiments to evaluate \techname{}, performing more than 720 CPU days of fuzzing on \ntarget{} real-world targets and comparing it against seven baselines, including two state-of-the-art CSS tools \aflcmin{} and \optimin{}.
Our extensive evaluation shows that, with a practical time budget of two hours, \techname{} selects valuable additional seeds that can help the downstream MGF mildly outperform other baselines in terms of code coverage and bug discovery.
A longer \techname{} run of 120 hours does not result in more code coverage, but it helps \techname{} select more interesting seeds that can enable downstream MGF to find a different bug.
Although this work reveals some limitations of \techname{}, it offers an alternative seed selection strategy that identifies and includes additional promising seeds rather than simply reducing the seed corpus.
We hope this idea can inspire further research in this direction.

\section{Data-Availability Statement}
We publish \techname{} artifacts and experimental data through Zenodo \cite{pocoartifact}.

\section*{Acknowledgments}
We thank all anonymous reviewers for their thorough reading and insightful comments. 
Their feedback has not only greatly improved this paper but also motivated us to pursue higher-quality research. 
This work is partially supported by the National Natural Science Foundation of China (U24A20337, 62372228).
We also gratefully acknowledge the support from the Huawei--Nanjing University Joint Laboratory for Software New Technology.

\appendix

\section{Appendix: Complete Experimental Results}
\label{app:appendix}

We report complete experimental results in this appendix, including the code coverage and bug discovery results.
The full evaluation consists of experiments with \lineno{1} \nseedsets{} seed sets, namely \allss{}, \optimin{}, \cmin{}, \cminplus{}, \cminpA{}, \techname{}, and \techA{}, where \cminpA{} is constructed by supplementing \cmin{} with $|\Delta|_\text{A}$ seeds randomly selected from \allss{}; and \lineno{2} nine fuzz campaigns, i.e., \nseedsets{} \camphours{}-hour setups with the aforementioned seed sets plus two ``24h+$\delta~(\delta\in\{t_\mathcal{P}(|S|_\text{2h}), 2\text{h}\})$'' campaigns seeded with \cmin{}. 
In the main text, we simplify the presentation of coverage and bug metrics due to space constraints.
Here, we provide the omitted descriptions:

\textbf{Code coverage metrics}.
In addition to the average final coverage and the Vargha-Delaney $\vga$, we analyze the unique coverage.
To do this, we first extract edge IDs using \showmap{}, the coverage tracing tool provided by AFL++, and then analyze coverage overlaps based on edge IDs.

\textbf{Bug discovery metrics}.
We define $t_\mathcal{B}$ as the time to the bug.
To measure how fast a specific bug is covered, we calculate the median value of $t_\mathcal{B}(\text{bug})$ across the repetitions in which the bug is hit.
We also quantify the consistency of discovering a specific bug using the success rate $\gamma(\text{bug})$, which is calculated as:
\begin{equation}
    \gamma(\text{bug}) = \frac{\text{\# bug discoveries}}{\text{\# repeats}} = \frac{\text{\# bug discoveries}}{\nrepeat}
\end{equation}

\subsection{Full Code Coverage Results}
\tabcap{\ref{tab:full-cov-final}} shows the complete final code coverage, which includes the nine campaign setups, while \tabcap{\ref{tab:full-a12-stats}} shows the complete $\vga$ results.
In \tabcap{\ref{tab:full-cov-final}}, columns 2\rangeline{}8 mark results that outperform \cmin{} in gray, while columns 9 and 10 highlight results that fall behind \techname{} in red.
In \tabcap{\ref{tab:full-a12-stats}}, the baseline method is \cmin{} (24h); the results that show small or greater effects are colored gray.
Apart from the results discussed in the main text, \tabcap{\ref{tab:full-cov-final}} and \tabcap{\ref{tab:full-a12-stats}} report two additional observations:
\begin{enumerate}[leftmargin=*, topsep=3pt]
    \item \cminpA{} performs worse than \cmin{} by achieving a similar or even lower final coverage; it also shows no improvement over \cmin{} on any of the eight targets in terms of $\vga$.
    These results suggest that \cminpA{} degrades the quality of the seed set by randomly introducing too many suboptimal seeds.

    \item \techname{} outperforms the setup ``\cmin{}(24h+2h)'' on four targets in terms of averaged final coverage.
    Moreover, \techname{} outperforms ``\cmin{}(24h+2h)'' on three targets in terms of $\vga$, performs similarly to it on three other targets, and underperforms it on the remaining two targets.
    These results suggest that when the fuzzing campaign seeded with \cmin{} is extended to 26 hours, \techname{} (using a practical two-hour budget) can still achieve comparable final coverage.

\end{enumerate}

\tabcap{\ref{tab:full-uniq-cov}} shows the unique coverage results in two groups, where Group-1 is displayed in the left three columns (\cmin{}, \cminplus{}, and \techname{}) and Group-2 is displayed in the right three columns (\cmin{}, \cminpA{}, and \techA{}).
As shown in Group-1, by including additional seeds, \techname{} enables MGF to cover 3\rangeline{}215 edges missed by the other two techniques among the \ntarget{} targets.
As shown in Group-2, \techA{} covers 6\rangeline{}215 edges missed by both \cmin{} and \cminpA{}.
These results indicate that the additional seeds selected by \techname{} allow MGF to explore code regions unseen with \cmin{}, even though the seed sets (i.e., the \techA{} setup) do not yield significant improvements over \cmin{} in terms of final coverage. 

\begin{table}
  \small
  \centering
  \caption{
  Final edge coverage achieved on the \ntarget{} targets, averaged across \nrepeat{} repetitions.
  $t_\mathcal{F}(S)$ denotes the fuzzing duration spent on the specific seed set $S$ produced by one of the studied seed selection techniques.
  Columns 2\rangeline{}8 are results of 24-hour fuzzing, where the results outperforming \cmin{} are colored gray.
  Columns 9 and 10 are results of fuzzing \cmin{} for a longer duration $24\text{h}+\delta$, where the extra duration $\delta \in \{t_\mathcal{P}(|S|_\text{2h}),\text{2h}\}$.
  The results underperforming \techname{} in columns 9 and 10 are highlighted red.}
\resizebox{\textwidth}{!}{
\begin{tabular}{c|r|r|r|r|r|r|r|r|r}
\hline
\multirow{1}[4]{*}{\textbf{TID}} & \multicolumn{7}{c|}{\boldmath{}\textbf{$t_\mathcal{F}(S) = \text{24h}$}\unboldmath{}} & \multicolumn{2}{c}{\boldmath{}\textbf{$t_\mathcal{F}(\cmin{}) = \text{24h}$\tiny{$+\delta$}}\unboldmath{}} \\
\cline{2-10}      & \textbf{\cmin{}} & \textbf{\allss{}} & \textbf{\optimin{}} & \textbf{\cminplus{}} & \boldmath{}\textbf{\cminplus{}$_\text{A}$}\unboldmath{} & \textbf{\techname{}} & \boldmath{}\textbf{\techname{}$_\text{A}$}\unboldmath{} & \boldmath{}\textbf{\tiny{$\delta= t_\mathcal{P}(|S|_\text{2h})$}}\unboldmath{} & \boldmath{}\textbf{\tiny{$\delta=\text{2h}$}}\unboldmath{} \\
\hline
T1   & 2972.7 & 2953.4 & \cellcolor[rgb]{0.906, 0.902, 0.902} 2976.9 & 2967.3 & 2937.0 & \cellcolor[rgb]{0.906, 0.902, 0.902} 2989.1 & 2939.9 & \cellcolor[rgb]{0.961, 0.776, 0.773} 2967.2 & \cellcolor[rgb]{0.961, 0.776, 0.773} 2970.5 \\
\hline
T2   & 8017.8 & 7946.4 & 7891.4 & 7998.0 & 8004.0 & \cellcolor[rgb]{0.906, 0.902, 0.902} 8018.0 & \cellcolor[rgb]{0.906, 0.902, 0.902} 8019.1 & \cellcolor[rgb]{0.961, 0.776, 0.773} 8005.6 & \cellcolor[rgb]{0.961, 0.776, 0.773} 8014.2 \\
\hline
T3   & 8738.1 & 8650.9 & 8620.9 & 8736.0 & 8732.3 & \cellcolor[rgb]{0.906, 0.902, 0.902} 8755.6 & \cellcolor[rgb]{0.906, 0.902, 0.902} 8744.0 & \cellcolor[rgb]{0.961, 0.776, 0.773} 8749.3 & 8757.0 \\
\hline
T4   & 12614.1 & \cellcolor[rgb]{0.906, 0.902, 0.902} 14896.2 & 11161.1 & 12538.1 & 12538.1 & \cellcolor[rgb]{0.906, 0.902, 0.902} 12665.5 & \cellcolor[rgb]{0.906, 0.902, 0.902} 12665.5 & 12828.1 & 12834.2 \\
\hline
T5   & 4908.9 & 4857.0 & 4778.3 & 4885.0 & 4868.8 & \cellcolor[rgb]{0.906, 0.902, 0.902} 4930.9 & \cellcolor[rgb]{0.906, 0.902, 0.902} 4936.6 & \cellcolor[rgb]{0.961, 0.776, 0.773} 4916.6 & \cellcolor[rgb]{0.961, 0.776, 0.773} 4925.6 \\
\hline
T6   & 1462.2 & 1441.9 & 1450.4 & 1459.9 & \cellcolor[rgb]{0.906, 0.902, 0.902} 1462.4 & 1460.6 & \cellcolor[rgb]{0.906, 0.902, 0.902} 1464.9 & 1463.6 & 1463.8 \\
\hline
T7   & 4246.5 & 4209.1 & 4174.7 & \cellcolor[rgb]{0.906, 0.902, 0.902} 4264.0 & 4159.9 & \cellcolor[rgb]{0.906, 0.902, 0.902} 4276.9 & 4210.5 & \cellcolor[rgb]{0.961, 0.776, 0.773} 4237.7 & \cellcolor[rgb]{0.961, 0.776, 0.773} 4254.2 \\
\hline
T8   & 3850.2 & \cellcolor[rgb]{0.906, 0.902, 0.902} 3894.2 & \cellcolor[rgb]{0.906, 0.902, 0.902} 3854.0 & 3832.3 & \cellcolor[rgb]{0.906, 0.902, 0.902} 3854.5 & \cellcolor[rgb]{0.906, 0.902, 0.902} 3871.5 & \cellcolor[rgb]{0.906, 0.902, 0.902} 3862.4 & \cellcolor[rgb]{0.961, 0.776, 0.773} 3862.5 & \cellcolor[rgb]{0.961, 0.776, 0.773} 3862.5 \\
\hline
\end{tabular}%
}
  \label{tab:full-cov-final}%
\end{table}%

\begin{table}
\small 
  \centering
  \caption{
  Vargha–Delaney $A$ measures and one-sided $p$-values (proposed > baseline).
  Each cell is a pair of ``$\hat{A}_{12}$ \tiny{($p$-val)}\small''.
  Columns 2\rangeline{}7 are results of 24-hour fuzzing, where \cmin{} is the baseline method, and others are the proposed methods.
  Columns 8 and 9 are the $\vga$ results of fuzzing \cmin{} for 24h+$\delta$ ($\delta \in \{t_\mathcal{P}(|S|_\text{2h}),\text{2h}\}$), where \cmin{} results are the baselines and \techname{} is the proposed method.
  Results showing small or greater effects (i.e., $\hat{A}_{12}> 0.56$ \cite{arcuri2014hitchhiker}) are shaded gray.}
\begin{tabular}{c|r|r|r|r|r|r|r|r}
\hline
\multirow{1}[4]{*}{\textbf{TID}} & \multicolumn{6}{c|}{\boldmath{}\textbf{$t_\mathcal{F}(|S|) = \text{24h}$}\unboldmath{}} & \multicolumn{2}{c}{\boldmath{}\textbf{$t_\mathcal{F}(\cmin{}) = \text{24h}$\tiny{$+\delta$}}\unboldmath{}} \\
\cline{2-9}      & \textbf{\allss{}} & \textbf{\optimin{}} & \textbf{\cminplus{}} & \boldmath{}\textbf{\cminplus{}$_\text{A}$}\unboldmath{} & \textbf{\techname{}} & \boldmath{}\textbf{\techname{}$_\text{A}$}\unboldmath{} & \boldmath{}\textbf{\tiny{$\delta= t_\mathcal{P}(|S|_\text{2h})$}}\unboldmath{} & \boldmath{}\textbf{\tiny{$\delta=\text{2h}$}}\unboldmath{} \\
\hline
T1   & 0.46 \tiny{(0.63)} & \cellcolor[rgb]{0.906, 0.902, 0.902} 0.57 \tiny{(0.31)} & 0.49 \tiny{(0.53)} & 0.30 \tiny{(0.93)} & \cellcolor[rgb]{0.906, 0.902, 0.902} 0.68 \tiny{(0.10)} & 0.29 \tiny{(0.94)} & \cellcolor[rgb]{0.906, 0.902, 0.902} 0.74 \tiny{(0.03)} & \cellcolor[rgb]{0.906, 0.902, 0.902} 0.71 \tiny{(0.06)} \\
\hline
T2   & 0.21 \tiny{(0.99)} & 0.02 \tiny{(1.00)} & 0.35 \tiny{(0.88)} & 0.44 \tiny{(0.69)} & 0.53 \tiny{(0.43)} & 0.54 \tiny{(0.40)} & \cellcolor[rgb]{0.906, 0.902, 0.902} 0.68 \tiny{(0.10)} & \cellcolor[rgb]{0.906, 0.902, 0.902} 0.56 \tiny{(0.34)} \\
\hline
T3   & 0.03 \tiny{(1.00)} & 0.00 \tiny{(1.00)} & 0.42 \tiny{(0.74)} & 0.42 \tiny{(0.74)} & \cellcolor[rgb]{0.906, 0.902, 0.902} 0.61 \tiny{(0.21)} & 0.54 \tiny{(0.41)} & 0.54 \tiny{(0.40)} & 0.47 \tiny{(0.59)} \\
\hline
T4   & \cellcolor[rgb]{0.906, 0.902, 0.902} 1.00 \tiny{(0.00)} & 0.00 \tiny{(1.00)} & 0.42 \tiny{(0.74)} & 0.52 \tiny{(0.45)} & 0.52 \tiny{(0.45)} & 0.52 \tiny{(0.45)} & 0.35 \tiny{(0.88)} & 0.35 \tiny{(0.88)} \\
\hline
T5   & 0.40 \tiny{(0.79)} & 0.23 \tiny{(0.98)} & 0.49 \tiny{(0.55)} & 0.41 \tiny{(0.76)} & 0.47 \tiny{(0.59)} & \cellcolor[rgb]{0.906, 0.902, 0.902} 0.58 \tiny{(0.29)} & 0.48 \tiny{(0.56)} & 0.46 \tiny{(0.63)} \\
\hline
T6   & 0.07 \tiny{(1.00)} & 0.15 \tiny{(1.00)} & 0.45 \tiny{(0.68)} & 0.48 \tiny{(0.56)} & 0.43 \tiny{(0.70)} & 0.54 \tiny{(0.41)} & 0.38 \tiny{(0.84)} & 0.38 \tiny{(0.84)} \\
\hline
T7   & 0.41 \tiny{(0.76)} & 0.29 \tiny{(0.95)} & \cellcolor[rgb]{0.906, 0.902, 0.902} 0.56 \tiny{(0.34)} & 0.31 \tiny{(0.93)} & \cellcolor[rgb]{0.906, 0.902, 0.902} 0.64 \tiny{(0.16)} & 0.40 \tiny{(0.79)} & \cellcolor[rgb]{0.906, 0.902, 0.902} 0.66 \tiny{(0.12)} & \cellcolor[rgb]{0.906, 0.902, 0.902} 0.59 \tiny{(0.26)} \\
\hline
T8   & \cellcolor[rgb]{0.906, 0.902, 0.902} 0.66 \tiny{(0.13)} & 0.53 \tiny{(0.43)} & 0.44 \tiny{(0.69)} & 0.50 \tiny{(0.52)} & \cellcolor[rgb]{0.906, 0.902, 0.902} 0.57 \tiny{(0.31)} & 0.55 \tiny{(0.37)} & 0.54 \tiny{(0.41)} & 0.54 \tiny{(0.41)} \\
\hline
\end{tabular}%
  \label{tab:full-a12-stats}%
\end{table}%

\begin{table}
\small
  \centering
  \caption{
  Statistics on the unique edges covered by \cmin{}, \cminplus{} (\cminplus{}$_\text{A}$), and \techname{} (\techname{}$_\text{A}$) in \nrepeat{} repetitions.}
\begin{tabular}{c|r|r|r||r|r|r}
\hline
\textbf{TID} & \textbf{\cmin{}} & \textbf{\cminplus{}} & \textbf{\techname{}} & \textbf{\cmin{}} & \boldmath{}\textbf{\cminplus{}$_\text{A}$}\unboldmath{} & \boldmath{}\textbf{\techname{}$_\text{A}$}\unboldmath{} \\
\hline
T1   & 28    & 11    & 52    & 13    & 24    & 54 \\
\hline
T2   & 22    & 30    & 43    & 18    & 31    & 51 \\
\hline
T3   & 24    & 53    & 55    & 44    & 52    & 54 \\
\hline
T4   & 159   & 137   & 215   & 159   & 137   & 215 \\
\hline
T5   & 24    & 21    & 33    & 59    & 22    & 90 \\
\hline
T6   & 4     & 10    & 3     & 3     & 4     & 6 \\
\hline
T7   & 79    & 57    & 136   & 70    & 40    & 59 \\
\hline
T8   & 29    & 53    & 92    & 23    & 79    & 51 \\
\hline
\end{tabular}%
  \label{tab:full-uniq-cov}%
\end{table}%

\subsection{Full Bug Discovery Results}

\begin{table}
\small
  \centering
  \caption{
  Statistics on triggering \mg{} bugs. 
  Each cell displays a ``$t_\mathcal{B}$ \tiny{($\gamma$)}\small'', where $t_\mathcal{B}$ is the median time to bug and $\gamma$ is the success rate of finding a bug in \nrepeat{} repetitions.
  The placeholder ``--'' denotes that a bug is not found in all repetitions (i.e., $\gamma=0$).
  Column $\mathcal{B}$ lists the abbreviated IDs of \mg{} bugs \cite{magmabugs}.
  Cells with results of success rates better than \cmin{} are shaded gray, while the worse ones are marked red.}
\resizebox{\textwidth}{!}{  
\begin{tabular}{c|c|r|r|r|r|r|r|r|r|r}
\hline
\multirow{1}[4]{*}{\textbf{T.}} & \multirow{1}[4]{*}{\boldmath{}\textbf{$\mathcal{B}$}\unboldmath{}} & \multicolumn{7}{c|}{\boldmath{}\textbf{$t_\mathcal{F}(S) = \text{24h}$}\unboldmath{}} & \multicolumn{2}{c}{\boldmath{}\textbf{$t_\mathcal{F}(\textsc{Cmin}) = \text{24h}$\tiny{$+\delta$}}\unboldmath{}} \\
\cline{3-11}      &       & \textbf{\textsc{Cmin}} & \textbf{\allss{}} & \textbf{\textsc{Opt.}} & \textbf{\textsc{Cmin+}} & \boldmath{}\textbf{\textsc{Cmin+}$_\text{A}$}\unboldmath{} & \textbf{\techname{}} & \boldmath{}\textbf{\techname{}$_\text{A}$}\unboldmath{} & \boldmath{}\textbf{$t_\mathcal{P}(|S|_\text{2h})$}\unboldmath{} & \textbf{2h} \\
\hline
\multirow{3.5}[14]{*}{T1} & S01   & 800 \tiny{(1.0)} & 480 \tiny{(1.0)} & 418 \tiny{(1.0)} & 652 \tiny{(1.0)} & 282 \tiny{(1.0)} & 838 \tiny{(1.0)} & 775 \tiny{(1.0)} & 800 \tiny{(1.0)} & 800 \tiny{(1.0)} \\
\cline{2-11}      & S05   & 508 \tiny{(1.0)} & 892 \tiny{(1.0)} & \cellcolor[rgb]{0.961, 0.776, 0.773} 485 \tiny{(0.9)} & 1632 \tiny{(1.0)} & 442 \tiny{(1.0)} & 1168 \tiny{(1.0)} & 1418 \tiny{(1.0)} & 508 \tiny{(1.0)} & 508 \tiny{(1.0)} \\
\cline{2-11}      & S06   & 1258 \tiny{(1.0)} & \cellcolor[rgb]{0.961, 0.776, 0.773} 1440 \tiny{(0.9)} & \cellcolor[rgb]{0.961, 0.776, 0.773} 1025 \tiny{(0.9)} & \cellcolor[rgb]{0.961, 0.776, 0.773} 1620 \tiny{(0.9)} & \cellcolor[rgb]{0.961, 0.776, 0.773} 435 \tiny{(0.9)} & 1380 \tiny{(1.0)} & 1272 \tiny{(1.0)} & 1258 \tiny{(1.0)} & 1258 \tiny{(1.0)} \\
\cline{2-11}      & S07   & 1120 \tiny{(1.0)} & 790 \tiny{(1.0)} & 915 \tiny{(1.0)} & 1510 \tiny{(1.0)} & 852 \tiny{(1.0)} & 1080 \tiny{(1.0)} & 1392 \tiny{(1.0)} & 1120 \tiny{(1.0)} & 1120 \tiny{(1.0)} \\
\cline{2-11}      & S17   & 262 \tiny{(1.0)} & 1375 \tiny{(1.0)} & 270 \tiny{(1.0)} & 188 \tiny{(1.0)} & 375 \tiny{(1.0)} & 422 \tiny{(1.0)} & 272 \tiny{(1.0)} & 262 \tiny{(1.0)} & 262 \tiny{(1.0)} \\
\cline{2-11}      & S20   & 1195 \tiny{(1.0)} & 1380 \tiny{(1.0)} & 635 \tiny{(1.0)} & 1282 \tiny{(1.0)} & 905 \tiny{(1.0)} & 1638 \tiny{(1.0)} & 1018 \tiny{(1.0)} & 1195 \tiny{(1.0)} & 1195 \tiny{(1.0)} \\
\cline{2-11}      & S24   & 1120 \tiny{(1.0)} & 718 \tiny{(1.0)} & 935 \tiny{(1.0)} & 1342 \tiny{(1.0)} & 475 \tiny{(1.0)} & 1075 \tiny{(1.0)} & 1038 \tiny{(1.0)} & 1120 \tiny{(1.0)} & 1120 \tiny{(1.0)} \\
\hline
\multirow{2.5}[10]{*}{T2} & X01   & 50740 \tiny{(0.4)} & \cellcolor[rgb]{0.906, 0.902, 0.902} 65475 \tiny{(0.8)} & \cellcolor[rgb]{0.906, 0.902, 0.902} 56530 \tiny{(0.7)} & \cellcolor[rgb]{0.906, 0.902, 0.902} 65125 \tiny{(0.5)} & \cellcolor[rgb]{0.906, 0.902, 0.902} 53440 \tiny{(0.5)} & \cellcolor[rgb]{0.906, 0.902, 0.902} 42330 \tiny{(0.5)} & \cellcolor[rgb]{0.906, 0.902, 0.902} 41312 \tiny{(0.6)} & 50740 \tiny{(0.4)} & \cellcolor[rgb]{0.906, 0.902, 0.902} 50975 \tiny{(0.5)} \\
\cline{2-11}      & X02   & ---   & ---   & ---   & ---   & ---   & \cellcolor[rgb]{0.906, 0.902, 0.902} 75835 \tiny{(0.1)} & \cellcolor[rgb]{0.906, 0.902, 0.902} 38690 \tiny{(0.1)} & ---   & --- \\
\cline{2-11}      & X09   & 1730 \tiny{(1.0)} & 908 \tiny{(1.0)} & 1368 \tiny{(1.0)} & 980 \tiny{(1.0)} & 1568 \tiny{(1.0)} & 1498 \tiny{(1.0)} & 1212 \tiny{(1.0)} & 1730 \tiny{(1.0)} & 1730 \tiny{(1.0)} \\
\cline{2-11}      & X12   & 40990 \tiny{(0.1)} & \cellcolor[rgb]{0.961, 0.776, 0.773} --- & \cellcolor[rgb]{0.906, 0.902, 0.902} 35765 \tiny{(0.3)} & 35230 \tiny{(0.1)} & \cellcolor[rgb]{0.961, 0.776, 0.773} --- & \cellcolor[rgb]{0.906, 0.902, 0.902} 45732 \tiny{(0.2)} & \cellcolor[rgb]{0.961, 0.776, 0.773} --- & 40990 \tiny{(0.1)} & 40990 \tiny{(0.1)} \\
\cline{2-11}      & X17   & 38 \tiny{(1.0)} & 20 \tiny{(1.0)} & 22 \tiny{(1.0)} & 38 \tiny{(1.0)} & 20 \tiny{(1.0)} & 38 \tiny{(1.0)} & 28 \tiny{(1.0)} & 38 \tiny{(1.0)} & 38 \tiny{(1.0)} \\
\hline
\multirow{3}[12]{*}{T3} & X01   & 1815 \tiny{(1.0)} & 1302 \tiny{(1.0)} & 2130 \tiny{(1.0)} & 1338 \tiny{(1.0)} & 1578 \tiny{(1.0)} & 1315 \tiny{(1.0)} & 2468 \tiny{(1.0)} & 1815 \tiny{(1.0)} & 1815 \tiny{(1.0)} \\
\cline{2-11}      & X02   & ---   & \cellcolor[rgb]{0.906, 0.902, 0.902} 19550 \tiny{(0.1)} & ---   & ---   & \cellcolor[rgb]{0.906, 0.902, 0.902} 48790 \tiny{(0.1)} & ---   & ---   & ---   & --- \\
\cline{2-11}      & X03   & ---   & \cellcolor[rgb]{0.906, 0.902, 0.902} 6750 \tiny{(1.0)} & \cellcolor[rgb]{0.906, 0.902, 0.902} 68800 \tiny{(0.2)} & \cellcolor[rgb]{0.906, 0.902, 0.902} 81625 \tiny{(0.1)} & \cellcolor[rgb]{0.906, 0.902, 0.902} 29475 \tiny{(1.0)} & \cellcolor[rgb]{0.906, 0.902, 0.902} 8590 \tiny{(1.0)} & \cellcolor[rgb]{0.906, 0.902, 0.902} 15900 \tiny{(1.0)} & ---   & --- \\
\cline{2-11}      & X09   & 1832 \tiny{(1.0)} & 1660 \tiny{(1.0)} & 2468 \tiny{(1.0)} & 1955 \tiny{(1.0)} & 1635 \tiny{(1.0)} & 2042 \tiny{(1.0)} & 1370 \tiny{(1.0)} & 1832 \tiny{(1.0)} & 1832 \tiny{(1.0)} \\
\cline{2-11}      & X12   & ---   & \cellcolor[rgb]{0.906, 0.902, 0.902} 49325 \tiny{(0.1)} & \cellcolor[rgb]{0.906, 0.902, 0.902} 64118 \tiny{(0.4)} & ---   & \cellcolor[rgb]{0.906, 0.902, 0.902} 77955 \tiny{(0.3)} & ---   & \cellcolor[rgb]{0.906, 0.902, 0.902} 71975 \tiny{(0.1)} & ---   & --- \\
\cline{2-11}      & X17   & 32 \tiny{(1.0)} & 30 \tiny{(1.0)} & 20 \tiny{(1.0)} & 25 \tiny{(1.0)} & 20 \tiny{(1.0)} & 22 \tiny{(1.0)} & 22 \tiny{(1.0)} & 32 \tiny{(1.0)} & 32 \tiny{(1.0)} \\
\hline
\multirow{4}[16]{*}{T4} & Q02   & 4308 \tiny{(1.0)} & 1740 \tiny{(1.0)} & 3312 \tiny{(1.0)} & 2860 \tiny{(1.0)} & 2838 \tiny{(1.0)} & 4328 \tiny{(1.0)} & 4328 \tiny{(1.0)} & 4308 \tiny{(1.0)} & 4308 \tiny{(1.0)} \\
\cline{2-11}      & Q10   & ---   & \cellcolor[rgb]{0.906, 0.902, 0.902} 51205 \tiny{(0.1)} & ---   & ---   & ---   & ---   & ---   & ---   & --- \\
\cline{2-11}      & Q12   & 36855 \tiny{(0.7)} & \cellcolor[rgb]{0.961, 0.776, 0.773} 33455 \tiny{(0.2)} & \cellcolor[rgb]{0.961, 0.776, 0.773} 37105 \tiny{(0.3)} & \cellcolor[rgb]{0.961, 0.776, 0.773} 39785 \tiny{(0.6)} & \cellcolor[rgb]{0.961, 0.776, 0.773} 40875 \tiny{(0.3)} & \cellcolor[rgb]{0.961, 0.776, 0.773} 37685 \tiny{(0.6)} & \cellcolor[rgb]{0.961, 0.776, 0.773} 37685 \tiny{(0.6)} & 36855 \tiny{(0.7)} & 36855 \tiny{(0.7)} \\
\cline{2-11}      & Q13   & 73575 \tiny{(0.1)} & \cellcolor[rgb]{0.961, 0.776, 0.773} --- & \cellcolor[rgb]{0.961, 0.776, 0.773} --- & \cellcolor[rgb]{0.961, 0.776, 0.773} --- & \cellcolor[rgb]{0.961, 0.776, 0.773} --- & \cellcolor[rgb]{0.961, 0.776, 0.773} --- & \cellcolor[rgb]{0.961, 0.776, 0.773} --- & 73575 \tiny{(0.1)} & 73575 \tiny{(0.1)} \\
\cline{2-11}      & Q14   & 18380 \tiny{(0.9)} & \cellcolor[rgb]{0.906, 0.902, 0.902} 11142 \tiny{(1.0)} & \cellcolor[rgb]{0.906, 0.902, 0.902} 7482 \tiny{(1.0)} & 11740 \tiny{(0.9)} & 11390 \tiny{(0.9)} & \cellcolor[rgb]{0.906, 0.902, 0.902} 12178 \tiny{(1.0)} & \cellcolor[rgb]{0.906, 0.902, 0.902} 12178 \tiny{(1.0)} & 18380 \tiny{(0.9)} & 18380 \tiny{(0.9)} \\
\cline{2-11}      & Q15   & ---   & \cellcolor[rgb]{0.906, 0.902, 0.902} 54535 \tiny{(0.1)} & ---   & ---   & ---   & ---   & ---   & ---   & --- \\
\cline{2-11}      & Q18   & 14118 \tiny{(1.0)} & 2962 \tiny{(1.0)} & 10948 \tiny{(1.0)} & 8005 \tiny{(1.0)} & 6010 \tiny{(1.0)} & 8978 \tiny{(1.0)} & 8978 \tiny{(1.0)} & 14118 \tiny{(1.0)} & 14118 \tiny{(1.0)} \\
\cline{2-11}      & Q20   & 81005 \tiny{(0.1)} & \cellcolor[rgb]{0.906, 0.902, 0.902} 40700 \tiny{(0.7)} & \cellcolor[rgb]{0.961, 0.776, 0.773} --- & 38090 \tiny{(0.1)} & \cellcolor[rgb]{0.961, 0.776, 0.773} --- & \cellcolor[rgb]{0.906, 0.902, 0.902} 80542 \tiny{(0.2)} & \cellcolor[rgb]{0.906, 0.902, 0.902} 80542 \tiny{(0.2)} & 81005 \tiny{(0.1)} & 81005 \tiny{(0.1)} \\
\hline
\multirow{1.5}[6]{*}{T5} & L02   & ---   & ---   & ---   & ---   & ---   & ---   & \cellcolor[rgb]{0.906, 0.902, 0.902} 71365 \tiny{(0.1)} & ---   & --- \\
\cline{2-11}      & L03   & ---   & \cellcolor[rgb]{0.906, 0.902, 0.902} 54690 \tiny{(0.7)} & ---   & ---   & \cellcolor[rgb]{0.906, 0.902, 0.902} 43420 \tiny{(0.1)} & \cellcolor[rgb]{0.906, 0.902, 0.902} 32910 \tiny{(0.1)} & \cellcolor[rgb]{0.906, 0.902, 0.902} 27572 \tiny{(0.2)} & ---   & --- \\
\cline{2-11}      & L04   & 28670 \tiny{(0.5)} & \cellcolor[rgb]{0.961, 0.776, 0.773} 30178 \tiny{(0.2)} & \cellcolor[rgb]{0.961, 0.776, 0.773} --- & \cellcolor[rgb]{0.906, 0.902, 0.902} 37645 \tiny{(0.7)} & \cellcolor[rgb]{0.961, 0.776, 0.773} 64010 \tiny{(0.3)} & \cellcolor[rgb]{0.961, 0.776, 0.773} 22685 \tiny{(0.3)} & \cellcolor[rgb]{0.906, 0.902, 0.902} 50235 \tiny{(0.7)} & \cellcolor[rgb]{0.906, 0.902, 0.902} 48868 \tiny{(0.6)} & \cellcolor[rgb]{0.906, 0.902, 0.902} 48868 \tiny{(0.6)} \\
\hline
\multirow{2}[8]{*}{T6} & P01   & ---   & ---   & ---   & \cellcolor[rgb]{0.906, 0.902, 0.902} 43060 \tiny{(0.1)} & ---   & ---   & ---   & ---   & --- \\
\cline{2-11}      & P03   & 22 \tiny{(1.0)} & 102 \tiny{(1.0)} & 15 \tiny{(1.0)} & 25 \tiny{(1.0)} & 25 \tiny{(1.0)} & 25 \tiny{(1.0)} & 25 \tiny{(1.0)} & 22 \tiny{(1.0)} & 22 \tiny{(1.0)} \\
\cline{2-11}      & P06   & 125 \tiny{(1.0)} & 205 \tiny{(1.0)} & 50 \tiny{(1.0)} & 98 \tiny{(1.0)} & 170 \tiny{(1.0)} & 100 \tiny{(1.0)} & 135 \tiny{(1.0)} & 125 \tiny{(1.0)} & 125 \tiny{(1.0)} \\
\cline{2-11}      & P07   & 11215 \tiny{(1.0)} & \cellcolor[rgb]{0.961, 0.776, 0.773} 1790 \tiny{(0.3)} & 9538 \tiny{(1.0)} & \cellcolor[rgb]{0.961, 0.776, 0.773} 20590 \tiny{(0.9)} & 13038 \tiny{(1.0)} & \cellcolor[rgb]{0.961, 0.776, 0.773} 14515 \tiny{(0.7)} & \cellcolor[rgb]{0.961, 0.776, 0.773} 13965 \tiny{(0.9)} & 11215 \tiny{(1.0)} & 11215 \tiny{(1.0)} \\
\hline
\multirow{4}[16]{*}{T7} & F02   & 22920 \tiny{(0.1)} & \cellcolor[rgb]{0.961, 0.776, 0.773} --- & \cellcolor[rgb]{0.961, 0.776, 0.773} --- & 955 \tiny{(0.1)} & \cellcolor[rgb]{0.961, 0.776, 0.773} --- & 59720 \tiny{(0.1)} & \cellcolor[rgb]{0.961, 0.776, 0.773} --- & 22920 \tiny{(0.1)} & 22920 \tiny{(0.1)} \\
\cline{2-11}      & F05   & 88 \tiny{(1.0)} & 292 \tiny{(1.0)} & \cellcolor[rgb]{0.961, 0.776, 0.773} 65 \tiny{(0.9)} & 148 \tiny{(1.0)} & \cellcolor[rgb]{0.961, 0.776, 0.773} 280 \tiny{(0.9)} & 192 \tiny{(1.0)} & 170 \tiny{(1.0)} & 88 \tiny{(1.0)} & 88 \tiny{(1.0)} \\
\cline{2-11}      & F06   & 88 \tiny{(1.0)} & 212 \tiny{(1.0)} & 78 \tiny{(1.0)} & 148 \tiny{(1.0)} & 350 \tiny{(1.0)} & 192 \tiny{(1.0)} & 172 \tiny{(1.0)} & 88 \tiny{(1.0)} & 88 \tiny{(1.0)} \\
\cline{2-11}      & F07   & 18 \tiny{(1.0)} & 32 \tiny{(1.0)} & 42 \tiny{(1.0)} & 30 \tiny{(1.0)} & 48 \tiny{(1.0)} & 25 \tiny{(1.0)} & 25 \tiny{(1.0)} & 18 \tiny{(1.0)} & 18 \tiny{(1.0)} \\
\cline{2-11}      & F08   & 48270 \tiny{(0.1)} & 27985 \tiny{(0.1)} & 13525 \tiny{(0.1)} & 67110 \tiny{(0.1)} & \cellcolor[rgb]{0.906, 0.902, 0.902} 38542 \tiny{(0.2)} & \cellcolor[rgb]{0.906, 0.902, 0.902} 59120 \tiny{(0.3)} & \cellcolor[rgb]{0.906, 0.902, 0.902} 34795 \tiny{(0.3)} & 48270 \tiny{(0.1)} & 48270 \tiny{(0.1)} \\
\cline{2-11}      & F09   & 4900 \tiny{(0.9)} & 7440 \tiny{(0.9)} & \cellcolor[rgb]{0.961, 0.776, 0.773} 7890 \tiny{(0.7)} & \cellcolor[rgb]{0.906, 0.902, 0.902} 5812 \tiny{(1.0)} & \cellcolor[rgb]{0.961, 0.776, 0.773} 10440 \tiny{(0.8)} & 8655 \tiny{(0.9)} & 5270 \tiny{(0.9)} & 4900 \tiny{(0.9)} & 4900 \tiny{(0.9)} \\
\cline{2-11}      & F12   & 768 \tiny{(1.0)} & 628 \tiny{(1.0)} & 428 \tiny{(1.0)} & 775 \tiny{(1.0)} & 1288 \tiny{(1.0)} & 665 \tiny{(1.0)} & 790 \tiny{(1.0)} & 768 \tiny{(1.0)} & 768 \tiny{(1.0)} \\
\cline{2-11}      & F14   & 1740 \tiny{(1.0)} & 760 \tiny{(1.0)} & 1130 \tiny{(1.0)} & 1620 \tiny{(1.0)} & 1250 \tiny{(1.0)} & 1380 \tiny{(1.0)} & 952 \tiny{(1.0)} & 1740 \tiny{(1.0)} & 1740 \tiny{(1.0)} \\
\hline
\multirow{2.5}[10]{*}{T8} & F02   & 16665 \tiny{(0.5)} & \cellcolor[rgb]{0.961, 0.776, 0.773} 23250 \tiny{(0.4)} & \cellcolor[rgb]{0.906, 0.902, 0.902} 12672 \tiny{(0.8)} & \cellcolor[rgb]{0.906, 0.902, 0.902} 23165 \tiny{(0.7)} & \cellcolor[rgb]{0.961, 0.776, 0.773} 21100 \tiny{(0.3)} & 15320 \tiny{(0.5)} & 7325 \tiny{(0.5)} & \cellcolor[rgb]{0.906, 0.902, 0.902} 20865 \tiny{(0.6)} & \cellcolor[rgb]{0.906, 0.902, 0.902} 20865 \tiny{(0.6)} \\
\cline{2-11}      & F07   & 30 \tiny{(1.0)} & 35 \tiny{(1.0)} & 32 \tiny{(1.0)} & 28 \tiny{(1.0)} & 55 \tiny{(1.0)} & 50 \tiny{(1.0)} & 35 \tiny{(1.0)} & 30 \tiny{(1.0)} & 30 \tiny{(1.0)} \\
\cline{2-11}      & F08   & 46055 \tiny{(0.1)} & \cellcolor[rgb]{0.906, 0.902, 0.902} 17798 \tiny{(0.2)} & 7525 \tiny{(0.1)} & \cellcolor[rgb]{0.906, 0.902, 0.902} 57575 \tiny{(0.3)} & \cellcolor[rgb]{0.961, 0.776, 0.773} --- & 14105 \tiny{(0.1)} & \cellcolor[rgb]{0.906, 0.902, 0.902} 30135 \tiny{(0.3)} & 46055 \tiny{(0.1)} & 46055 \tiny{(0.1)} \\
\cline{2-11}      & F12   & 1255 \tiny{(1.0)} & 1305 \tiny{(1.0)} & 952 \tiny{(1.0)} & 918 \tiny{(1.0)} & 708 \tiny{(1.0)} & 642 \tiny{(1.0)} & 402 \tiny{(1.0)} & 1255 \tiny{(1.0)} & 1255 \tiny{(1.0)} \\
\cline{2-11}      & F14   & 2165 \tiny{(1.0)} & 1290 \tiny{(1.0)} & 2490 \tiny{(1.0)} & 2718 \tiny{(1.0)} & 2058 \tiny{(1.0)} & 1808 \tiny{(1.0)} & 2128 \tiny{(1.0)} & 2165 \tiny{(1.0)} & 2165 \tiny{(1.0)} \\
\hline
\end{tabular}}
  \label{tab:full-bug-stats}%
\end{table}%

\tabcap{\ref{tab:full-bug-stats}} exhibits the detailed bug discovery results; the results with $\gamma$ better (or worse) than \cmin{} are colored gray (or red).
As shown, the bug discovery results of the two \cmin{}(24h+$\delta$) setups are identical to \cmin{}(24h), indicating that the limited additional fuzzing time (e.g., two hours) does not contribute further to bug discovery.
compared with \cmin{}, \techname{} improves $\gamma$ on seven bugs while lowering $\gamma$ on four other bugs, suggesting that \techname{} changes the nature of bug exposure by including additional seeds.
\techA{} performs similarly to \techname{} in the total number of bugs found, with minor differences in the specific bugs identified.
Specifically, \techA{} exposes two bugs (i.e., T3-X12 and T5-L02) that \techname{} fails to reveal, while also missing two bugs (i.e., T2-X12\footnote{X12 (XML012) is a bug planted in LibXML2, the backbone library of T2 and T3; Similar situations also exist in LibTIFF.} and T7-F02) that \techname{} finds.
These differences indicate that incorporating additional seeds can be a double-edged sword: while doing so can promote MGF to discover some new bugs, it can also lower MGF's success rate in uncovering other bugs.
We also observe that \techA{},  \techname{}, \cminpA{}, all of them expose bugs that \cmin{} fails to do on T5 (\texttt{lua}), where \techA{} hits two additional bugs while \techname{} and \cminpA{} each obtain one.
This observation suggests that T5 is a target that can benefit from suppressed golden seeds, confirming our initial hypothesis behind \techname{}.

Similar results are also observed on \allss{}---the baseline stands for the non-selection strategy, which uses the entire seed corpus to initialize the downstream MGF.
The fuzz campaigns seeded with \allss{} achieve better $\gamma$ on 10 of the bugs while decreasing $\gamma$ on the other eight bugs.
The results on \techname{}, \techA{}, and \allss{} imply that including additional seeds can both \lineno{1} improve the chances that MGF finds certain bugs that can be hard to find by fully-minimized seed sets (e.g., \cmin{} and \optimin{} in our evaluations), and \lineno{2} lower the possibilities of uncovering another set of bugs.
Combined with the results on unique coverage (\tabcap{\ref{tab:full-uniq-cov}}), this observation suggests that additional seeds, such as those included by \techname{}, \techA{}, and \allss{}, provide data that can steer MGF toward distinct code regions, even if they do not directly contribute new coverage by themselves.

\begin{figure}
    \centering
    \includegraphics[width=.5\linewidth]{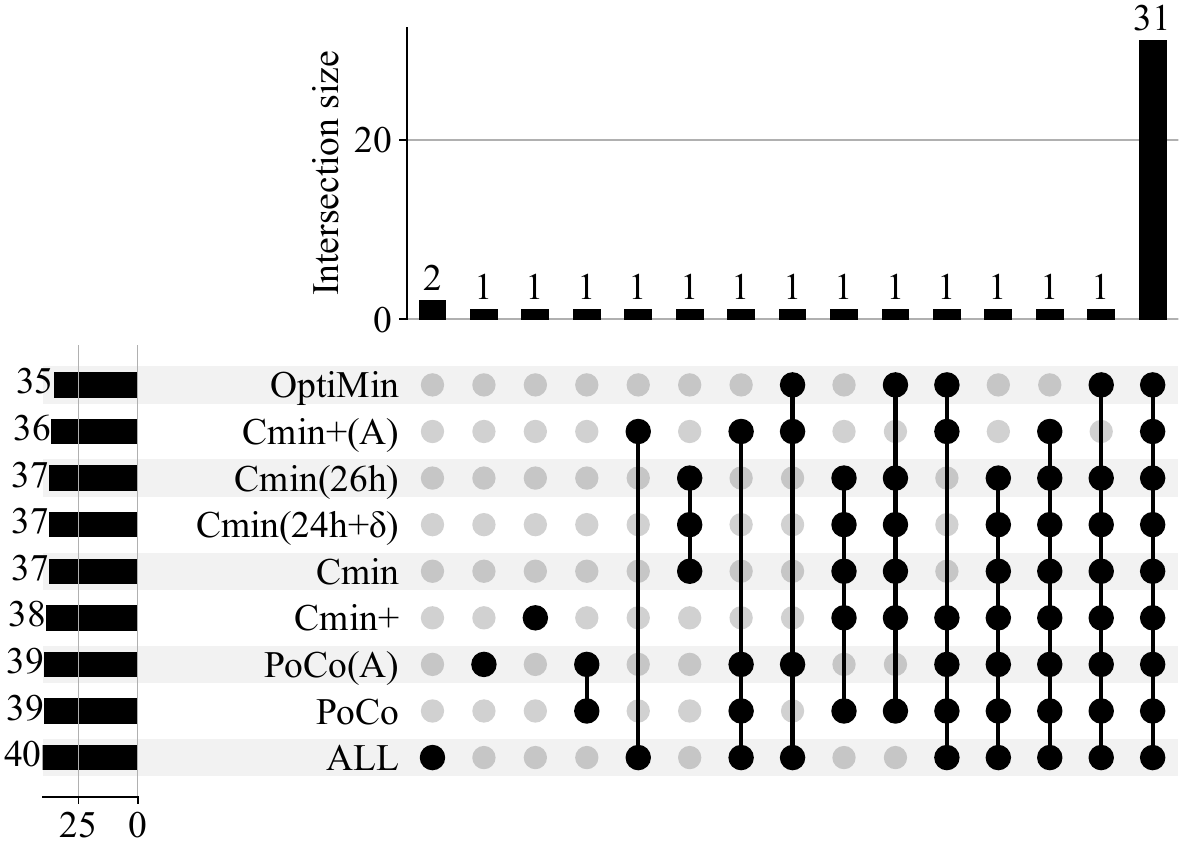}
    \caption{UpSet plots illustrating the overlaps of the bugs triggered by all campaigns in \nrepeat{} repetitions.}
    \label{fig:full-bugs-upset}
    \Description{Full Bug overlaps.}
\end{figure}

\figcap{\ref{fig:full-bugs-upset}} displays the rankings (where lower positions indicate more bugs found) and overlaps of the bug discovery results across the nine campaign setups.
As shown, \techname{}, \allss{}, and \cminplus{} reveal more bugs than \cmin{} does, including unique bugs that \cmin{} fails to uncover, suggesting that the inclusion of additional seeds can benefit MGF in bug discovery.
\techname{} and \techA{} trigger 39 bugs and are ranked behind \allss{}, which finds a total of 40 bugs.
Although the two \techname{} variants slightly underperform in terms of the number of bugs found, they still expose bugs (one for \techname{} and two for \techA{}) that \allss{} fails to detect; this highlights the value of the extra seeds contributed by \techname{}.
\cminpA{} performs poorly in bug discovery: Not only does it underperform \cmin{} by finding one less bug in total, but also it fails to find any unique bug solely.
This result suggests that random seeds can ``pollute'' the seed sets produced by a CSS tool like \cmin{}, which in turn reflects the value of the additional seeds selected by \techA{} (\techname{}).
Regarding \optimin{}, although it minimizes the seed corpora the most (\tabcap{\ref{tab:stats-seed-selection}}), it ranks the lowest among the studied techniques in terms of the number of bugs found.
This result implies that \optimin{} may make an ``over-minimization'' that impairs MGF's bug-finding capabilities.

\bibliographystyle{ACM-Reference-Format}
\bibliography{bibfile}
\end{document}